\documentclass[a4paper,11pt]{article}
\usepackage{amsmath,amssymb,amsthm,amsfonts}    % AMS math packages
\usepackage{mathrsfs,bm}                        % Bold math symbols
\usepackage{dsfont}
\usepackage{bm}
\usepackage{eurosym}                            % Euro symbol
\usepackage{xcolor}                             % Color text management
\usepackage{url,hyperref}                       % Urls and hypertexts (for online files) management
\usepackage{float}                              % Floating objects management
\usepackage[small,bf]{caption}                  % Advanced caption management
\usepackage{graphicx}
\usepackage{epstopdf}                           % Figures management
\usepackage{subfigure}
\graphicspath{{./Figures/}}                     % Figures path.
\usepackage{psfrag}                             % Allows to overlay EPS figures with arbitrary LaTeX constructions
\usepackage{epsfig}                             % Encapsulated postscript management
\usepackage{booktabs,multirow,longtable}        % Advanced table management
\usepackage[titletoc,page]{appendix}
\usepackage[english]{babel}

\makeatletter \@addtoreset{equation}{section} \makeatother

\begin{document}
% --- Definitions of new Latex commands ---
%
% REMIND: always keep a single file with all definitions, beware of multiple copies.

% --- Text ---
\newcommand{\ie}{i.e.~} 	                    % i.e. with non-breakable space
\newcommand{\eg}{e.g.~} 	                    % e.g. with non-breakable space
\newcommand{\rhs}{r.h.s.~} 	                    % r.h.s. with non-breakable space
\newcommand{\wrt}{w.r.t.~} 	                    % w.r.t. with non-breakable space
\newcommand{\old}[1]{\textcolor{blue}{#1}}      % old text, blue, strike-through tex
\newcommand{\new}[1]{\textcolor{red}{#1}}       % new text, red, strike-through tex
\newcommand{\comment}[1]{\textcolor{green}{#1}} % comments, green, strike-through tex
\newcommand{\textbox}[1]{\mbox{\textit{#1}}} % text inside math

% --- Mathematical Definitions ---
\newtheorem{theorem}{Theorem}[section]
\newtheorem{definition}[theorem]{Definition}
\newtheorem{definitions}[theorem]{Definition}
\newtheorem{proposition}[theorem]{Proposition}
\newtheorem{remark}[theorem]{Remark}
\newtheorem{corollary}[theorem]{Corollary}
\newtheorem{example}[theorem]{Example}
\newtheorem{lemma}[theorem]{Lemma}

% --- Environments ---
\newenvironment{system}{\left\lbrace\begin{array}{@{}l@{}}}{\end{array}\right.}

% --- Double delimitators ---
\newcommand{\Parenthesis}[1]{\left( #1 \right)}         % (xxx)
\newcommand{\Brack}[1]{\left\lbrack #1 \right\rbrack}   % [xxx]
\newcommand{\Brace}[1]{\left\lbrace #1 \right\rbrace}   % {xxx}
\newcommand{\Abs}[1]{\left\lvert #1 \right\rvert}       % |xxx|
\newcommand{\Norm}[1]{\left\lVert #1 \right\rVert}      % ||xxx||
\newcommand{\Mean}[1]{\left\langle #1 \right\rangle}    % <xxx>
\newcommand{\QuoteDouble}[1]{``#1''}                    % double quotation marks
\newcommand{\QuoteSingle}[1]{`#1'}                      % single quotation marks

% --- Math operators ---
\newcommand{\beq}{\begin{equation}}         % \begin{equation}
\newcommand{\eeq}{\end{equation}}           % \end{equation}
\newcommand{\Max}{\mbox{Max}}                                   % Max operator
\newcommand{\Min}{\mbox{Min}}                                   % Min operator
\newcommand{\Derp}[2]{\frac{\partial #1}{\partial #2}}          % partial derivative, 1st order
\newcommand{\DerpXX}[2]{\frac{\partial^2 #1}{\partial #2 ^2}}   % partial derivative, 2nd order
\newcommand{\DerpXY}[3]{\frac{\partial^2 #1}{\partial #2 \partial #3}}    % partial derivative, 2nd order, cross
\newcommand{\half}[0]{\frac{1}{2}}                                                  % one half
\newcommand{\ind}[1]{\mathbf{1}_{#1}} 	                                            % indicator
\newcommand{\E}[1]{\mathbb{E}\left[#1\right]}                                       % expectation
\newcommand{\Econd}[2]{\mathbb{E}\left[#1\;\middle \vert\;\mathcal{F}_{#2}\right]}  % conditional expectation

% --- Sets ---
\newcommand{\Nset}{\mathbb{N}}      % Natural numbers set
\newcommand{\Zset}{\mathbb{Z}}      % Integer numbers set
\newcommand{\Qset}{\mathbb{Q}}      % Rational numbers set
\newcommand{\Rset}{\mathbb{R}}      % Real numbers set

% --- Finance ---
\newcommand{\Black}{\mbox{Black}}				    % Black's formula

% --- Financial Instruments ---
\newcommand{\Depo}{\mbox{\textbf{Depo}}}			% Deposit
\newcommand{\FRA}{\mbox{\textbf{FRA}}}			    % Forward Rate Agreement
\newcommand{\Futures}{\mbox{\textbf{Futures}}}      % Futures
\newcommand{\ZCB}{\mbox{ZCB}}					    % Zero Coupon Bond
\newcommand{\Swap}{\mbox{\textbf{Swap}}}			% Swap
\newcommand{\Swaplet}{\mbox{\textbf{Swaplet}}}		% Swaplet
\newcommand{\IRS}{\mbox{\textbf{IRS}}}			    % Interest Rate Swap
\newcommand{\IRSlet}{\mbox{\textbf{IRSlet}}}        % Interest Rate Swaplet
\newcommand{\OIS}{\mbox{\textbf{OIS}}}              % OIS (Overnight Indexed Swap)
\newcommand{\OISlet}{\mbox{\textbf{OISlet}}}        % OISlet
\newcommand{\BSwap}{\mbox{\textbf{BSwap}}}          % Basis Swap
\newcommand{\BSwaplet}{\mbox{\textbf{BSwaplet}}}    % Basis Swaplet
\newcommand{\IRBS}{\mbox{\textbf{IRBS}}}            % Interest Rate Basis Swap
\newcommand{\IRBSlet}{\mbox{\textbf{IRBSlet}}}      % Interest Rate Basis Swaplet
\newcommand{\CCS}{\mbox{\textbf{CCS}}}              % Cross Currency Swap
\newcommand{\CCSwap}{\mbox{\textbf{CCSwap}}}        % Cross Currency Swap
\newcommand{\CCSwaplet}{\mbox{\textbf{CCSwaplet}}}  % Cross Currency Swaplet
\newcommand{\CCSlet}{\mbox{\textbf{CCSlet}}}        % Cross Currency Swaplet
\newcommand{\Caplet}{\mbox{\textbf{Caplet}}}        % Caplet
\newcommand{\Floorlet}{\mbox{\textbf{Floorlet}}}    % Floorlet
\newcommand{\cf}{\mbox{\textbf{cf}}}                % Caplet/Floorlet
\newcommand{\CAP}{\mbox{\textbf{Cap}}}              % Cap
\newcommand{\Floor}{\mbox{\textbf{Floor}}}          % Floor
\newcommand{\CF}{\mbox{\textbf{CF}}}                % Cap/Floor
\newcommand{\Swaption}{\mbox{\textbf{Swaption}}}    % Swaption
\newcommand{\CMSlet}{\mbox{\textbf{CMSlet}}}        % CMSlet
\newcommand{\CMS}{\mbox{\textbf{CMS}}}              % CMS
\newcommand{\CMScf}{\mbox{\textbf{CMScf}}}          % CMS caplet/floorlet
\newcommand{\FXFwd}{\mbox{\textbf{FXFwd}}}          % Forex forward

\title{\textbf{Pricing and Risk Management\\
with High-Dimensional Quasi Monte Carlo \\
and Global Sensitivity Analysis}}

\author{Marco Bianchetti\footnote{Market Risk Management, Banca Intesa Sanpaolo, Piazza G. Ferrari 10, 20121, Milan, Italy, \texttt{marco.bianchetti@intesasanpaolo.com}, corresponding author.}, Sergei Kucherenko\footnote{Imperial College, London, UK, \texttt{s.kucherenko@imperial.ac.uk}}, and Stefano Scoleri\footnote{Iason Ltd, Italy, \texttt{stefano.scoleri@gmail.com}}}

\date{\today}
\maketitle

\begin{abstract}
\noindent We review and apply Quasi Monte Carlo (QMC) and Global
Sensitivity Analysis (GSA) techniques to pricing and risk
management (greeks) of representative financial instruments of
increasing complexity. We compare QMC vs standard Monte Carlo (MC)
results in great detail, using high-dimensional Sobol' low
discrepancy sequences, different discretization methods, and
specific analyses of convergence, performance, speed up,
stability, and error optimisation for finite differences greeks.
We find that our QMC outperforms MC in most cases, including the
highest-dimensional simulations and greeks calculations, showing
faster and more stable convergence to exact or almost exact
results. Using GSA, we are able to fully explain our findings in
terms of reduced effective dimension of our QMC simulation,
allowed in most cases, but not always, by Brownian bridge
discretization. We conclude that, beyond pricing, QMC is a very
promising technique also for computing risk figures, greeks in
particular, as it allows to reduce the computational effort of
high-dimensional Monte Carlo simulations typical of modern risk
management.
\end{abstract}

\newpage
\tableofcontents

\vspace{2cm} \noindent \textbf{JEL classifications}: C63, G12,
G13.

\vspace{1cm} \noindent \textbf{Keywords}: derivative, option,
European, Asian, barrier, knock-out, cliquet, greeks, Brownian
bridge, global sensitivity analysis, Monte Carlo, Quasi Monte
Carlo, random, pseudo random, quasi random, low discrepancy,
Sobol', convergence, speed-up

\vspace{1cm} \noindent \textbf{Acknowledgements}: M.B.
acknowledges fruitful discussions with many colleagues at
international conferences, in Risk Management of Intesa Sanpaolo,
and in Financial Engineering and Trading Desks of Banca IMI.

\vspace{1cm} \noindent \textbf{Disclaimer}: the views expressed
here are those of the authors and do not represent the opinions of
their employers. They are not responsible for any use that may be
made of these contents.
\newpage

\section{Introduction}
\label{SecIntro} Nowadays market and counterparty risk measures,
based on multi-dimensional, multi-step Monte Carlo simulations,
are very important tools for managing risk, both on the front
office side, for sensitivities (greeks) and credit, funding,
capital valuation adjustments (CVA, FVA, KVA, generically called
XVAs) and on the risk management side, for risk measures and
capital allocation. Furthermore, they are typically required for
regulatory risk internal models and validated by regulators. The
daily production of prices and risk measures for large portfolios
with multiple counterparties is a computationally intensive task,
which requires a complex framework and an industrial approach. It
is a typical high budget, high effort project in banks.
\par
In the past decades, much effort was devoted to the application of
Monte Carlo techniques\footnote{The Monte Carlo method was coined
in the 1940s by John von Neumann, Stanislaw Ulam and Nicholas
Metropolis, working on nuclear weapons (Manhattan Project) at Los
Alamos National Laboratory \cite{Neu1951}. Metropolis suggested
the name Monte Carlo, referring to the Monte Carlo Casino, where
Ulam's uncle often gambled away his money \cite{Met1987}. Enrico
Fermi is believed to have used some kind of ``manual simulation''
in the 1930s, working out numerical estimates of nuclear reactions
induced by slow neutrons, with no computers
\cite{LASL1966,Met1987}.} to derivatives pricing
\cite{Boy1977,Boy1997,Jac01,Gla03}. The main reason is that
complex financial instruments usually cannot be priced through
analytical formulas, and the computation of high-dimensional
integrals is required. Monte Carlo simulation is, then, a common
way to tackle such problems, since it reduces integration to
function evaluations at many random points and to averaging on
such values. As a result, virtually any product can be easily
priced in any dimension. However, this method is rather time
consuming and the convergence rate is slow, since the root mean
square error decays as $N^{-1/2}$, where $N$ is the number of
sampled points. Various ``variance reduction'' techniques exist,
which can improve the efficiency of the simulation, but they don't
modify the convergence rate \cite{Jac01,Gla03}.
\par
Quasi Monte Carlo represents a very efficient alternative to
standard Monte Carlo, capable to achieve, in many cases, a faster
convergence rate and, hence, higher accuracy
\cite{Jac01,Gla03,MonFer1999,SobKuc05a,SobKuc05b,Wan09,KucBal11,SobAso12}.
The idea behind Quasi Monte Carlo methods is to use, instead of
pseudo-random numbers (PRN), low discrepancy sequences (LDS, also
known as quasi-random numbers) for sampling points. Such LDS are
designed in such a way that the integration domain is covered as
uniformly as possible, while PRN are known to form clusters of
points and always leave some empty areas. Indeed, the very random
nature of PRN generators implies that there is a chance that newly
added points end up near to previously sampled ones, thus they are
wasted in already probed regions which results in rather low
convergence. On the contrary, LDS ``know'' about the positions of
previously sampled points and fill the gaps between them. Among
several known LDS, Sobol' sequences have been proven to show
better perfomance than others and for this reason they are widely
used in Finance \cite{Jac01,Gla03}. However, it is also known that
construction of efficient Sobol' sequences heavily depends on the
so-called initial numbers and therefore very few Sobol' sequence
generators show good efficiency in practical tests \cite{SobAso12}.
\par
Compared to Monte Carlo, Quasi Monte Carlo techniques also have
some disadvantages. Firstly, there is no ``in sample'' estimation
of errors: since LDS are deterministic, there is not a notion of
probabilistic error. There have been developed some techniques,
known under the name of randomized Quasi Monte Carlo, which
introduce appropriate randomizations in the construction of LDS,
opening up the possibility of measuring errors through a
confidence interval while preserving the convergence rate of Quasi
Monte Carlo \cite{Gla03}. The drawback is the sacrifice of
computational speed and, often, of some precision. Secondly,
effectiveness of Quasi Monte Carlo depends on the integrand
function,  and, most importantly, the convergence rate can depend
on the dimensionality of the problem. The latter can be seen as a
big obstacle, since many problems in financial engineering
(especially in risk management) are known to be high-dimensional.
However, many financial applications have been reported where
Quasi Monte Carlo outperforms standard Monte Carlo even in the
presence of very high dimensions
\cite{PasTrau1995,PapPas1999,CafMor1997,KreMer1998a,KreMer1998b,KucSha07,SobAso12}.
This fact is usually explained by a reduced effective dimension of
the problem, with respect to its nominal dimension. The concept of
effective dimensions was introduced in \cite{CafMor1997}. It was
suggested that QMC is superior to MC if the effective dimension of
an integrand is not too large. The notion is based on the ANalysis
Of VAriances (ANOVA). In \cite{Lemieux2000} it was shown how the
ANOVA components are linked to the effectiveness of QMC
integration methods. It is important to measure the effective
dimension in order to predict the efficiency of a Quasi Monte
Carlo algorithm. Moreover, various techniques can be used to
reduce effective dimension and, thus, improve efficiency: this is
possible because the effective dimension\footnote{Actually, the
effective dimension in the truncation sense can be reduced in this
way. See Section \ref{SecGSA} for the formal definition of
effective dimensions.} can vary by changing the order in which the
variables are sampled. The optimal way to achieve this can be a
hard task, it could depend on the specific model and a general
solution is not known at present. One popular choice in the
financial literature on path-dependent option pricing
\cite{CafMor1997,KucSha07} is to apply the Brownian bridge
discretization to the simulation of the underlying stochastic
process, which is based on the use of conditional distributions.
Unlike the standard discretization, which generates values of the
Brownian motion sequentially along the time horizon, the Brownian
bridge discretization first generates the Brownian motion value at
the terminal point, then it fills a midpoint using the value
already found at the terminal point and then subsequent values at
the successive midpoints using points already simulated at
previous steps. In terms of QMC sampling, this simulation scheme
means that the first coordinate of the QMC vector is used to
simulate the terminal value of the Brownian motion, while
subsequent coordinates are used to generate intermediate points.
There are many studies which show that superior performance of the
QMC approach with the Brownian bridge discretization in comparison
with the standard discretization using MC or QMC sampling, in
application \eg to Asian options \cite{CafMor1997,KucSha07}.
However, it was pointed in \cite{Pap01} that, in some cases, the
Brownian bridge can perform worse than the standard discretization
in QMC simulation. The big question is how to know with certainty
which numerical scheme will provide superior efficiency in QMC
simulation. Global Sensitivity Analysis (GSA) is the answer.
\par
GSA is a very powerful tool in the analysis of complex models as
it offers a comprehensive approach to model analysis. Traditional
sensitivity analysis, called \textit{local} within the present
context, applied to a function $f(x)$ is based on specifying a
point $x_0$ in the function domain and then computing a derivative
$\Derp{f}{x}$ at $x = x_0$. GSA instead does not require to
specify a particular point $x_0$ in the domain, since it explores
the whole domain (hence the name \textit{global}). It also
quantifies the effect of varying a given input (or set of inputs)
while all other inputs are varied as well, providing a measure of
interactions among variables. GSA is used to identify key
parameters whose uncertainty most affects the output. This
information can be used to rank variables, fix unessential
variables and decrease problem dimensionality. Reviews of GSA can
be found in \cite{SobKuc05b} and \cite{Sal10}. The variance-based
method of global sensitivity indices developed by Sobol' \cite{Sob01}
became very popular among practitioners due to its efficiency and
easiness of interpretation. There are two types of Sobol'
sensitivity indices: the main effect indices, which estimate the
individual contribution of each input parameter to the output
variance, and the total sensitivity indices, which measure the
total contribution of a single input factor or a group of inputs.
\par
For modelling and complexity reduction purposes, it is important
to distinguish between the model \textit{nominal} dimension and
its \textit{effective} dimension. The notions of effective
dimension in the truncation and superposition sense were
introduced in \cite{CafMor1997}. Further, Owen added
the notion of ``average dimension'' which is more practical from
the computational point of view \cite{LiuOwe06}. Definitions and evaluations of
effective dimensions are based on the knowledge of Sobol'
sensitivity indices. Quite often complex mathematical models have
effective dimensions much lower than their nominal dimensions. The
knowledge of model effective dimensions is very important since it
allows to apply various complexity reduction techniques. In the
context of quantitative Finance, GSA can be used to estimate
effective dimensions of a given problem. In particular, it can
assess the efficiency of a particular numerical scheme (such as
the Brownian bridge or standard discretizations).
\par
The paper is organized as follows: Section \ref{SecMC} contains a brief review on Quasi Monte Carlo methodology and on Low Discrepancy Sequences, with particular emphasis on financial applications. Section \ref{SecGSA} introduces GSA and
the notions of effective dimensions, establishing a link with QMC efficiency. In Section \ref{SecRes} we present the
results of prices and sensitivities (greeks) computation for selected payoffs: both GSA and convergence analysis are
performed, with the purpose to compare MC and QMC efficiencies via a thorough error analysis. Finally, conclusions and directions of future work are given in Section \ref{SecConclusions}.
In particular we propose to apply our methodology to risk management issues, where a faster and smoother convergence would represent a great advantage in terms of both computational effort and budget.
Some technical details are discussed in the Appendices.

\section{Monte Carlo and Quasi Monte Carlo Methods in Finance}
\label{SecMC}

\subsection{General motivation}
\label{SecGM} In Finance, many quantities of interest, such as
prices and greeks, are defined as expectation values under a given
probability measure, so their evaluation requires the computation
of multidimensional integrals of a (generally complicated)
function.
\par
Let's consider a generic financial instrument written on a single
asset $S$ with a single payment date $T$. We denote the
instrument's payoff at time $T$ as $\mathcal{P}(S_t,\bm{\theta})$,
where $S_t$ is the underlying asset value at time $t\in [0,T]$,
and $\bm{\theta}$ is a set of relevant parameters, including
\emph{instrument} parameters, such as strikes, barriers, fixing
dates of the underlying $S$, callable dates, payment dates, etc.,
described in the contract, and \emph{pricing} parameters, such as
interest rates, volatilities, correlations, etc., associated with
the pricing model.
\par
Using standard no-arbitrage pricing theory, see \eg \cite{Duf01},
the price of the instrument at time $t=0$ is given by
\begin{gather}\label{opt:price}
V_0(\bm{\theta})=\mathbb{E}^Q[D(0,T)\mathcal{P}(S_t,\bm{\theta})|\mathcal{F}_0],\\
D(0,T)= \exp{\Parenthesis{-\int_0^T r(t)dt}},
\end{gather}
where $\Parenthesis{\Omega,\mathcal{F},Q}$ is a probability space with risk-neutral probability measure $Q$ and filtration $\mathcal{F}_t$ at time $t$, $\mathbb{E}^Q[\,\cdot\,]$ is the expectation with respect to $Q$, $D(0,T)$ is the stochastic discount factor, and $r(t)$ is the risk-neutral short spot interest rate.
Notice that the values of $S$ at intermediate times $t$ before final payment date $T$ may enter into the definition of the payoff
$\mathcal{P}$.
\par
In order to price the financial instrument,
we assume a generic Wiener diffusion model for the dynamics of the underlying asset $S$,
\begin{equation}
\label{SDE}
dS_t = \mu\Parenthesis{t,S_t}dt+\sigma\Parenthesis{t,S_t}dW^P_t,
\end{equation}
with initial condition $S_0$, where $P$ is the real-world probability measure, $\mu$ is the real-world drift, $\sigma$ is the volatility, and $W^P_t$ is a Brownian motion under $P$, such that $dW_t\sim Z\sqrt{dt}$, where $Z\sim N(0,1)$ is a standard normal random variable.
The solution to eq. (\ref{SDE}) is given by
\begin{equation}
\label{SDEsolution}
S_t = S_0 + \int_0^t\mu\Parenthesis{u,S_u}du + \int_0^t\sigma\Parenthesis{u,S_u}dW^P_u\, ,
\end{equation}
see \eg \cite{Oks1992}. In particular, in the Black-Scholes model
the underlying asset $S_t$ follows a simple log-normal stochastic
process
\begin{equation}\label{BS}
dS_t=\mu\, S_t\, dt+\sigma\, S_t\, dW^P_t,
\end{equation}
with constant $\mu$ and $\sigma$. The solution to equation
(\ref{BS}) in a risk-neutral world (under the risk-neutral
probability measure $Q$) is given by\footnote{We assume a constant
interest rate $r$ for simplicity. See \eg \cite{BriMer06},
appendix B, for a generalization to stochastic interest rates.}
\begin{equation}
\label{BSsol}
S_t=S_0\exp{\Brack{\Parenthesis{r-\frac{1}{2}\sigma^2}t+\sigma W^Q_t}}.
\end{equation}
\QuoteDouble{Greeks} are derivatives of the price
$V_0(\bm{\theta})$ \wrt specific parameters $\bm{\theta}$. They
are very important quantities which need to be computed for
hedging and risk management purposes. In the present work, we will
consider in particular the following greeks:
\begin{align}
\label{Greeks}
\Delta &= \frac{\partial V_0}{\partial S_0},\\
\Gamma &= \frac{\partial^2 V_0}{\partial S^2_0},\\
\mathcal{V} &= \frac{\partial V_0}{\partial\sigma},
\end{align}
called delta, gamma and vega, respectively. Notice that, in the
Black-Scholes model, delta is exactly the hedge of the financial
instrument \wrt the risky underlying $S$, and vega is a derivative
\wrt a \emph{model} parameter (the constant volatility $\sigma$ in
the Black-Scholes SDE (\ref{BS})).
\par
The solution to the pricing equation (\ref{opt:price}) requires
the knowledge of the values of the underlying asset $S$ at the
relevant contract dates $\Brace{T_1,\ldots,T_n}$. Such values may
be obtained by solving the SDE (\ref{SDEsolution}). If the SDE
cannot be solved explicitly, we must resort to a discretization
scheme, computing the  values of $S$ on a time grid
$\Brace{t_1,\ldots,t_D}$, where $t_1<t_2<\cdots<t_D$, and $D$ is
the number of time steps. Notice that the contract dates must be
included in the time grid,
$\Brace{T_1,\ldots,T_n}\subset\Brace{t_1,\ldots,t_D}$. For
example, the Euler discretization scheme consists of approximating
the integral equation (\ref{SDEsolution}) by
\begin{equation}
\label{SDEdiscrete}
S_j = S_{j-1} + \mu\Parenthesis{t_{j-1},S_{j-1}}\Delta t_j + \sigma\Parenthesis{t_{j-1},S_{j-1}}\Delta W_j,\quad j=1,\ldots,D,
\end{equation}
where $\Delta t_j = t_j - t_{j-1}$, $\Delta W_j = W_j-W_{j-1}$, and $t_0 =0$.
In particular, the discretization of the Black-Scholes solution (\ref{BSsol}) leads to
\begin{equation}\label{BSsolDiscr}
S_j=S_{j-1}\exp{\Brack{\Parenthesis{r-\frac{\sigma^2}{2}}\Delta t_j+\sigma\Delta W_j}},\quad j=1,\ldots,D.
\end{equation}
Clearly, the price in eq. (\ref{opt:price}) will depend on the
discretization scheme adopted. See \cite{KloPla1995} for the order
of convergence of Euler and other discretization schemes.
\par
We consider two discretization schemes in eq. (\ref{BSsolDiscr}): standard discretization (SD) and Brownian bridge discretization (BBD).
In the SD scheme the Brownian motion is discretized as follows:
\begin{equation}
\label{std}
\Delta W_j = \sqrt{\Delta t_j}Z_j,\quad j=1,\ldots,D.
\end{equation}
In the BBD scheme the first variate is used to generate the terminal value of the Brownian motion, while subsequent variates are used to generate intermediate points, conditioned to points already simulated at earlier and later time steps, according to the following formula,
\begin{equation}
\label{BB}
\begin{split}
&W_0=0,\\
&W_D=\sqrt{\Delta t_{D0}} Z_1,\\
&W_j=\frac{\Delta t_{kj}}{\Delta t_{ki}}\,W_i + \frac{\Delta
t_{ji}}{\Delta t_{ki}}\,W_k+\sqrt{\frac{\Delta t_{kj}\Delta
t_{ji}}{\Delta t_{ki}}} Z_l, \quad t_i<t_j<t_k\,
,\;\;\;l=2,\ldots,D,
\end{split}
\end{equation}
where $\Delta t_{ab}=t_a-t_b$.
Unlike the SD scheme, which generates the Brownian motion sequentially across time steps, the BBD scheme uses different orderings: as a result, the variance in the stochastic part of (\ref{BB}) is smaller than that in (\ref{std}) for the same time steps, so that the first few points contain most of the variance. Both schemes have the same variance, hence their MC convergence rates are the same, but QMC sampling shows different efficiencies for SD and BBD, which will be discussed in the following sections.
\par
The number $D$ of time steps required in the discretization of the SDE (\ref{SDEdiscrete}) is the nominal dimension of the computational problem: indeed, the expectation value in (\ref{opt:price}) is formally an integral of the payoff, regarded as a function of $Z_1,\ldots,Z_D$.
In general, financial instruments may depend on multiple underlying assets $S^1,\ldots,S^M$: in this case, the dimension of the problem is given by $D\times M$.
In conclusion, the pricing problem (\ref{opt:price}) is reduced to the evaluation of high-dimensional integrals. This motivates the use of Monte Carlo techniques.
\par
Throughout this work, we will focus on the \textit{relative}
effects of the dimension $D$ and of the discretization schemes on
the MC and QMC simulations. Thus, we will assume a simple
Black-Scholes underlying dynamics for simplicity. This choice will
be also useful as a reference case to interpret further results
based on more complex dynamics\footnote{For example, we could
introduce jumps or Heston dynamics, see \eg \cite{Wil06}.}. We
stress that using simple and solvable dynamics is an approximation
often used in risk management practice for risk measures
calculation on large portfolios with multiple underlying risk
factors, because of computational bottlenecks.

\subsection{Pseudo Random Numbers and Low Discrepancy Sequences}
\label{SecRN}
Standard gaussian numbers ${Z_j}$ are computed using transformation of uniform variates $x_j\sim\ i.i.d. \;U(0,1)$,
\begin{equation}\label{conversion}
Z_j = \Phi^{-1}(x_j),\quad j=1,\ldots,D,
\end{equation}
where $\Phi^{-1}$ is the inverse cumulative distribution function of the standard normal distribution.
Hence the pricing problem (\ref{opt:price}) can be reduced to the evaluation of integrals of the following generic form
\begin{equation}\label{integral}
V=\int_{H^D}f(\bm{x})d^D\bm{x},
\end{equation}
where $H^D=[0,1]^D$ is the $D-$dimensional unit hypercube.
The standard Monte Carlo estimator of (\ref{integral}) has the form
\begin{equation}
\label{estint}
V_N\simeq \frac{1}{N}\sum_{k=1}^Nf(\bm{x}_k),
\end{equation}
where $\Brace{\bm{x}_k}_{k=1}^N$ is a sequence of $N$ random
points in $H^D$. Sequences $\Brace{\bm{x}_k}_{k=1}^N$ are produced
by appropriate Random Number Generators (RNGs). In particular,
Pseudo Random Number Generators (PRNGs) are computer algorithms
that produce \emph{deterministic} sequences of pseudo random
numbers (PRNs) mimicking the properties of true random sequences.
Such sequences are completely determined by a set of initial
values, called the PRNG's state. Thus, pseudo random sequences are
reproducible, using the same set of state variables. PRNGs are
characterized by the seed, \ie a random number used to initialize
the PRNG, the period, \ie the maximum length, over all possible
state variables, of the sequence without repetition, and the
distribution of the generated random numbers, which is generally
uniform $[0,1)$. The most famous PRNG is the Mersenne Twister
\cite{MatNis1998}, with the longest period of $2^{19937}-1$ and
good equidistribution properties guaranteed up to, at least, 623
dimensions. Pseudo random sequences are known to be plagued by
clustering: since new points are added randomly, they don't
necessarily fill the gaps among previously sampled points. This
fact causes a rather slow convergence rate. Consider an
integration error
\begin{equation}\label{interr}
\varepsilon=|V-V_N|.
\end{equation}
By the Central Limit Theorem the root mean square error of the Monte Carlo method is
\begin{equation}\label{MC:err}
\varepsilon_{MC}=\Brack{ \mathbb{E}(\varepsilon^2)}^{1/2} =
\frac{\sigma_f}{\sqrt{N}}\, ,
\end{equation}
where $\sigma_f$ is the standard deviation of $f(x)$. Although
$\varepsilon_{MC}$ does not depend on the dimension $D$, as in the
case of lattice integration on a regular grid, it decreases slowly
with increasing $N$. Variance reduction techniques, such as
antithetic variables \cite{Jac01,Gla03}, only affect the numerator
in (\ref{MC:err}).
\par
In order to increase the rate of convergence, that is to increase
the power of $N$ in the denominator of (\ref{MC:err}), one has to
resort to Low Discrepancy Sequences (LDS), also called Quasi
Random Numbers (QRNs), instead of PRNs. The discrepancy of a
sequence $\Brace{\bm{x}_k}_{k=1}^N$ is a measure of how
inhomogeneously the sequence is distributed inside the unit
hypercube $H^D$. Formally, it is defined by \cite{Jac01}
\begin{equation}\begin{split}
&\mathcal{D}^D_N(\bm{x}_1,\ldots,\bm{x}_N)
= \underset{\bm{\xi}\in H^D}{\sup} \left|\frac{n\left[\mathcal{S}^D(\bm{\xi}),\bm{x}_1,\ldots,\bm{x}_N\right]}{N} - m(\bm{\xi})\right|,\\
&\mathcal{S}^D(\bm{\xi}) = [0,\xi_1)\times\cdots\times
[0,\xi_D)\subset H^D,\quad
m(\bm{\xi}) = \prod_{j=1}^D \xi_j,\\
\end{split}\end{equation}
where
\begin{equation}\begin{split}
&n\left[\mathcal{S}^D(\bm{\xi}),\bm{x}_1,\ldots,\bm{x}_N\right] =
\sum_{k=1}^N \mathds{1}_{\left\{\bm{x}_k \in
\mathcal{S}^D(\bm{\xi})\right\}} = \sum_{k=1}^N \prod_{j=1}^D
\mathds{1}_{\left\{x_{k,j}\leq \xi_j\right\}}\,
\end{split}\end{equation}
is the number of sampled points that are contained in hyper-rectangle $\mathcal{S}^D\subset H^D$. It can be shown that the expected discrepancy of a pseudo random sequence is of the order of $\ln(\ln N)/\sqrt{N}$.
A Low Discrepancy Sequence is a sequence $\Brace{\bm{x}_k}_{k=1}^N$ in $H^D$ such that, for any $N>1$, the
first $N$ points ${\bm{x}_1,\ldots,\bm{x}_N}$ satisfy inequality
\begin{equation}
\mathcal{D}^D_N(\bm{x}_1,\ldots,\bm{x}_N) \leq c(D)\frac{\ln^D
N}{N}\, ,
\end{equation}
for some constant $c(D)$ depending only on $D$ \cite{Nid88}.
Unlike PRNGs, Low Discrepancy Sequences are deterministic sets of
points. They are typically constructed using number theoretical
methods. They are designed to cover the unit hypercube as
uniformly as possible. In the case of sequential sampling, new
points take into account the positions of already sampled points
and fill the gaps between them. Notice that a regular grid of
points in $H^D$ does not ensure low discrepancy, since projecting
adjacent dimensions easily produces overlapping points.
\par
A Quasi Monte Carlo (QMC) estimator of the integral (\ref{integral}) is of the form (\ref{estint}) with the only
difference that the sequence $\Brace{\bm{x}_k}_{k=1}^N$ is sampled using LDS instead of PRNs. An upper bound for the QMC integration error is given by the Koksma-Hlawka inequality
\begin{equation}\label{QMC:errbound}
\varepsilon_{QMC}\le
V(f)\mathcal{D}^D_N=\mathcal{O}\left(\frac{\ln^D N}{N}\right)\, ,
\end{equation}
where $V(f)$ is the variation of the integrand function in the
sense of Hardy and Krause, which is finite for functions of
bounded variation \cite{KucBal11}. The convergence rate of
(\ref{QMC:errbound}) is asymptotically faster than (\ref{MC:err}),
but it is rather slow for feasible $N$. Moreover, it depends on
the dimensionality $D$. However, eq. (\ref{QMC:errbound}) is just
an upper bound: what is observed in most numerical tests
\cite{KucBal11,CafMor1997} is a power law
\begin{equation}
\label{err:QMC}
\varepsilon_{QMC}\sim \frac{c}{N^\alpha}\, ,
\end{equation}
where the value of $\alpha$ depends on the model function and, therefore, is not \emph{a priori} determined as for MC.
When $\alpha>0.5$ the QMC method outperforms standard MC: this situation turns out to be quite common in financial problems.
We will measure $\alpha$ for some representative financial instruments, showing that its value can be very close to 1 when the \emph{effective} dimension of $f$ is low, irrespective of the nominal dimension $D$. The concept of effective dimension, and the methodology to compute it, will be introduced in the following sections.
\par
We stress that, since LDS are deterministic, there are no
statistical measures like variances associated with them. Hence,
the constant $c$ in (\ref{err:QMC}) is not a variance and
(\ref{err:QMC}) does not have a probabilistic interpretation as
for standard MC. To overcome this limitation, Owen
suggested to introduce randomization into LDS at the same time
preserving their superiority to PRN uniformity properties \cite{Owe93}. Such
LDS became known as \emph{scrambled} (see also \cite{Gla03}). In
practice, the integration error for both MC and QMC methods for
any fixed $N$ can be estimated by computing the following error
averaged over $L$ independent runs:
\begin{equation}\label{error}
\varepsilon_N=\sqrt{\frac{1}{L}\,\sum_{\ell=1}^L\left(V-V_N^{(\ell)}\right)^2},
\end{equation}
where $V$ is the exact, or estimated at a very large extreme value
of $N\rightarrow\infty$, value of the integral and $V_N^{(\ell)}$
is the simulated value for the $\ell$th run, performed using $N$
PRNs, LDS, or scrambled LDS. For MC and QMC based on scrambled
LDS, runs based on different seed points are statistically
independent. In the case of QMC, different runs are obtained using
non overlapping sections of the LDS. Actually, scrambling LDSs
weakens the smoothness and stability properties of the Monte Carlo
convergence, as we will see in Section \ref{SecStability}. Hence,
in this paper we will use the approach based on non-overlapping
LDSs, as in \cite{SobKuc05a}.
\par
The most known LDS are Halton, Faure, Niederreiter and Sobol'
sequences. Sobol' sequences, also called $LP\tau$ sequences or
$(t, s)$ sequences in base 2 \cite{Nid88}, became the most
known and widely used LDS in finance due to their efficiency
\cite{Jac01,Gla03}. Sobol' sequences were constructed under the
following requirements \cite{Sob1967}:
\begin{enumerate}
    \item Best uniformity of distribution as $N\to\infty$.
    \item Good distribution for fairly small initial sets.
    \item A very fast computational algorithm.
\end{enumerate}
The efficiency of Sobol' LDS depend on the so-called
initialisation numbers. In this work we used \texttt{SobolSeq8192}
generator provided by BRODA \cite{BRODA}. \texttt{SobolSeq} is an
implementation of the 8192 dimensional Sobol' sequences with
modified initialisation numbers. Sobol' sequences produced by
\texttt{SobolSeq8192} can be up to and including dimension
$2^{13}$, and satisfy additional uniformity properties: Property A
for all dimensions and Property A' for adjacent dimensions (see
\cite{SobAso12} for details\footnote{BRODA releases also
\texttt{SobolSeq32000} and \texttt{SobolSeq64000} }). It has been
found in \cite{SobAso12} that \texttt{SobolSeq} generator
outperforms all other known LDS generators both in speed and
accuracy.

\section{Global Sensitivity Analysis and Effective Dimensions}
\label{SecGSA}

As we mentioned in the Introduction and Section \ref{SecRN}, effective dimension is the key to explain the superior efficiency of QMC \wrt MC. Hence, it is crucial to develop techniques to estimate the effective dimension and to find the most important variables in a MC simulation.
\par
The variance-based method of global sensitivity indices developed
by Sobol' became very popular among practitioners due to its
efficiency and easiness of interpretation \cite{SobKuc05b,Sal10}.
There are two types of Sobol' sensitivity indices: the main effect
indices, which estimate the individual contribution of each input
parameter to the output variance, and the total sensitivity
indices, which measure the total contribution of a single input
factor or a group of inputs. Sobol' indices can be used to rank
variables in order of importance,  to identify non-important
variables, which can then be fixed at their nominal values to
reduce model complexity, and to analyze the efficiency of various
numerical schemes.
\par
Consider a mathematical model described by an integrable function $f(x)$, where the input $x=(x_1,\ldots,x_D)$ is taken in a $D$-dimensional domain $\Omega$ and the output is a scalar.
Without loss of generality, we choose $\Omega$ to be the unit hypercube $H^D$. The input variables $x_1,\ldots,x_D$ can, then, be regarded as independent uniform random variables each defined in the unit interval $[0,1]$.
The starting point of global sensitivity analysis (GSA) is the analysis of variance (ANOVA) decomposition of the model function,
\begin{equation}\label{ANOVA1}
f(x) = f_0 + \sum_i f_i(x_i) + \sum_{i<j}f_{ij}(x_i,x_j) + \ldots
+ f_{1\,2\cdots D}(x_1,\ldots,x_D)\, .
\end{equation}
The expansion (\ref{ANOVA1}) is unique, provided that
\begin{equation}\label{ANOVA2}
\int_0^1f_{i_1\cdots i_s}(x_{i_1},\ldots,x_{i_s})dx_{i_k}=0\,
,\;\;\;\forall k=1,\ldots,s\, .
\end{equation}
The ANOVA decomposition expands the function $f$ into a sum of terms, each depending on an increasing number of variables: a generic component $f_{i_1\cdots i_s}(x_{i_1},\ldots,x_{i_s})$, depending on $s$ variables, is called an $s$-order term. It follows from (\ref{ANOVA2}) that the ANOVA decomposition is orthogonal and that its terms can be explicitly found as follows,
\begin{equation}
\begin{split}
f_0=&\int_{H^D}f(x)d^Dx,\\
f_i(x_i)=&\int_{H^{D-1}}f(x)\prod_{k\neq i}dx_k\, -f_0,\\
f_{ij}(x_i,x_j)=&\int_{H^{D-2}}f(x)\prod_{k\neq i,j}dx_k\, -f_0-f_i(x_i)-f_j(x_j),
\end{split}
\end{equation}
and so on. If $f$ is square-integrable, its variance decomposes into a sum of partial variances:
\begin{equation}\label{totvar}
\sigma^2 = \sum_i\sigma^2_i+\sum_{i<j}\sigma^2_{ij}+\ldots+\sigma^2_{12\cdots D},
\end{equation}
where
\begin{equation}
\label{var}
\sigma^2_{i_1\cdots i_s} =\int_0^1f^2_{i_1\cdots i_s}(x_{i_1},\ldots,x_{i_s})dx_{i_1}\cdots dx_{i_s}.
\end{equation}
Sobol' sensitivity indices are defined as
\begin{equation}
\label{SobSensIndex}
S_{i_1\cdots i_s}=\frac{\sigma^2_{i_1\cdots i_s}}{\sigma^2}
\end{equation}
and measure the fraction of total variance accounted by each
$f_{i_1\cdots i_s}$ term of the ANOVA decomposition. From
(\ref{totvar}) it follows that all Sobol' indices are non negative
and normalized to 1. First order Sobol' indices $S_i$ measure the
effect of single variables $x_i$ on the output function; second
order Sobol' indices $S_{ij}$ measure the interactions between
pairs of variables, \ie the fraction of total variance due to
variables $x_i$ and $x_j$ which cannot be explained by a sum of
effects of single variables; higher order Sobol' indices
$S_{i_1\cdots i_s}$, with $s>2$, measure the interactions among
multiple variables, \ie the fraction of total variance due to
variables $x_{i_1},\ldots,x_{i_s}$ which cannot be explained by a
sum of effects of single variables or lower order interactions.
\par
The calculation of Sobol' sensitivity indices in eq. (\ref{SobSensIndex}) requires, in principle, $2^D$ valuations of the multi-dimensional integrals in eq. (\ref{var}), which is a very cumbersome, or even impossible, computational task.
Furthermore, for practical purposes, and in particular when the function $f$ has low order interactions, it is not actually necessary to know all the possible Sobol' indices, but just an appropriate selection of them.
Thus, it is very useful to introduce Sobol' indices for subsets of variables and total Sobol' indices.
Let $y=\Brace{x_{i_1},\ldots,x_{i_m}}\subseteq x, 1\le i_1\le\ldots,\le i_m\le D$, be a subset of $x$, and $z=y^\complement\subseteq x$ its complementary subset, and define
\begin{equation}
\begin{split}
&S_y = \sum_{s=1}^D\,\sum_{(i_1<\cdots<i_s)\in K}S_{i_1\cdots i_s},\\
&S_y^{tot} = \,1-S_z\,,
\end{split}
\end{equation}
where $K=\{i_1,\ldots,i_m\}$. Notice that $0\le S_y\le
S_y^{tot}\le 1$. The quantity $S_y^{tot}-S_y$ accounts for all the
interactions between the variables in subsets $y$ and $z$. It
turns out that there exist efficient formulas which allow to avoid
the knowledge of ANOVA components and to compute Sobol' indices
directly from the values of function $f$ \cite{Sob01}. These
formulas are based on the following integrals,
\begin{equation}
\label{SI}
\begin{split}
S_y=&\frac{1}{\sigma^2}\,\int_0^1[f(y',z')-f_0][f(y',z)-f(y,z)]dy\,dz\,dy'dz'\, ,\\
S_y^{tot}=&\frac{1}{2\sigma^2}\,\int_0^1[f(y,z)-f(y',z)]^2 dy\,dz\,dy'\, ,\\
\sigma^2=&\int_0^1f^2(y,z)dy\,dz\,-f_0^2\, ,\\
f_0=&\int_0^1f(y,z)dy\,dz\,,
\end{split}
\end{equation}
where the integration variables are the components of the vectors
$y,z,y',z'$, such that $x=y\cup z$, and the first two integrals
depend on the choice of $y$. Such integrals can be evaluated, in
general, via MC/QMC techniques \cite{KucBal11,Sal02}.
\par
Furthermore, usually enough information is already given by the first order indices $S_i$ and by corresponding total effect indices $S_i^{tot}$, linked to a single variable $y=\Brace{x_i}$.
For these Sobol' indices, it's easy to see that
\begin{itemize}
    \item $S_i^{tot}=0$: the output function does not depend on
    $x_i$,
    \item $S_i=1$: the output function depends only on $x_i$,
    \item $S_i=S_i^{tot}$: there is no interaction between $x_i$ and other variables.
\end{itemize}
Notice that just $D+2$ function evaluations for each MC trial are necessary to compute all $S_i$ and $S_i^{tot}$ indices in eqs. (\ref{SI}): one function evaluation at point $x=\Brace{y,z}$, one at point $x'=\Brace{y',z'}$, and $D$ evaluations at points  $x''=\Brace{y',z},\forall\; y' = \Brace{x_i},\;i = 1,\ldots,D$.
\par
We stress that the approach presented above is applicable only to
the case of independent input variables, which admits a unique
ANOVA decomposition. In the case of dependent (correlated) input
variables, the computation of variance-based global sensitivity
indices is more involved. A generalization of GSA to dependent
variables can be found in \cite{KucTar12}.
\par
We finally come to the notion of effective dimensions, firstly
introduced in \cite{CafMor1997}. Let $|y|$ be the cardinality of a
subset of variables $y$. The effective dimension in the
\emph{superposition sense}, for a function $f$ of $D$ variables,
is the smallest integer $d_S$ such that
\begin{equation}
\sum_{0<|y|<d_S }S_y\ge 1-\varepsilon
\end{equation}
for some threshold $\varepsilon$ (arbitrary and usually chosen to be less than $5\%$). If a function has an effective dimension $d_S$ in the superposition sense, it can be approximated by a sum of $d_S$-dimensional terms, with an approximation error below $\varepsilon$.
\par
The effective dimension in the \emph{truncation sense} is the smallest integer $d_T$ such that
\begin{equation}
\sum_{y\subseteq \{1,2,\ldots,d_T\}}S_y\ge 1-\varepsilon.
\end{equation}
The effective dimension $d_S$ does not depend on the order of sampling of variables, while $d_T$ does. In general, the following inequality holds,
\begin{equation}
\label{eqDimensions}
d_S\le d_T\le D.
\end{equation}
Effective dimensions can be estimated solely from indices $S_i$
and $S_i^{tot}$ using eqs. (\ref{SI}) with $y=i$, as described in
\cite{KucBal11}, where relationships among such indices are used
to classify functions in three categories according to their
dependence on variables. For the so-called type A functions,
variables are not all equally important and the effective
dimension in the truncation sense $d_T$ is small, such that
$d_S\leq d_T\ll D$. They are characterized by the following
relationship
\begin{equation}
\label{eqTypeA}
\frac{S^{tot}_y}{|y|}\gg \frac{S_z^{tot}}{|z|},
\end{equation}
where $y\subseteq x$ is a leading subset of variables, $z=y^\complement\subseteq x$ its complementary subset.
Functions with equally important variables have $d_T\simeq D$ and they can be further divided in two groups: type B and C functions. Type B functions have dominant low-order interactions and small effective dimension in the superposition sense $d_S$, such that $d_S\ll d_T\simeq D$. For such functions, Sobol' indices satisfy the following relationships:
\begin{equation}
\label{eqTypeB}
S_i\simeq S_i^{tot},\quad \forall\;i=1,\ldots,D, \quad\sum_{i=1}^DS_i\simeq 1.
\end{equation}
Type C functions have dominant higher-order interactions
\begin{equation}
\label{eqTypeC}
S_i\ll
S_i^{tot}\, ,\;\sum_{i=1}^DS_i\ll 1
\end{equation}
and effective dimensions $d_S\simeq d_T\simeq D$. This classification is summarized in Table \ref{tab:effdim}.
\begin{table}\small
\centering
\subtable{%
\begin{tabular}{cccc}
  \hline
  \textbf{Type} & \textbf{Description} & \textbf{Relationship between SI} & \textbf{Eff. dimensions} \\
  \hline
   A & Few important variables & $S_y^{tot}/|y|\gg S_z^{tot}/|z|$ & $d_S\leq d_T\ll D$ \\
   B & Low-order interactions & $S_i\simeq S_j,S_i\simeq S_i^{tot}\,, \forall\; i,j$  & $d_S\ll d_T\simeq D$ \\
   C & High-order interactions & $S_i\simeq S_j,S_i\ll S_i^{tot}\,, \forall\; i,j$ & $d_S\simeq d_T\simeq D$ \\
  \hline
\end{tabular}
}%\qquad\qquad
\caption{Classification of functions \wrt their dependence on variables, based on GSA.}
\label{tab:effdim}
\end{table}
\par
Owen introduced in \cite{Owe03} the notion of the average dimension $d_A$,
which can assume fractional values, defined as
\begin{equation} d_A := \sum_{0<|y|<D}|y|\,S_y\,,
\end{equation}
and showed that it can be rather straightforwardly computed as
\begin{equation}\label{d_A}
d_A=\sum_{i=1}^DS_i^{tot}\, .
\end{equation}
It has been suggested in \cite{SobShu14} that QMC should
outperform MC when $d_A\lesssim 3$. This is confirmed in our
findings, see Section \ref{SecGSAresults}.
\par
It has been proved in many works \cite{KucBal11,CafMor1997, Owe03}
that QMC outperforms MC regardless of the nominal dimension
whenever the effective dimension is low in one or more senses.
Hence, in the case of type A and type B functions (we assume that
functions are sufficiently smooth), QMC always outperform MC,
while for type C functions the two methods are expected to have
similar efficiency. Actually, type A and B functions are very
common in financial problems. We also note that the performance of
the QMC method for Type A functions sometimes, but not always, can
be greatly improved by using effective dimension reduction
techniques, such as Brownian bridge, which will be demonstrated in
the following section.

\section{Test Cases and Numerical Results}
\label{SecRes}
In this section we apply MC and QMC techniques to high-dimensional pricing problems. Our aim is to test the efficiency of QMC with respect to standard MC in computing prices and greeks (delta, gamma, vega) for selected payoffs $\mathcal{P}$ with increasing degree of complexity and path-dependency.

\subsection{Selected Payoffs and Test Set-Up}
\label{SecPayoffs}
We selected the following instruments as test cases.
\begin{enumerate}
    \item \textbf{European call}:
        \beq\mathcal{P} = \max(S_D-K,0)\label{eqPayoffCall}.\eeq
    \item \textbf{Asian call}:
        \beq\mathcal{P} = \max(\bar{S}-K,0),\quad \bar{S}=\left(\prod_{j=1}^D S_j\right)^{1/D}
        \label{eqPayoffAsian}.\eeq
    \item \textbf{Double knock-out}:
        \beq\mathcal{P} = \max(S_D-K,0)\,\mathds{1}_{\{B_l<S_j<B_u\}},\quad\forall j=1,\ldots,D
        \label{eqPayoffDKO}.\eeq
    \item \textbf{Cliquet}:
        \beq\mathcal{P} = \max\left\{\sum_{j=1}^D \max\left[0,\min\left(C,\frac{S_j-S_{j-1}}{S_{j-1}}\right)\right],F\right\}.\label{eqPayoffCliquet}\eeq
\end{enumerate}
In the above definitions, $K$ denotes the strike price, $B_l$ and $B_u$ are the values of the lower and upper barrier, respectively, $C$ is a local cap and $F$ is a global floor.
In all test cases we use the following payoff parameters:
\begin{itemize}
    \item maturity: $T=1$,
    \item strike: $K=100$,
    \item lower barrier: $B_l = 0.5\,S_0$,
    \item upper barrier: $B_u = 1.5\, S_0$,
    \item global floor: $F=0.16$,
    \item local cap: $C=0.08$.
\end{itemize}
Such selection guarantees an increasing level of complexity and
path-dependency. The European call is included just as a simple
reference case, for which analytical formulas are available for
price and greeks, see \eg \cite{Wil06}. The Asian call with
arithmetic average is the simplest and most diffused non-European
payoff; we choose geometric average payoff such that analytical
formulas are available\footnote{See \eg \cite{Wil06} and
references therein.}. The double barrier is another very diffused
payoff with stronger path-dependency. Finally, the Cliquet option
is a typical strongly path-dependent payoff based on the
performance of the underlying stock. Clearly, many other possible
payoffs could be added to the test (\eg autocallable), but we
think that such selection should be complete enough to cover most
of the path-dependency characteristics relevant in the Monte Carlo
simulation. We assume that the underlying process $S_t$ follows a geometric
Brownian motion as described in Section \ref{SecGM}, with the
following model parameters:
\begin{itemize}
    \item spot: $S_0=100$,
    \item volatility: $\sigma = 0.3$,
    \item number of time steps: $D=32$.
\end{itemize}
The process $S_t$ is discretized across $D$ time steps
$\Brace{t_1<\cdots<t_j<\cdots<t_D}$, so that $S_D$ is its value at
maturity. Recall that, in the single asset case, the number of
time simulation steps is equal to the dimension of the
path-dependent simulation. As discussed at the end of Section
\ref{SecMC}, we choose a simple dynamics for $S_t$ because our
main goal to compare MC and QMC simulations \wrt the effect of the
dimension $D$ and of the discretization schemes.
\par
The numerical computations are performed in \texttt{Matlab} using three different sampling techniques:
\begin{itemize}
    \item MC+SD + antithetic variables + Mersenne Twister generator,
    \item QMC+SD + \texttt{SobolSeq8192} generator,
    \item QMC+BBD + \texttt{SobolSeq8192} generator.
\end{itemize}
The notations for the simulation parameters are:
\begin{itemize}
    \item $N$: number of simulated paths for the underlying,
    \item $D$: number of time steps used to discretize each underlying's path,
    \item $L$: number of independent runs.
\end{itemize}
Notice that, using the Black-Scholes model, the number $D$ of time steps is also the nominal dimension of the MC simulation.
Following the specifics of Sobol' sequences, we take $N=2^p$, where $p$ is an integer, since this guarantees the lowest discrepancy properties.
\par
Simulation errors $\varepsilon_N$ are analyzed by computing the root mean square error (RMSE) as defined by (\ref{error}), where $V$ is a reference value of prices or greeks given by analytical formulas (for European and geometric Asian options) or simulated with a large number of scenarios ($N=2^{23}$) (for Double Knock-out and Cliquet options).
To assess and compare performance of MC and QMC methods with different discretization schemes, we compute the scaling of the RMSE as a function of $N$ by fitting the function $\varepsilon_N$ with a power law $c\,N^{-\alpha}$ (\ref{err:QMC}).
In the MC case, the value of $\alpha$ is expected to be $0.5$ in all situations, while in the QMC case it is expected to be higher than $0.5$ for Type A and B functions.
\par
Finally, greeks for the payoffs above are computed via finite differences, using central difference formulas for delta, gamma and vega, with shift parameter $\epsilon$,
\begin{align}
\label{GreeksFD} \Delta_\mathcal{P} &= \frac{\partial
V_0^\mathcal{P}}{\partial S_0}
    \simeq \frac{V^\mathcal{P}_0(S_0+h)-V^\mathcal{P}_0(S_0-h)}{2h},\nonumber\\
\Gamma_\mathcal{P} &= \frac{\partial^2 V_0^\mathcal{P}}{\partial
S^2_0}
    \simeq \frac{V^\mathcal{P}_0(S_0+h)-2V^\mathcal{P}_0(S_0)+V^\mathcal{P}_0(S_0-h)}
    {h^2},\nonumber\\
\mathcal{V}_\mathcal{P} &= \frac{\partial
V_0^\mathcal{P}}{\partial\sigma}
    \simeq \frac{V^\mathcal{P}_0(\sigma+h)-V^\mathcal{P}_0(\sigma-h)}{2h},
\end{align}
where the increment $h$ is chosen to be $h=\epsilon S_0$, for
delta and gamma, and $h=\epsilon$, for vega, for a given
``shift parameter'' $\epsilon$. Notice that the calculation of
price and three greeks using eqs. (\ref{GreeksFD}) and payoffs
(\ref{eqPayoffCall}-\ref{eqPayoffCliquet}) above requires $N_p =
5+5+5+3=18$ functions evaluations (the Cliquet has null delta and
gamma). In the MC simulations for greeks we use path recycling of
both pseudo random and LDS sequences to minimize the variance of
the greeks, as suggested in \cite{Jac01} and \cite{Gla03}. Notice
that the analysis of the RMSE for greeks is, in general, more
complex than that for prices, since the variance of the MC
simulation mixes with the bias due to the approximation of
derivatives with finite differences with shift $\epsilon$. We
discuss how to deal with this issue in Appendix
\ref{App:GreekErr}.

\subsection{Global Sensitivity Analysis for Prices and Greeks}
\label{SecGSAresults} Sobol' indices $S_i$ and $S_i^{tot}$ are
computed for both the standard and Brownian bridge discretizations
using eqs. (\ref{SI}), where $f$ is the relevant model function
(the instrument payoff or a greek with finite differences) and
$y=\Brace{x_i}$, $y'=\Brace{x'_i}$,
$z=\Brace{x_1,\ldots,x_{i-1},x_{i+1},\ldots,x_D}$,
$z'=\Brace{x'_1,\ldots,x'_{i-1},x'_{i+1},\ldots,x'_D}$. Here
${x_i}$ are the uniform variates $x_i\sim i.i.d.\ U[0,1]$ used in eq.
(\ref{conversion}). The integrals in eqs. (\ref{SI}) are computed
using QMC simulation with the following parameters:
\begin{itemize}
    \item number of simulations: $N=2^{17}$,
    \item shift parameter for finite difference\footnote{See Appendix \ref{App:GreekErr}.}: $\epsilon=10^{-4},10^{-3},10^{-2}$.
\end{itemize}
\par
\begin{figure}[ht]
\centering
\subfigure[Price]{\includegraphics[width=2.5in,height=1.9in,keepaspectratio=false]{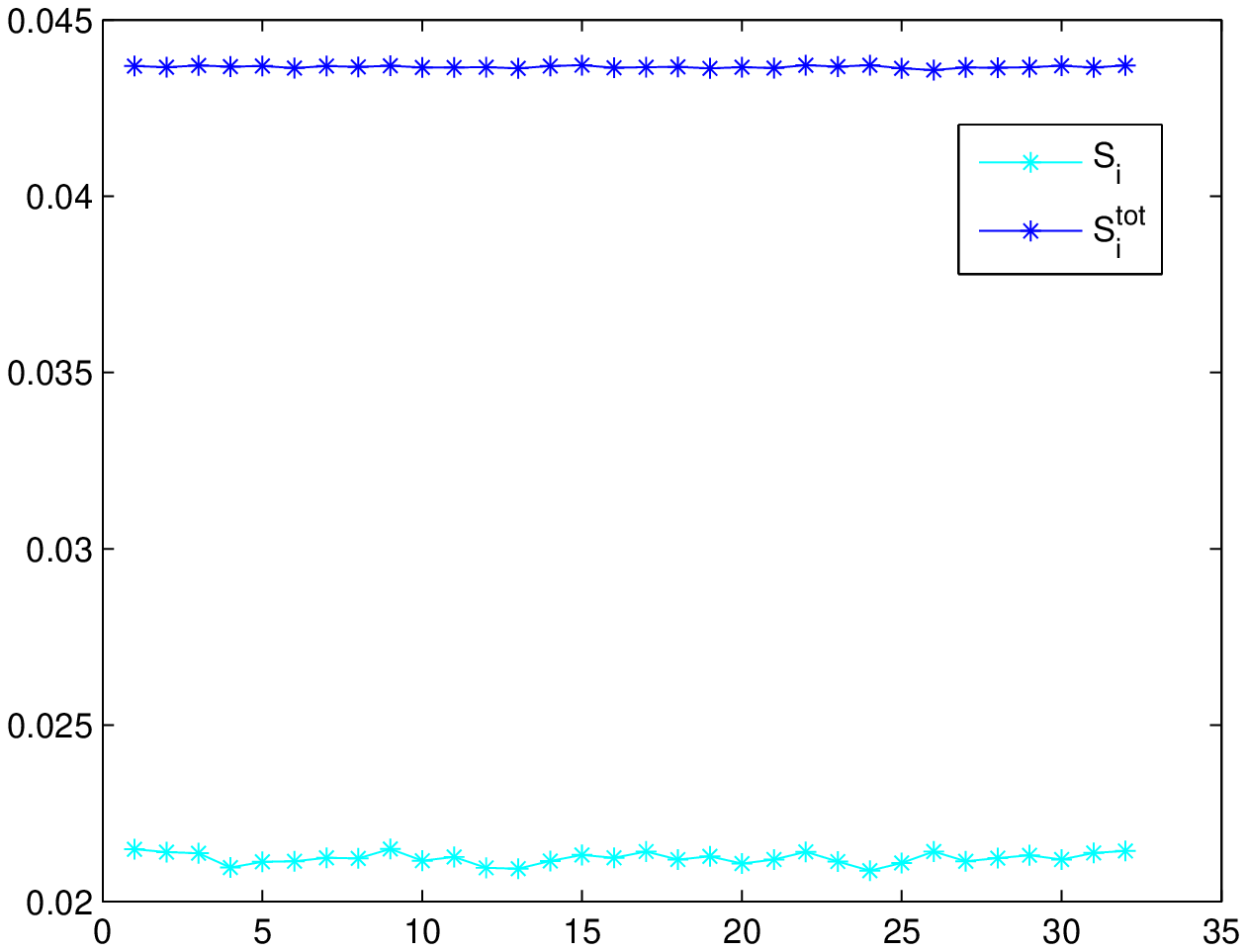}}
\subfigure[Delta]{\includegraphics[width=2.5in,height=1.9in,keepaspectratio=false]{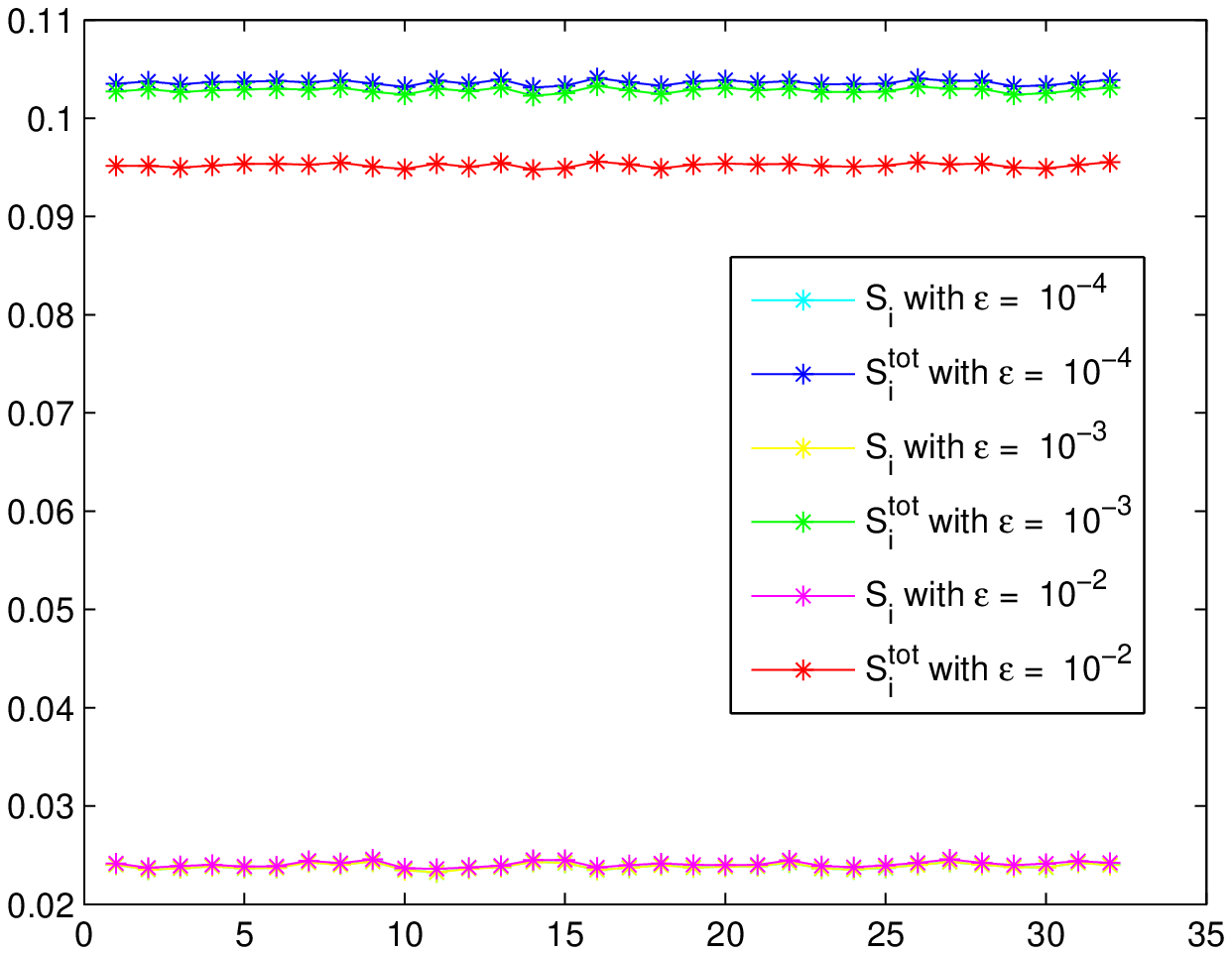}}
\subfigure[Gamma]{\includegraphics[width=2.5in,height=1.9in,keepaspectratio=false]{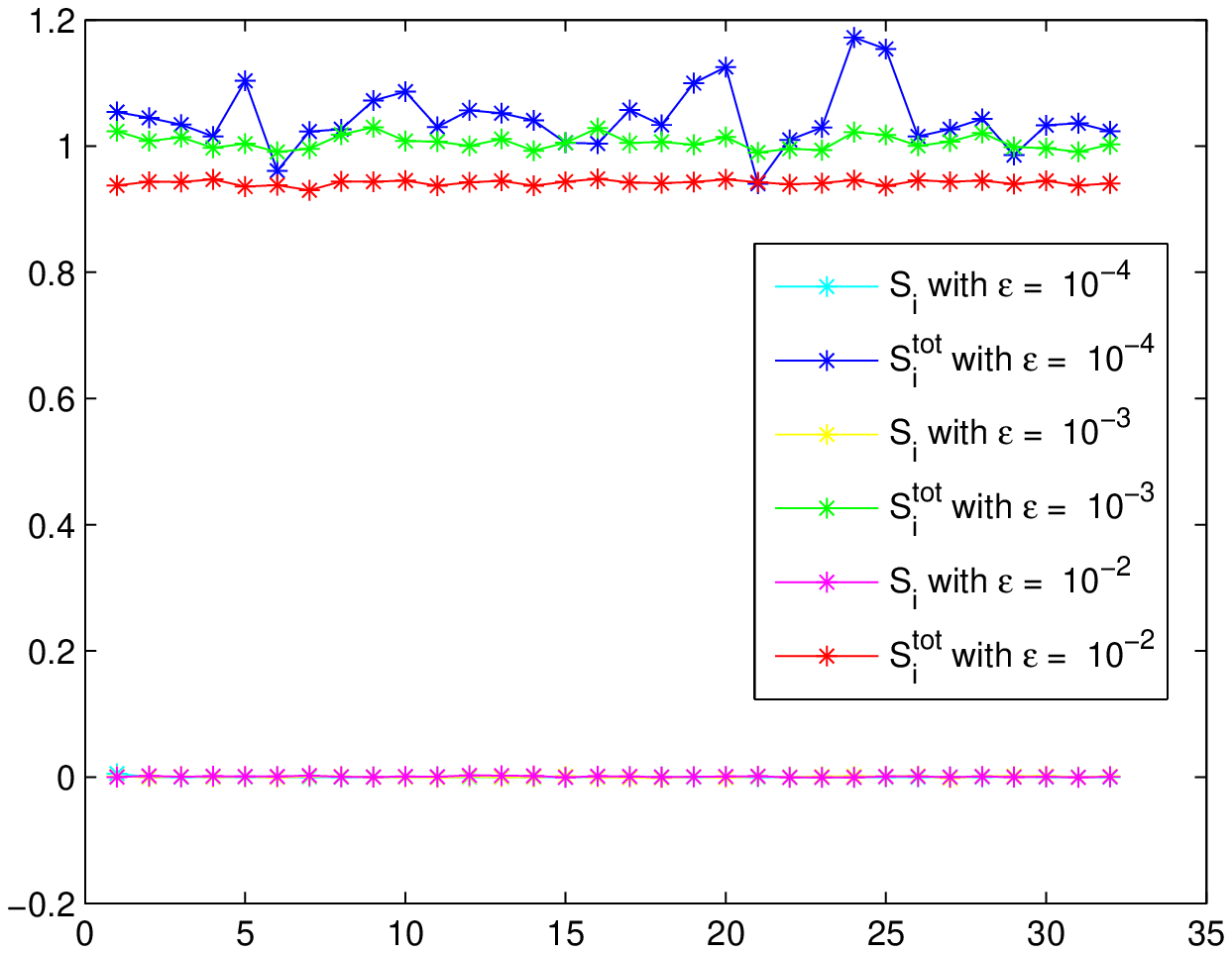}}
\subfigure[Vega]{\includegraphics[width=2.5in,height=1.9in,keepaspectratio=false]{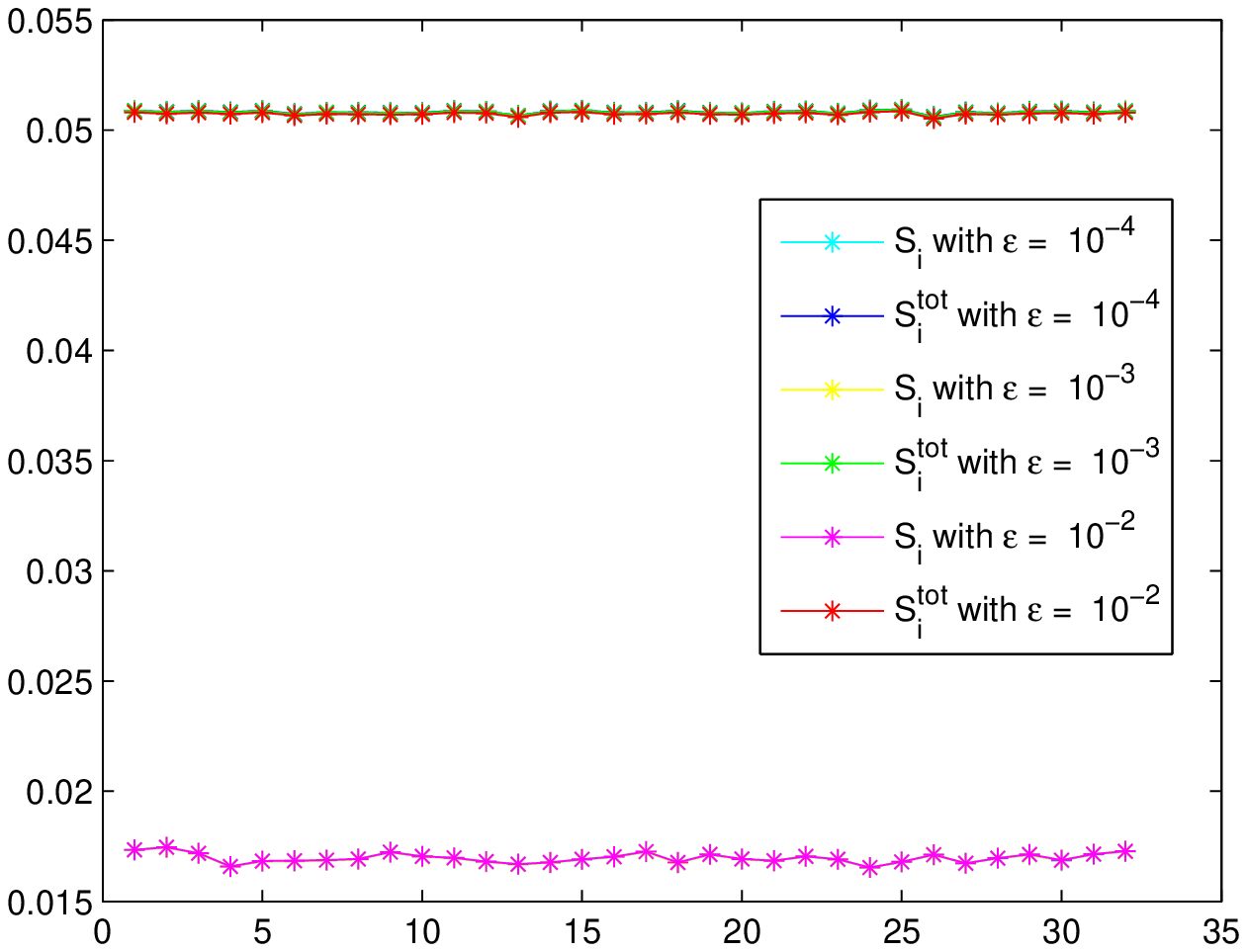}}
\caption{European call option price $(a)$ and greeks
$(b),(c),(d)$, SD, $D=32$. First order Sobol' indices $S_{i}$ and
total sensitivity indices $S_{i}^{tot}$ versus time step $i$.
}
\label{fig:1}
\end{figure}
\begin{figure}[ht]
\centering
\subfigure[Price]{\includegraphics[width=2.5in,height=1.9in,keepaspectratio=false]{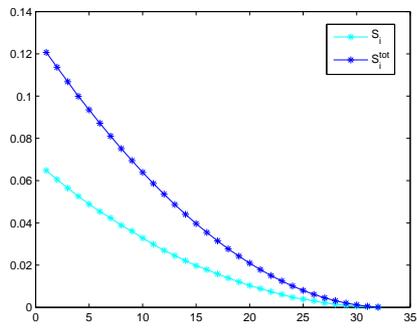}}
\subfigure[Delta]{\includegraphics[width=2.5in,height=1.9in,keepaspectratio=false]{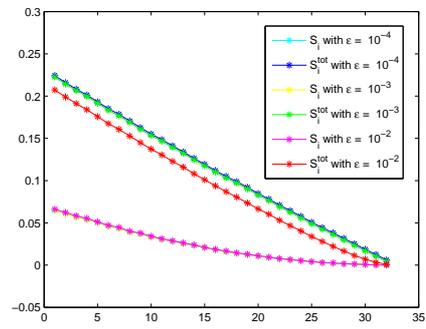}}
\subfigure[Gamma]{\includegraphics[width=2.5in,height=1.9in,keepaspectratio=false]{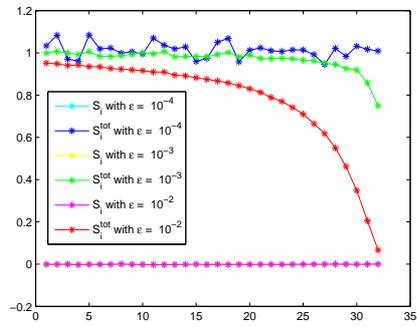}}
\subfigure[Vega]{\includegraphics[width=2.5in,height=1.9in,keepaspectratio=false]{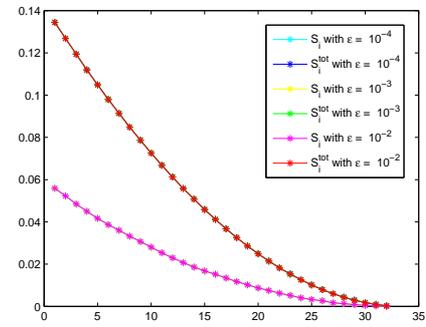}}
\caption{Asian call option. Parameters as in Figure \ref{fig:1}.}
\label{fig:2}
\end{figure}
\begin{figure}[ht]
\centering
\subfigure[Price]{\includegraphics[width=2.5in,height=1.9in,keepaspectratio=false]{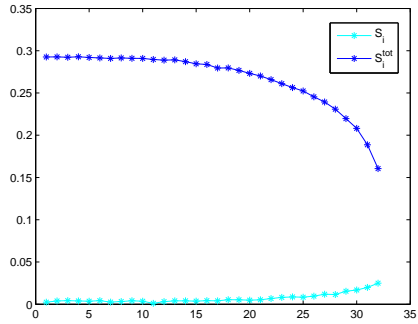}}
\subfigure[Delta]{\includegraphics[width=2.5in,height=1.9in,keepaspectratio=false]{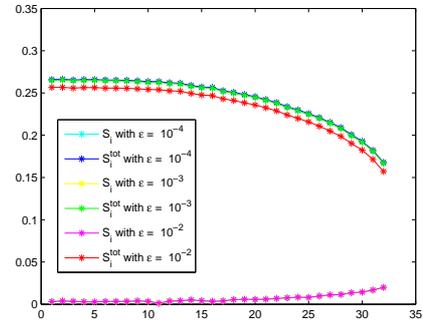}}
\subfigure[Gamma]{\includegraphics[width=2.5in,height=1.9in,keepaspectratio=false]{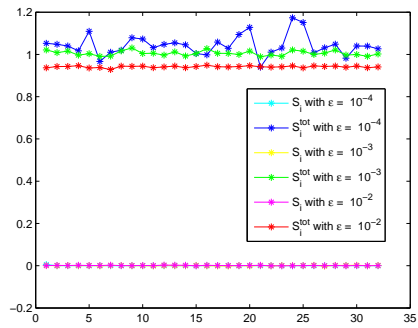}}
\subfigure[Vega]{\includegraphics[width=2.5in,height=1.9in,keepaspectratio=false]{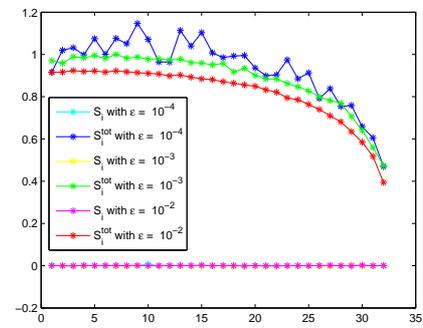}}
\caption{Double Knock-out call option. Parameters as in Figure \ref{fig:1}.}
\label{fig:3}
\end{figure}
\begin{figure}[ht]
\centering
\subfigure[Price]{\includegraphics[width=2.5in,height=1.9in,keepaspectratio=false]{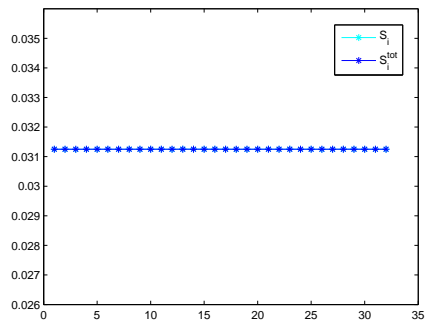}}
\subfigure[Vega]{\includegraphics[width=2.5in,height=1.9in,keepaspectratio=false]{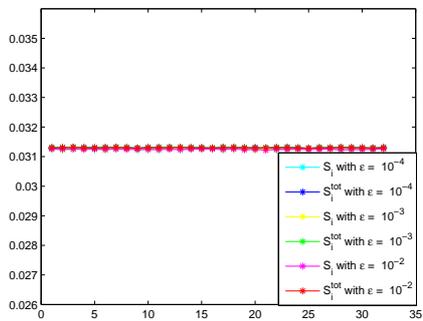}}
\caption{Cliquet option. Parameters as in Figure \ref{fig:1}. Delta and gamma are null for Cliquet options.}
\label{fig:4}
\end{figure}

Effective dimensions are estimated in the following way:
\begin{itemize}
  \item The effective dimension in the truncation sense $d_T$ is computed using inequality (\ref{eqTypeA}), looking for a minimal set of variables $y=\Brace{x_1,\ldots,x_{d_T}}$ such that the quantity $S_z^{tot}|y|/S_y^{tot}|z|$ is smaller than 1\%. Since the calculation of $d_T$ depends on the order of sampling variables, the result depends on the discretization scheme used, that is SD or BBD.
  \item The effective dimension in the superposition sense $d_S$ is estimated using dimension $d_T$ as an upper bound according to inequality (\ref{eqDimensions}). In order to distinguish between Type B and Type C functions, we look at ratios $S_i/S_i^{tot}$ and $\sum_i S_i$ according to eqs. (\ref{eqTypeB}), (\ref{eqTypeC}).
  \item The effective average dimension $d_A$ is computed according to eq. (\ref{d_A}).
\end{itemize}
The results of GSA for the SD are shown in Figures \ref{fig:1}-\ref{fig:4}.
Measures based on Sobol' indices are provided in Table \ref{tab:1}. These measures are used to compute effective
dimensions and to classify integrands in (\ref{integral}) corresponding to prices and greeks according to Table
\ref{tab:effdim}.
\begin{table}\small
\centering
\subtable{%
\begin{tabular}{c c c c c c c c}
  \toprule
  \textbf{Payoff}& \textbf{Function} & $\mathbf{S_i/S_i^{tot}}$ & $\mathbf{\sum_iS_i}$ & $\mathbf{d_T}$ & $\mathbf{d_S}$ & $\mathbf{d_A}$ & \textbf{Effect of} $\mathbf{\epsilon}$ \\
  \midrule
  European & Price & 0.49 & 0.68 & 32 & $< 32$ & 1.40 & - \\
                    & Delta & 0.26$\to$0.23 & 0.77 & 32 & $< 32$ & 3.2 & small \\
                    & Gamma & $10^{-4}\to 10^{-2}$ & $10^{-4}\to 10^{-2}$ & 32 & $ 32$ & 32 & high \\
                    & Vega & 0.33 & 0.543 & 32 & $< 32$ & 1.64 & no \\
  \hline
  Asian    & Price & 0.54$\to$0.43 & 0.714 & $< 32$ & $< 32$ & 1.38 & - \\
                   & Delta & 0.32$\to 10^{-2}$ & 0.71$\to$0.74 & 32 & $< 32$ & 3.5 & small \\
                   & Gamma & $10^{-4}\to 10^{-2}$ & $10^{-4}\to 10^{-2}$ & 32 & $ 32$ & $31\to 25$ & high \\
                   & Vega & 0.42$\to$0.01 & 0.611 & $< 32$ & $< 32$ & 1.57 & no \\
  \hline
  Double KO& Price & 0.01$\to$0.15 & 0.22 & 32 & $< 32$ & 8.5 & - \\
                   & Delta & 0.01$\to 0.12$ & 0.22 & 32 & $< 32$ & 7.6 & no \\
                   & Gamma & $10^{-5}\to 10^{-7}$ & $10^{-4}\to 10^{-2}$ & 32 & $ 32$ & $31.2\to 29.8$ & high \\
                   & Vega & $10^{-5}\to 10^{-8}$ & $10^{-4}\to 10^{-2}$ & 32 & $ 32$ & 28 & high \\
  \hline
  Cliquet  & Price & 1 & 1 & 32 & 1 & 1 & - \\
                   & Vega & 1 & 1 & 32 & 1 & 1 & no \\
  \bottomrule
\end{tabular}}
\caption{Summary of GSA metrics and effective dimensions of prices
and greeks for SD scheme. Arrow ``$\to$'' in the column for
${S_i/S_i^{tot}}$ denotes the change in the value with the
increase of index $i$ and/or with the increase of shift parameter
$\epsilon$; in the column for ${\sum_iS_i}$ it denotes the change
in the value with the increase of shift parameter $\epsilon$. The
numerical computation of the figures in this table required $N_p
\times (D+2)\times N_\epsilon \times N = 18 \times 34 \times 3
\times 2^{17} = 240,648,192$ function evaluations. We show
significant digits only, we do not show MC errors because of
limited space.} \label{tab:1}
\end{table}
\clearpage
From these results we draw the following conclusions.
\begin{enumerate}
\item European option (Figure \ref{fig:1}): price, delta and vega are type B functions, while gamma is type C function.
\item Asian option (Figure \ref{fig:2}): price, and vega are type B functions, while delta and gamma are type C function.
\item Double KO option (Figure \ref{fig:3}): price and all greeks are type C functions.
\item Cliquet option (Figure \ref{fig:4}): price and vega are type B functions with $d_S=1$ (delta and gamma for a Cliquet option are null). We recall that $d_S=1$ means that there are no interactions among variables.
\end{enumerate}
\par
\begin{figure}[ht]
\centering
\subfigure[Price]{\includegraphics[width=2.5in,height=1.9in,keepaspectratio=false]{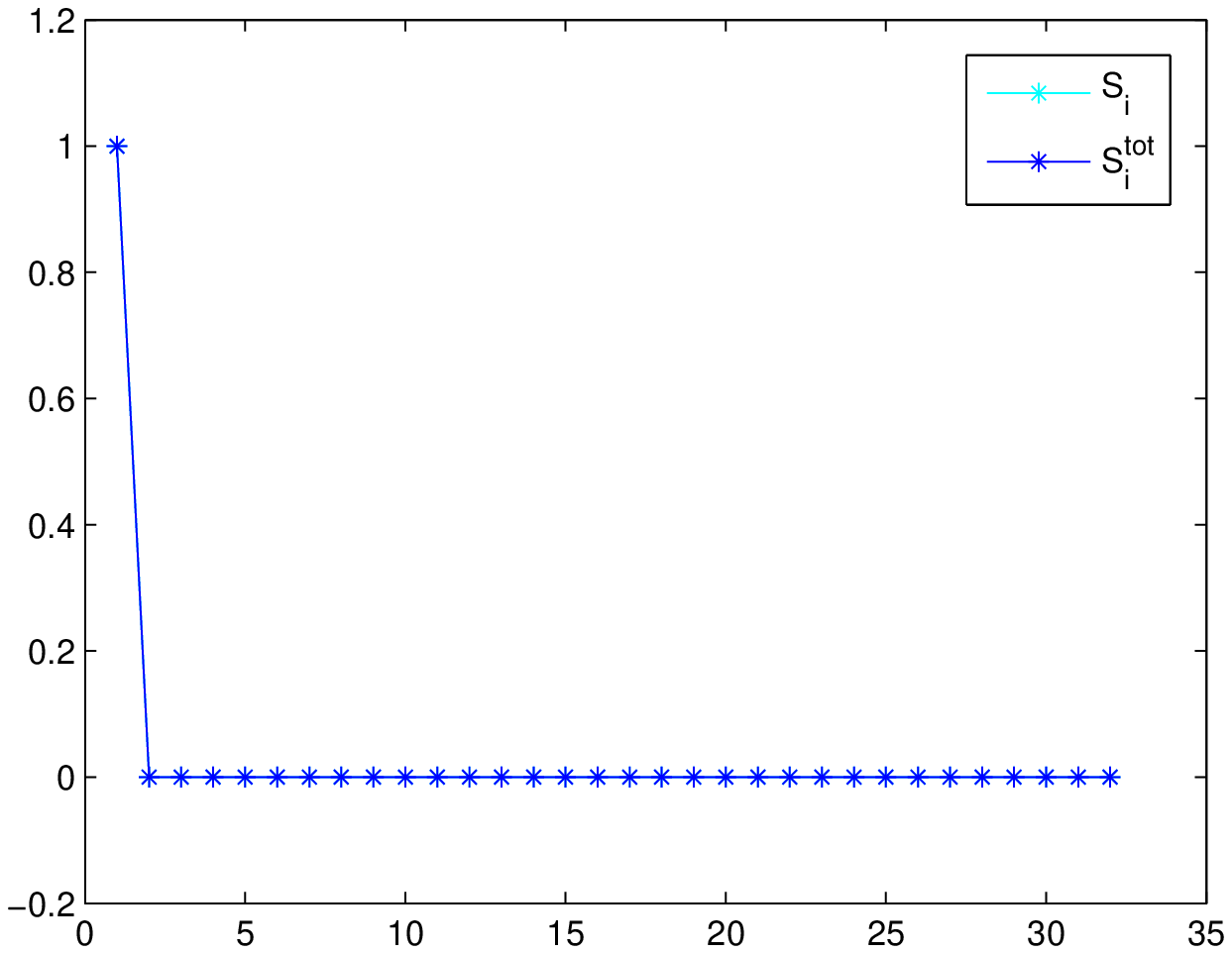}}
\subfigure[Delta]{\includegraphics[width=2.5in,height=1.9in,keepaspectratio=false]{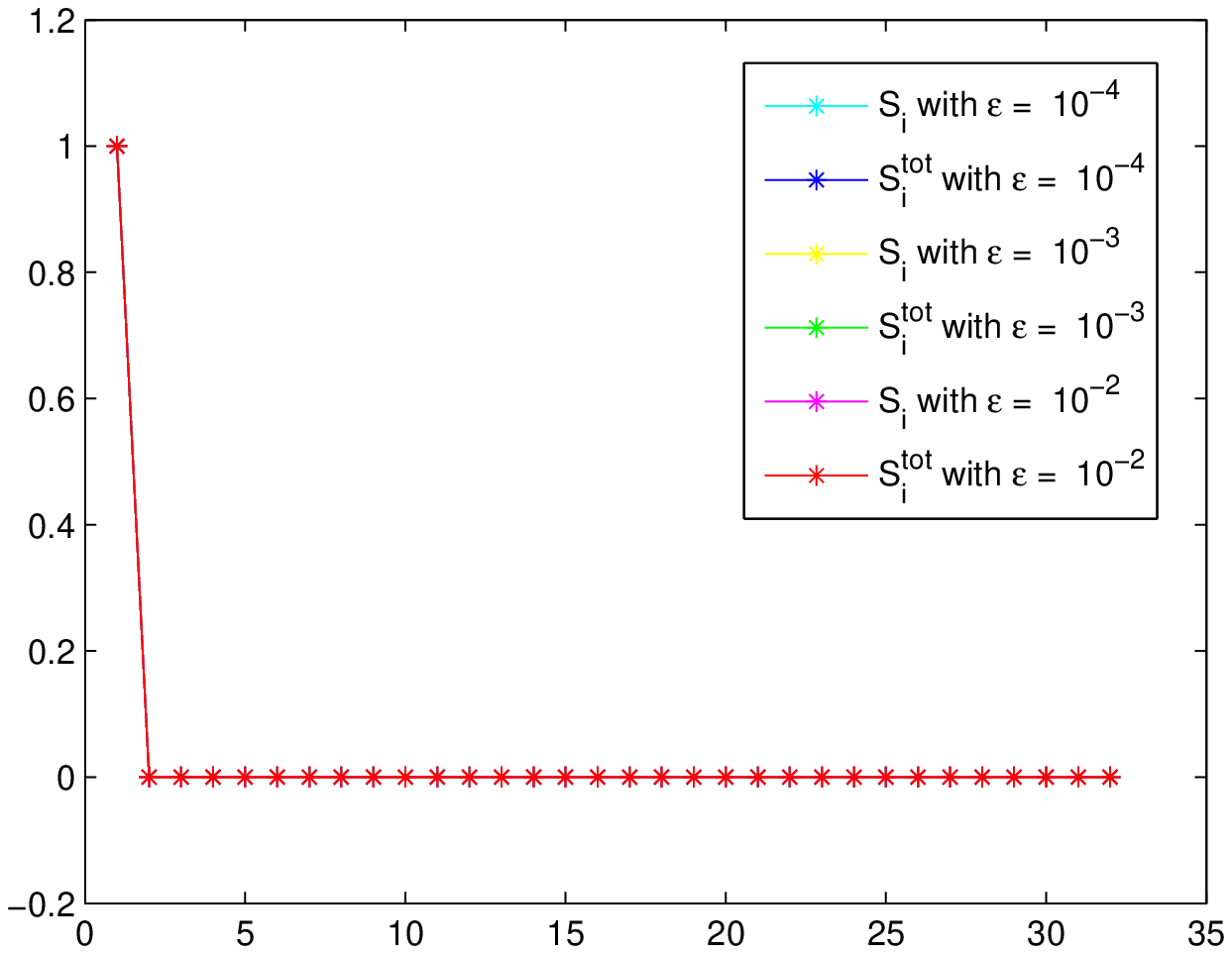}}
\subfigure[Gamma]{\includegraphics[width=2.5in,height=1.9in,keepaspectratio=false]{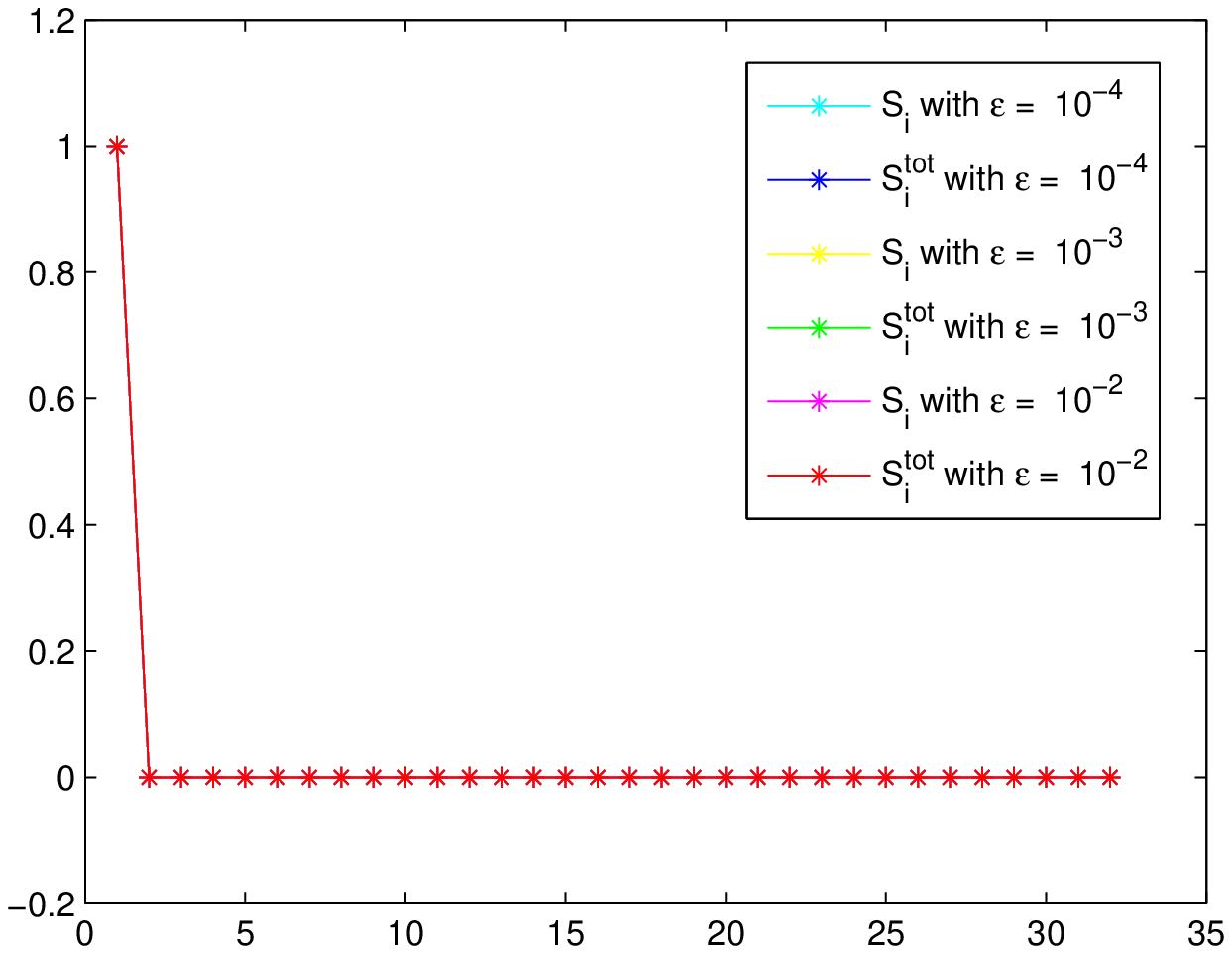}}
\subfigure[Vega]{\includegraphics[width=2.5in,height=1.9in,keepaspectratio=false]{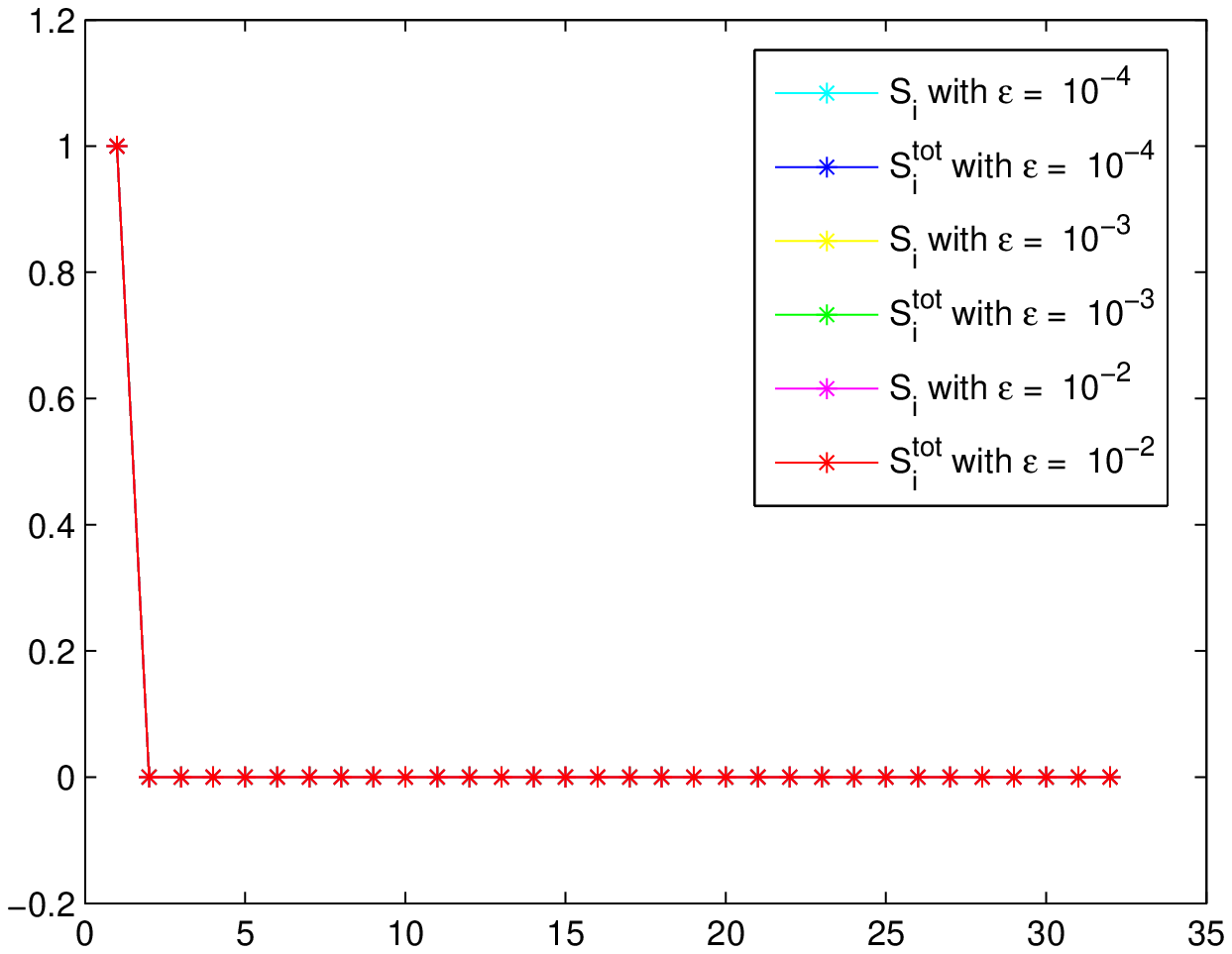}}
\caption{European call option price $(a)$ and greeks $(b),(c),(d)$, BBD, $D=32$. First order Sobol' indices $S_{i}$ and total sensitivity indices $S_{i}^{tot}$ versus time step $i$. }
\label{fig:5}
\end{figure}
\begin{figure}[ht]
\centering
\subfigure[Price]{\includegraphics[width=2.5in,height=1.9in,keepaspectratio=false]{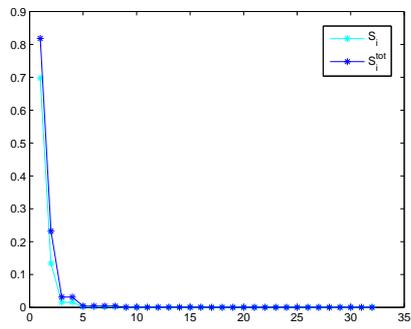}}
\subfigure[Delta]{\includegraphics[width=2.5in,height=1.9in,keepaspectratio=false]{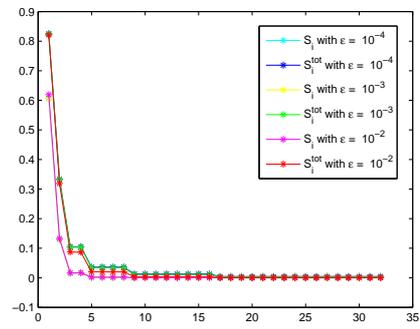}}
\subfigure[Gamma]{\includegraphics[width=2.5in,height=1.9in,keepaspectratio=false]{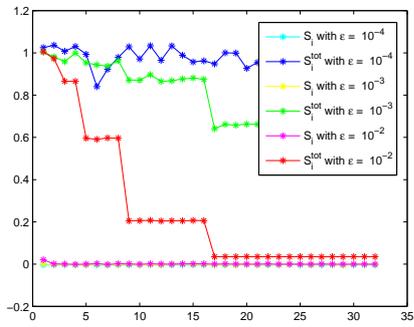}}
\subfigure[Vega]{\includegraphics[width=2.5in,height=1.9in,keepaspectratio=false]{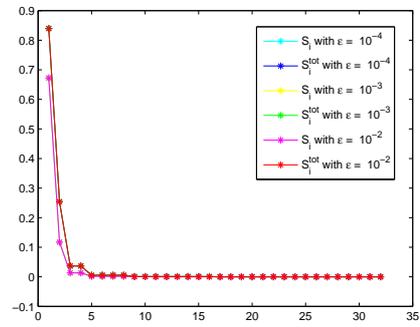}}
\caption{Asian call option. Details as in Figure \ref{fig:5}.}
\label{fig:6}
\end{figure}
\begin{figure}[ht]
\centering
\subfigure[Price]{\includegraphics[width=2.5in,height=1.9in,keepaspectratio=false]{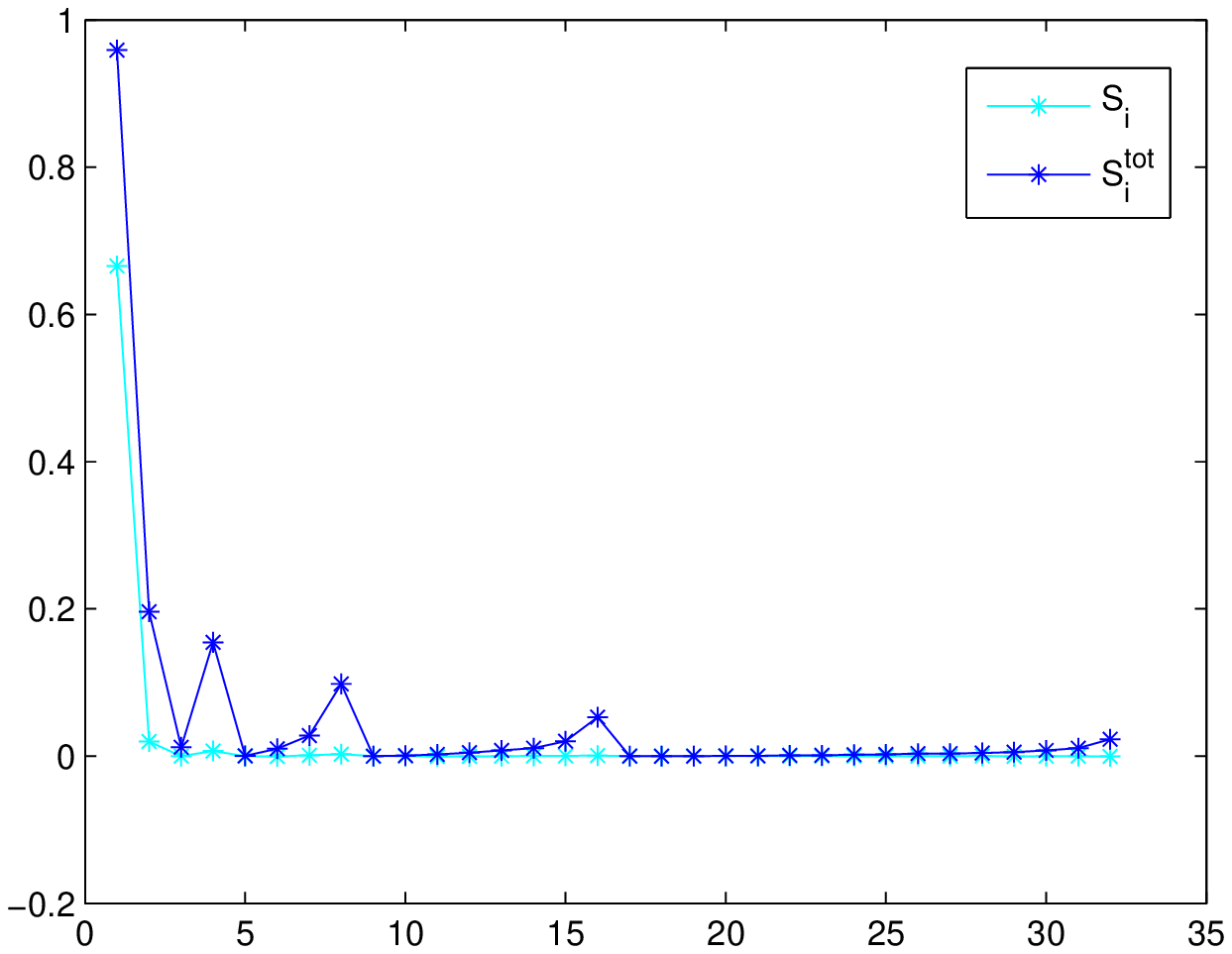}}
\subfigure[Delta]{\includegraphics[width=2.5in,height=1.9in,keepaspectratio=false]{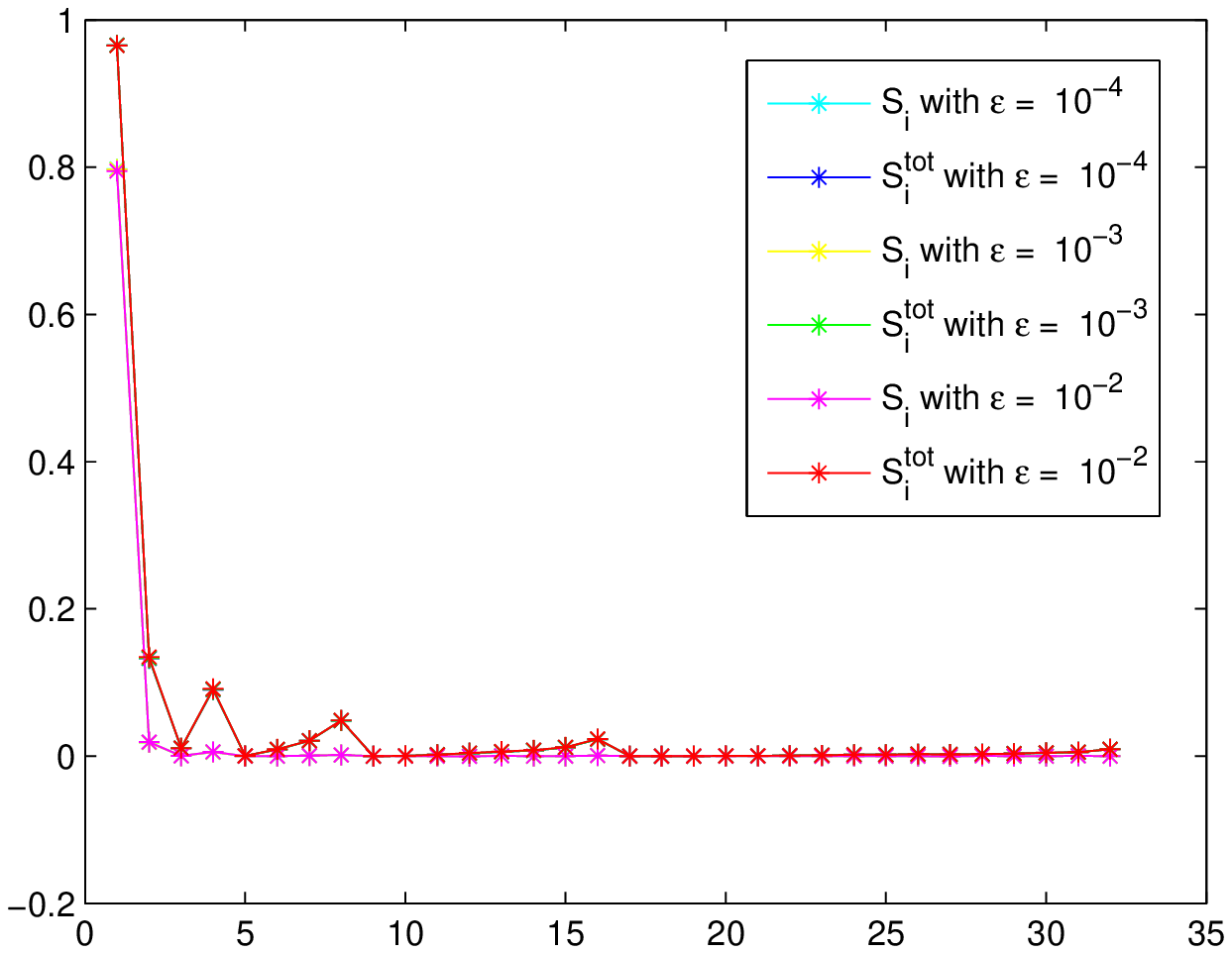}}
\subfigure[Gamma]{\includegraphics[width=2.5in,height=1.9in,keepaspectratio=false]{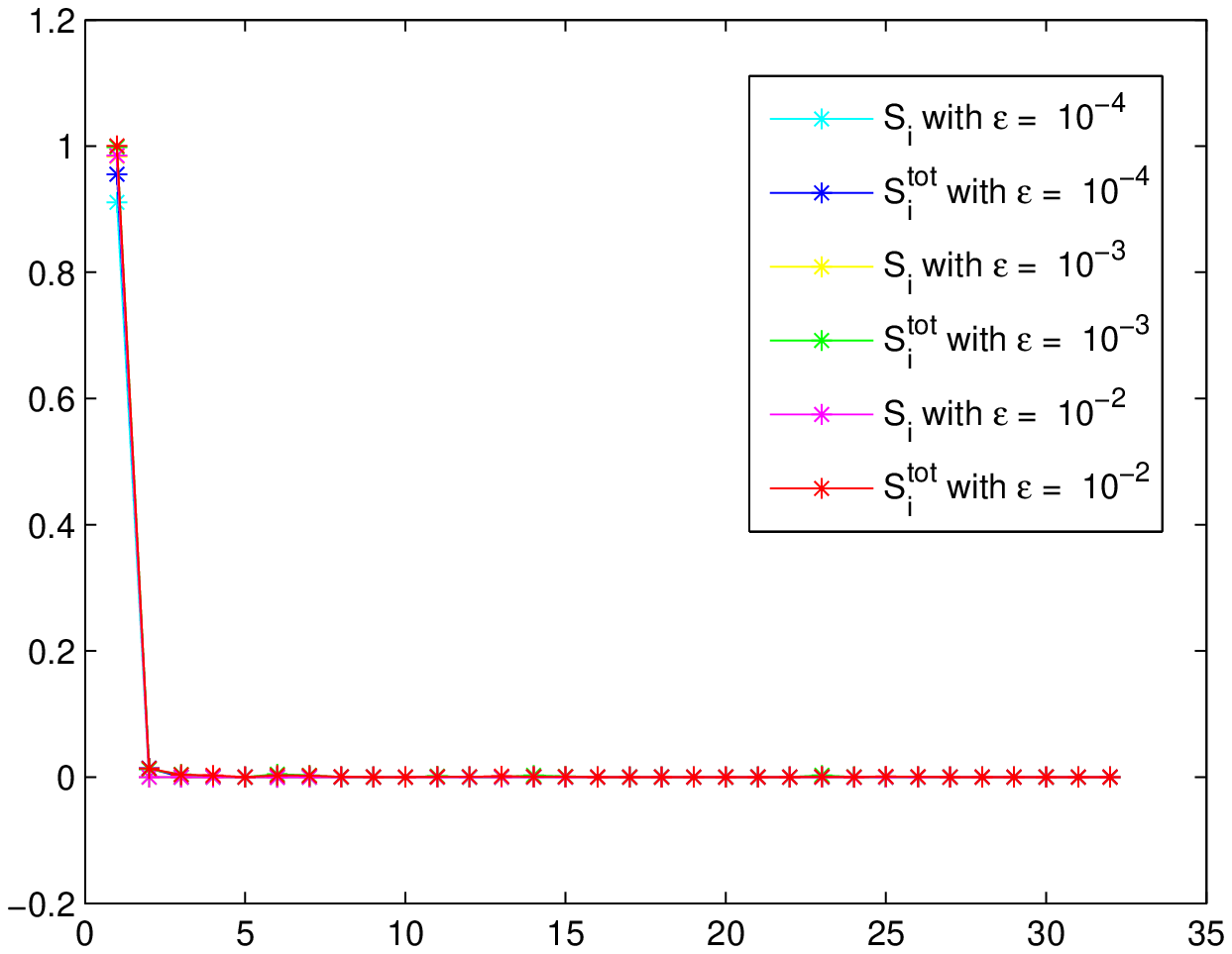}}
\subfigure[Vega]{\includegraphics[width=2.5in,height=1.9in,keepaspectratio=false]{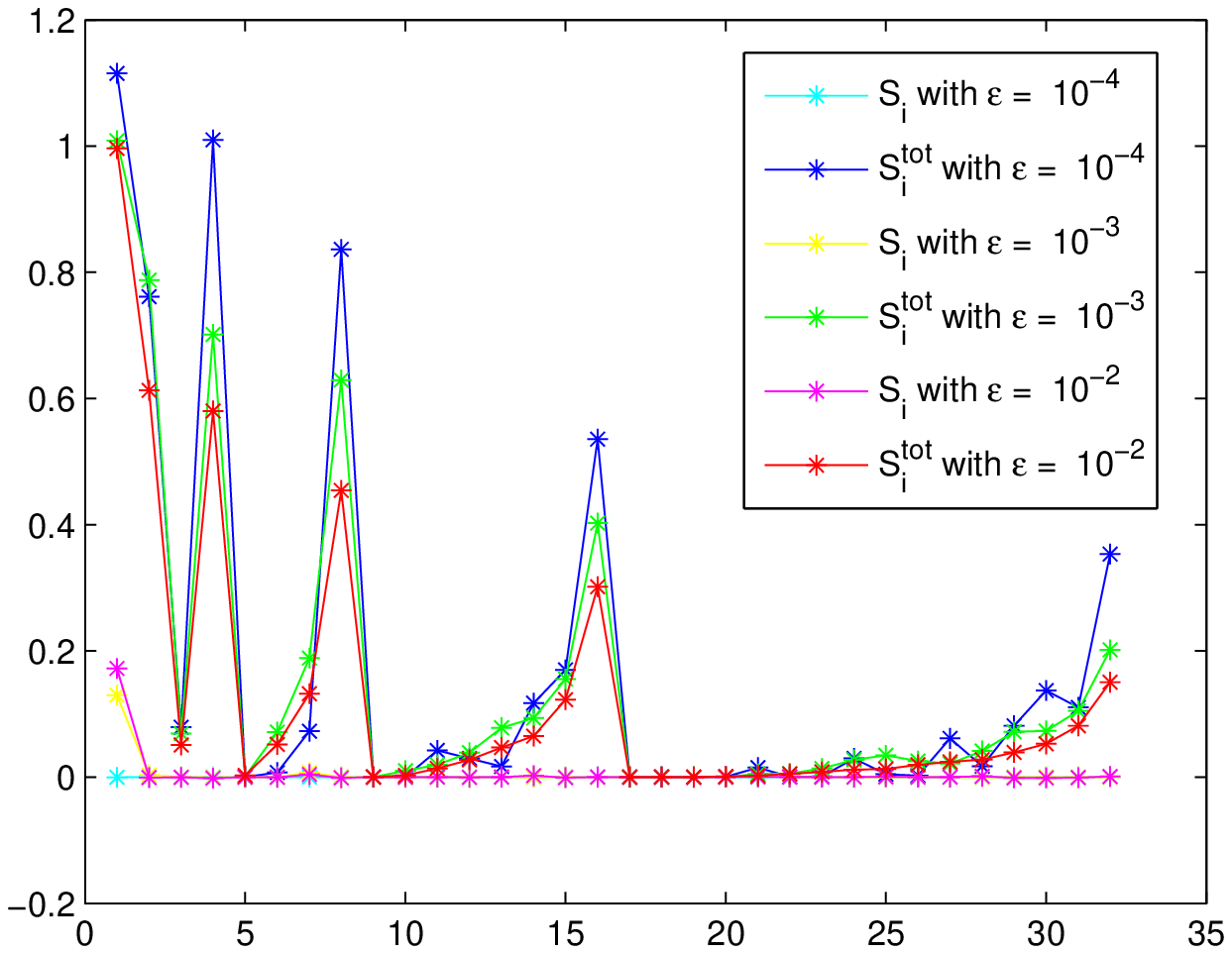}}
\caption{Double Knock-out call option. Details as in Figure \ref{fig:5}.}
\label{fig:7}
\end{figure}
\begin{figure}[ht]
\centering
\subfigure[Price]{\includegraphics[width=2.5in,height=1.9in,keepaspectratio=false]{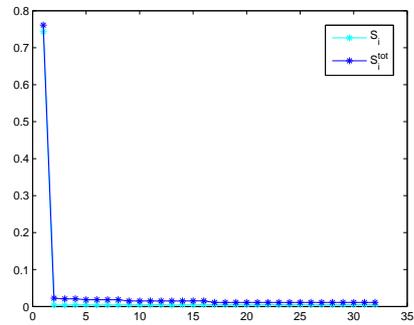}}
\subfigure[Vega]{\includegraphics[width=2.5in,height=1.9in,keepaspectratio=false]{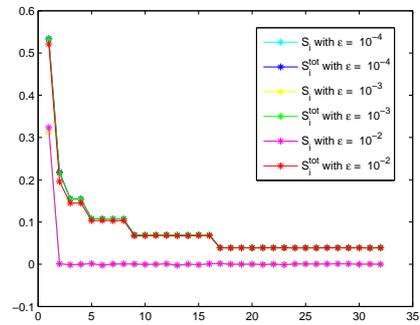}}
\caption{Cliquet option. Details as in Figure \ref{fig:5}.}
\label{fig:8}
\end{figure}

The analogous results of GSA for BBD are shown in Figures \ref{fig:5}-\ref{fig:8} and in Table \ref{tab:2}.
\begin{table}[h]\small
\centering
\subtable{%
\begin{tabular}{c c c c c c c c}
  \toprule
  \textbf{Payoff}& \textbf{Function} & $\mathbf{S_i/S_i^{tot}}$ & $\mathbf{\sum_iS_i}$ & $\mathbf{d_T}$ & $\mathbf{d_S}$ & $\mathbf{d_A}$ & \textbf{Effect of} $\mathbf{\epsilon}$ \\
  \midrule
  European & Price & 1 & 1 & 1 & 1 & 1 & - \\
           & Delta & 1 & 1 & 1 & 1 & 1 & no \\
           & Gamma & 1 & 1 & 1 & 1 & 1 & no \\
           & Vega & 1 & 1 & 1 & 1 & 1 & no \\
  \hline
  Asian    & Price & 0.853$\to$0.4 & 0.875 & 2 & $\leq$2 & 1.13 & - \\
           & Delta & 0.733$\to$0.01 & 0.778 & 4 & $\leq$4 & $1.68\to 1.43$ & small \\
           & Gamma & $10^{-2}\to 10^{-4}$ & $0.022\to 10^{-4}$ & 32 & 32 & $31\to 8$ & high \\
           & Vega & 0.802$\to$0.03 & 0.827 & 2 & $\leq$2 & 1.20 & no \\
  \hline
  Double KO& Price & 0.70$\to$0.01 & 0.70 & $\simeq 2$ & $\leq$2 & 1.63 & - \\
           & Delta & 0.83$\to$0.01 & 0.83 & 2 & $\leq$2 & 1.37 & no \\
           & Gamma & 1 & $1\to 0.95$ & 1 & 1 & 1.0 & small \\
           & Vega & $10^{-4}\to 0.2$ & $10^{-6}\to 10^{-4}$ & 32 & 32 & $4.8\to 3.9$ & high \\
  \hline
  Cliquet  & Price & 0.978$\to$0.2 & 0.892 & $\simeq 2$ & $\leq 2$ & 1.19 & - \\
           & Vega & 0.595$\to$0.001 & 0.32 & $\simeq 32$ & $\leq 32$ & 2.6 & no \\
  \bottomrule
\end{tabular}}
\caption{Summary of GSA metrics and effective dimensions of prices
and greeks for BBD scheme. Details as in Table \ref{tab:1}.}
\label{tab:2}
\end{table}
\clearpage
From these results we draw the following conclusions.
\begin{enumerate}
\item European option (Figure \ref{fig:5}): price and all greeks are type A functions with $d_S=1$. The value of sensitivity indexes for the first input, corresponding to the terminal value $t = T$, is $\simeq 1$, while the following variables have sensitivity indexes $\simeq 0$. Clearly, BBD is much more efficient than SD.
\item Asian option (Figure \ref{fig:6}): price, delta and vega are type A functions. Comments as for the European option above. Gamma remains a type C function as for SD.
\item Double KO option (Figure \ref{fig:7}): price, delta and gamma are type A functions. Comments as for the Asian option above. Vega remains a type C function as for SD.
\item Cliquet option (Figure \ref{fig:8}): price is a type A function. Similarly to the European option, the value of sensitivity indexes for the first input, corresponding to the terminal value $t = T$, is $\simeq 1$, while the following values of $S_i$ are $\simeq 0$. Vega is a type C function, since the ratio $S_i/S_i^{tot}$ reaches small values revealing interacting variables. Thus in this case BBD is much less efficient than SD.
\end{enumerate}
In conclusion, prices and greeks are always Type B or C functions
for QMC+SD (Table \ref{tab:1}), while they are predominantly Type
A functions, with a few exceptions, for QMC+BBD (Table
\ref{tab:2}). In most cases switching from SD to BBD reduces the
effective dimension in the truncation sense $d_T$.
\par
The different efficiency of QMC+BBD vs QMC+SD is completely
explained by the properties of Sobol' low discrepancy sequences.
The initial coordinates of Sobol' LDS are much better distributed
than the later high dimensional coordinates
\cite{Gla03,CafMor1997}. The BBD changes the order in which inputs
(linked with time steps) are sampled. As follows from GSA, in most
cases for  BBD the low index variables (terminal values of time
steps, mid-values and so on) are much more important than higher
index variables. The BBD uses lower index, well distributed
coordinates from each $D$-dimensional LDS vector to determine most
of the structure of a path, and reserves other coordinates to fill
in finer details. That is, well distributed coordinates are used
for important variables and other not so well distributed
coordinates are used for far less important variables. This
results in a significantly improved accuracy of QMC integration.
However, this technique does not always improve the efficiency of
the QMC method as \eg for Cliquet options: in this case GSA
reveals that for SD all inputs are equally important and,
moreover, there are no interactions among them, which is an ideal
case for application of Sobol' low discrepancy sequences; the BBD,
on the other hand, favoring higher index variables destroys
independence of inputs introducing interactions, which leads to
higher values of $d_S$ and $d_A$. As a result, we observe
degradation in performance of the QMC method.

\subsection{Performance Analysis}
\label{SecPerformance}
In this section we compare the relative performances of MC and QMC techniques. This analysis is crucial to establish if QMC outperforms MC, and in what sense.
\par
\begin{figure}[ht]
\centering
\subfigure[Price]{\includegraphics[width=3.1in,height=2.4in,keepaspectratio=false]{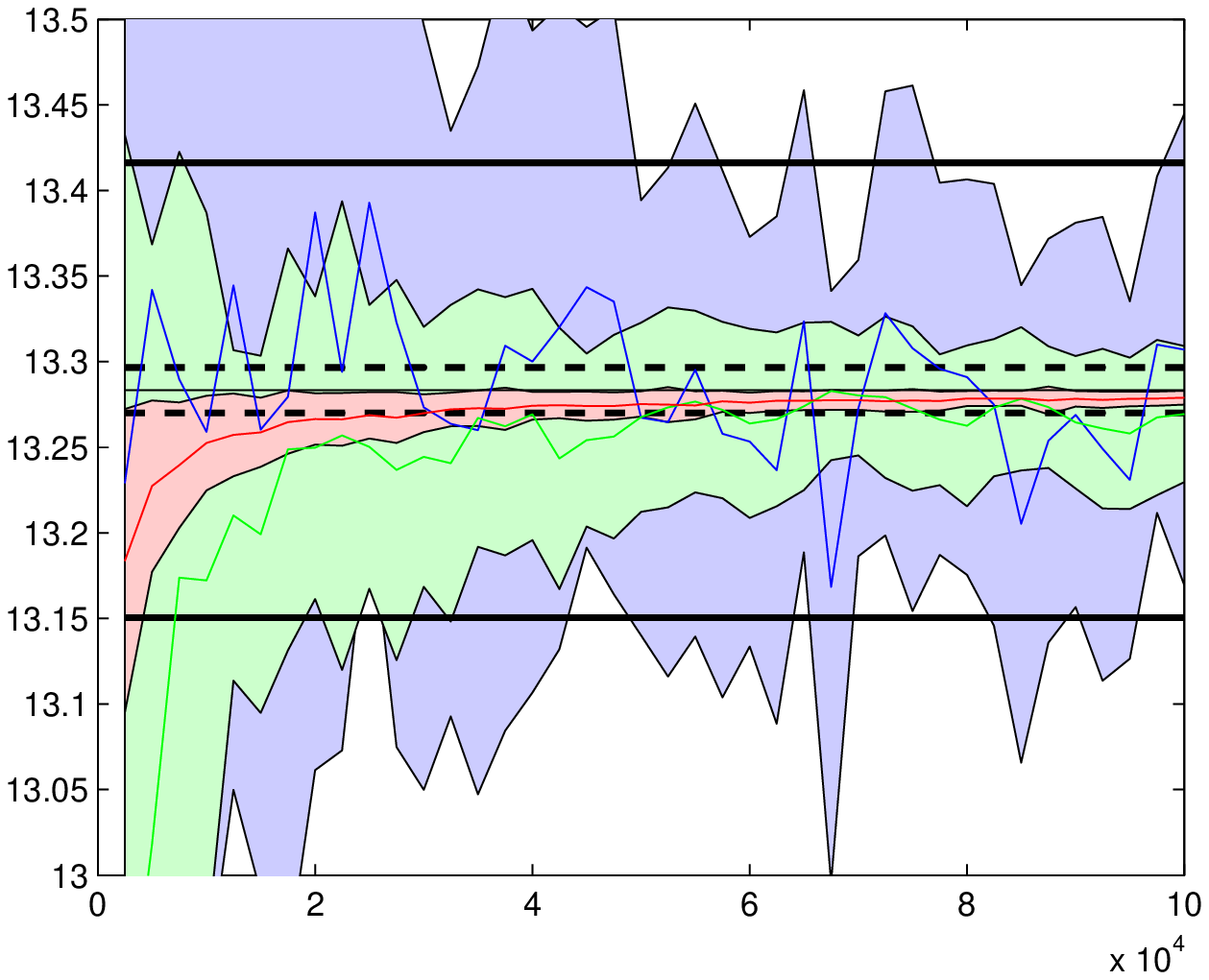}}
\subfigure[Delta]{\includegraphics[width=3.1in,height=2.4in,keepaspectratio=false]{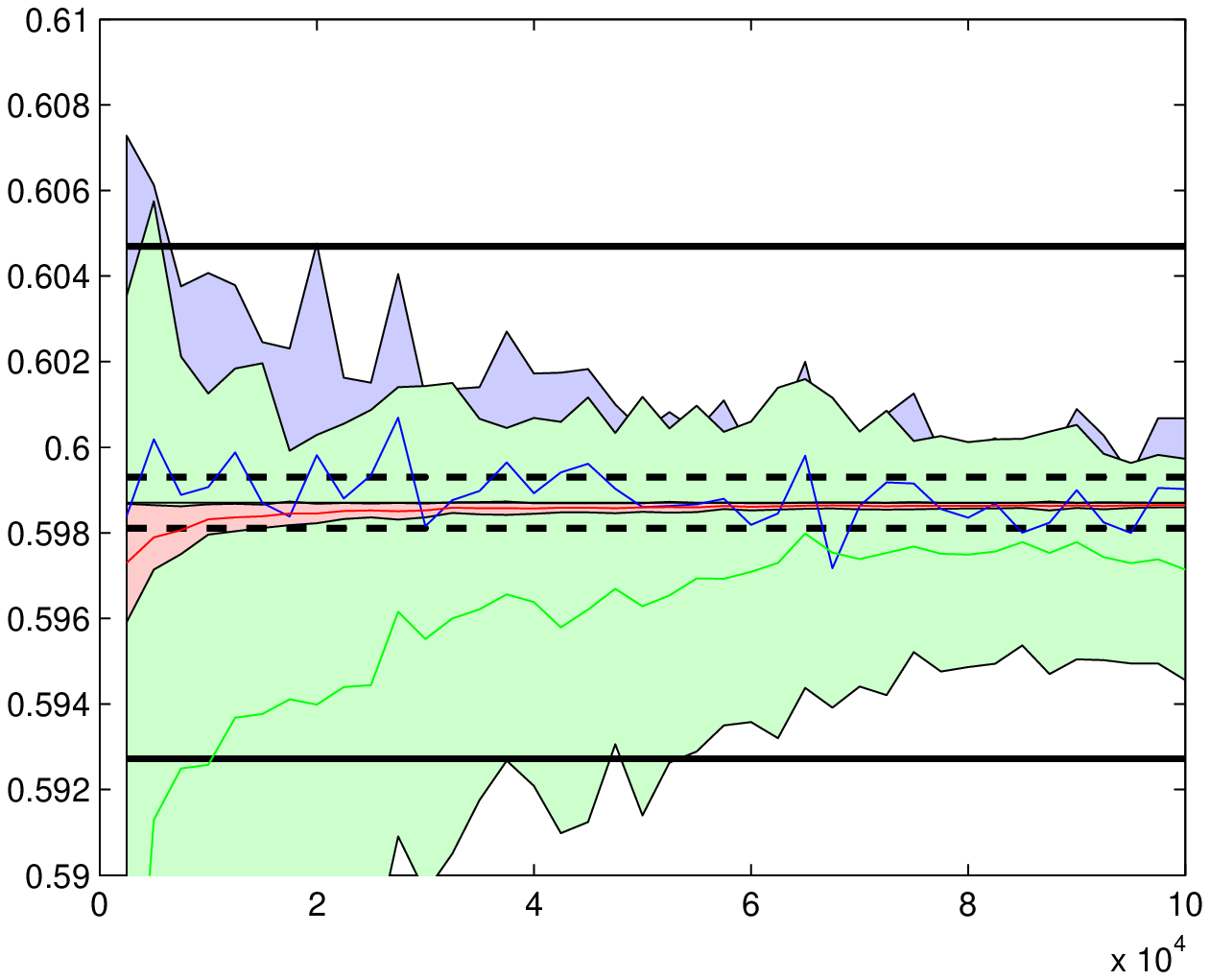}}
\subfigure[Gamma]{\includegraphics[width=3.1in,height=2.4in,keepaspectratio=false]{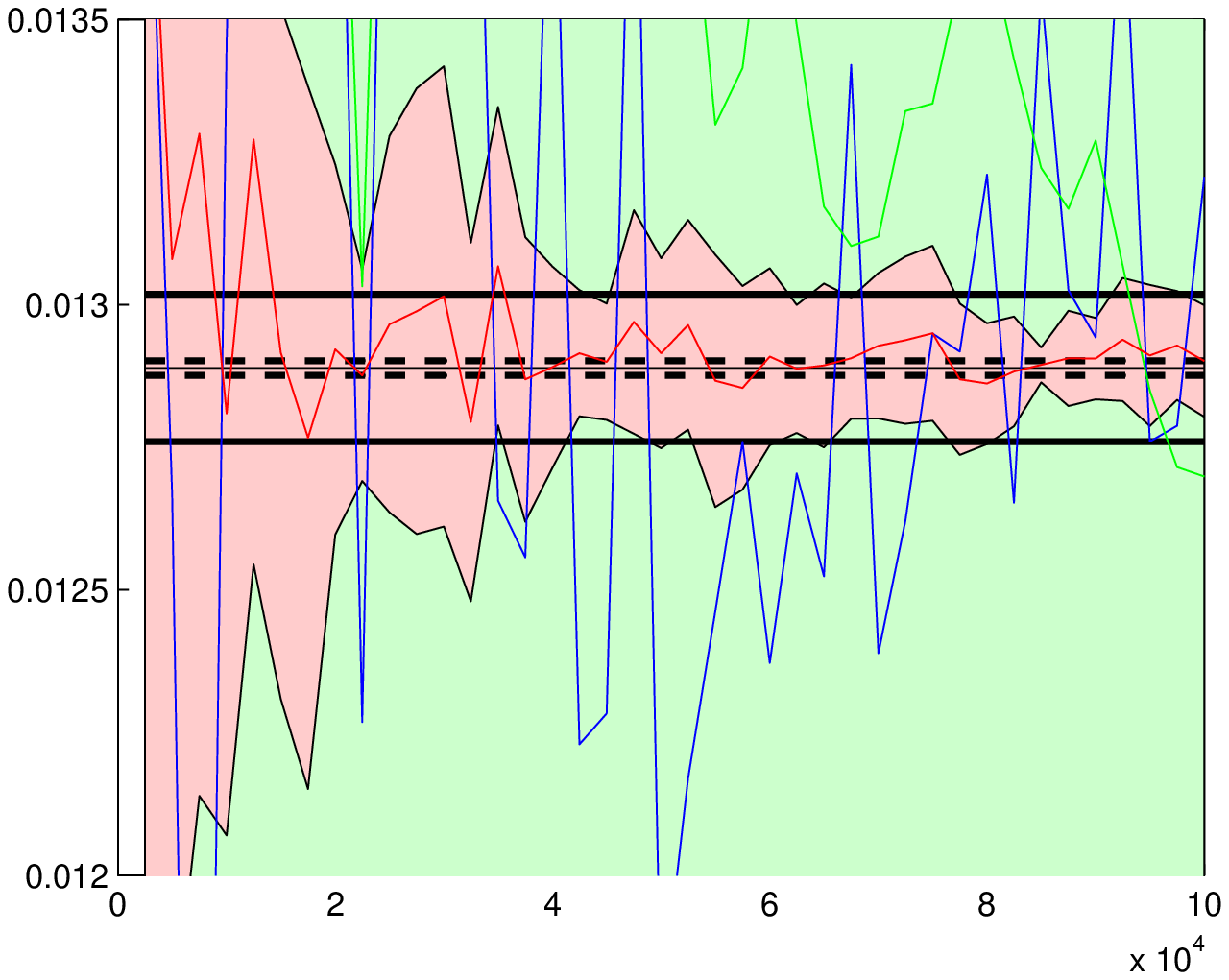}}
\subfigure[Vega]{\includegraphics[width=3.1in,height=2.4in,keepaspectratio=false]{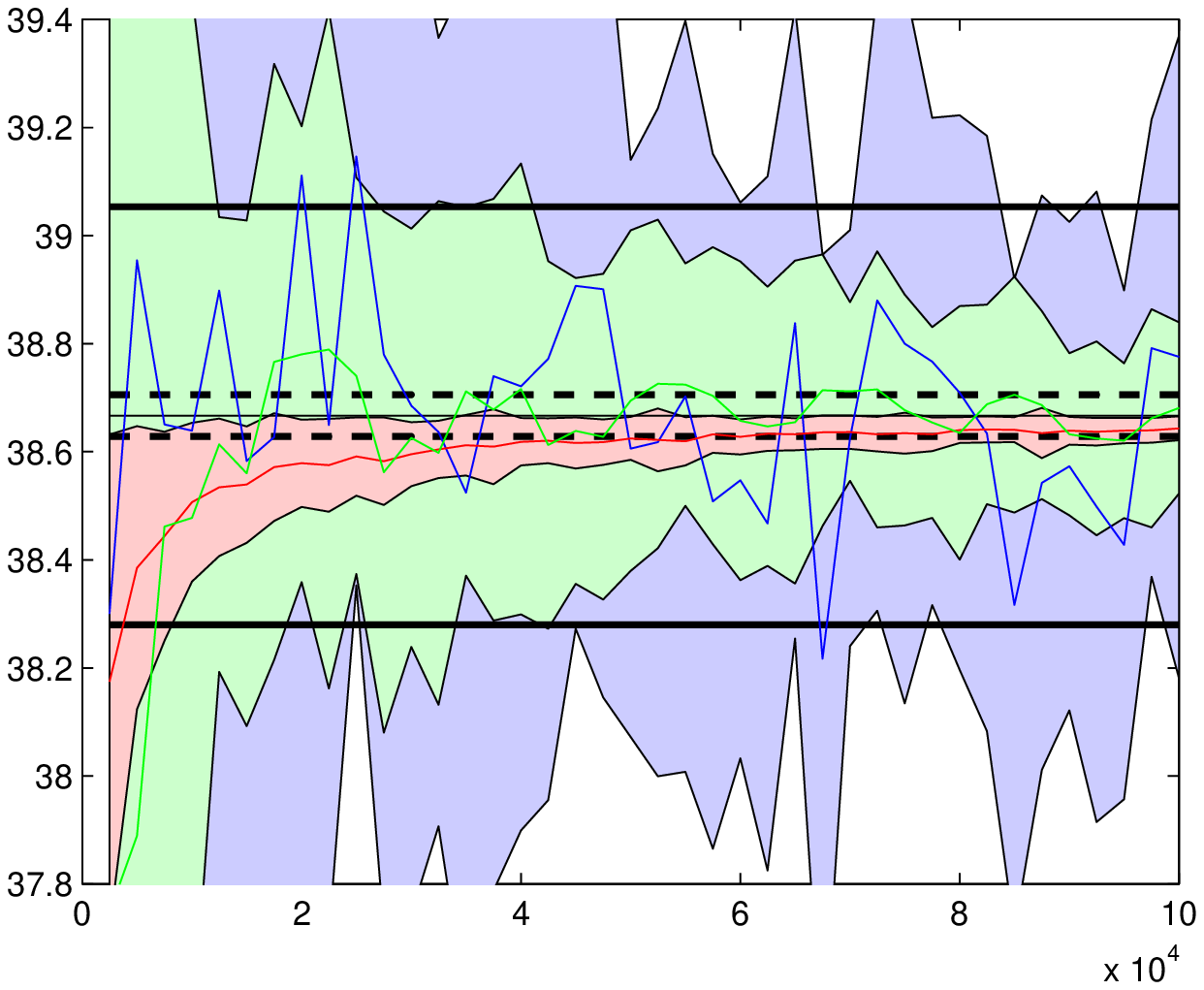}}
\caption{European call option price $(a)$ and greeks $(b),(c),(d)$
convergence diagrams versus number of simulated paths for MC+SD
with antithetic variables (solid blue line), QMC+SD (solid green
line) and QMC+BBD (solid red line). Shaded areas represent 3-sigma
errors around the corresponding run (solid line). 1\% and 0.1\%
accuracy regions are marked by horizontal black solid and dashed
lines, respectively. Number of dimensions is $D=32$. Shift
parameter is $\epsilon=10^{-3}$.} \label{fig:9}
\end{figure}
\begin{figure}[ht]
\centering
\subfigure[Price]{\includegraphics[width=3.1in,height=2.4in,keepaspectratio=false]{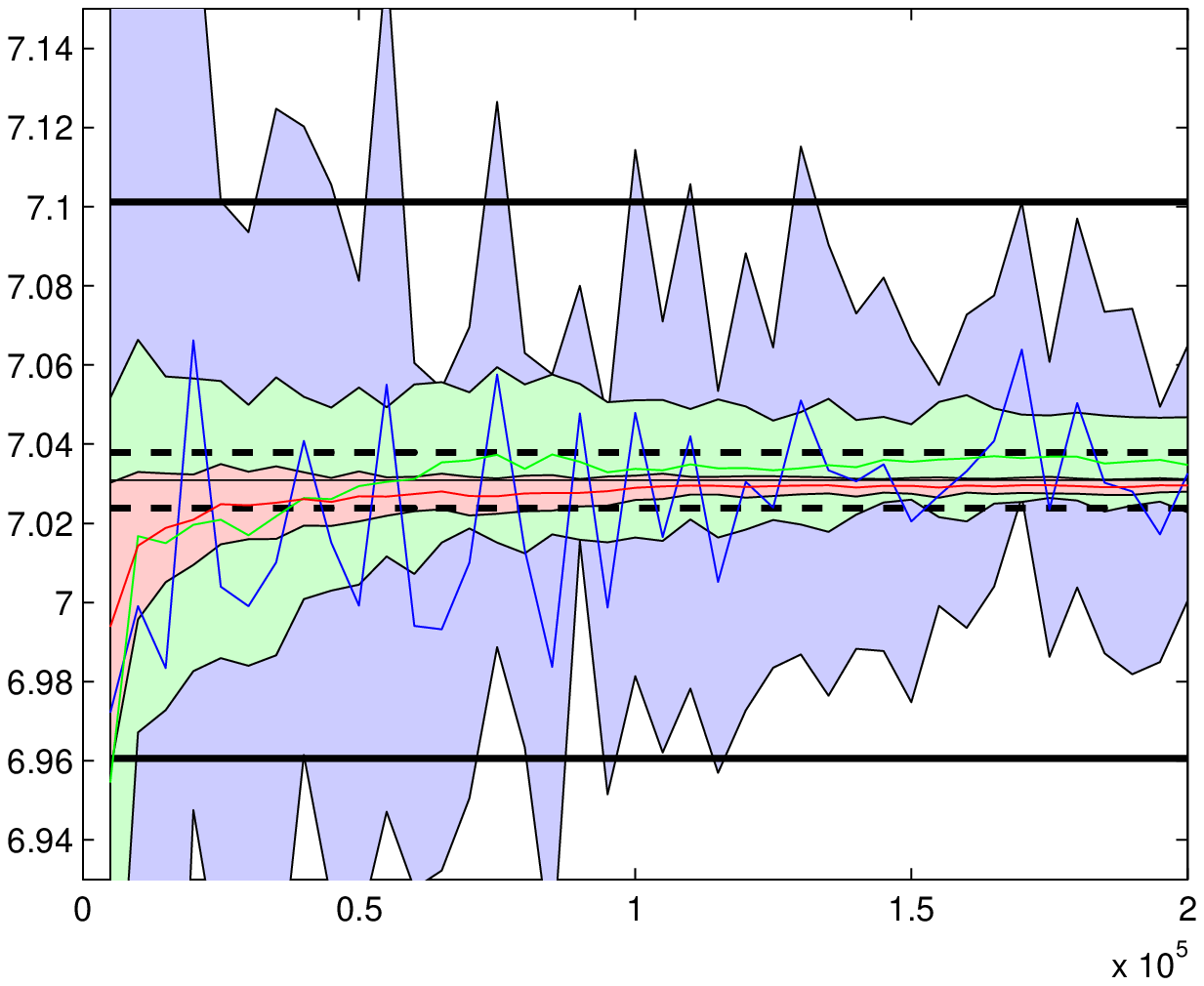}}
\subfigure[Delta]{\includegraphics[width=3.1in,height=2.4in,keepaspectratio=false]{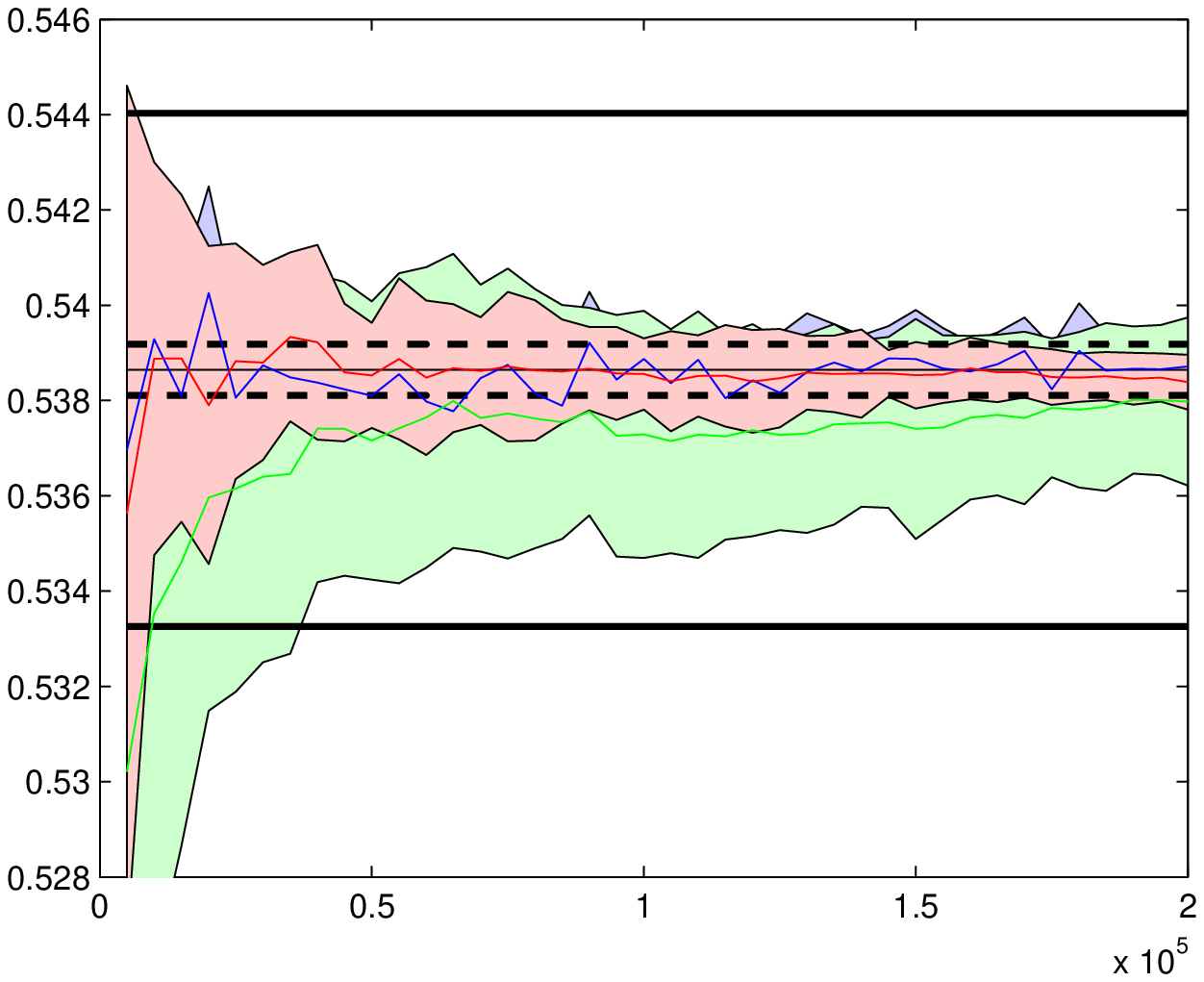}}
\subfigure[Gamma]{\includegraphics[width=3.1in,height=2.4in,keepaspectratio=false]{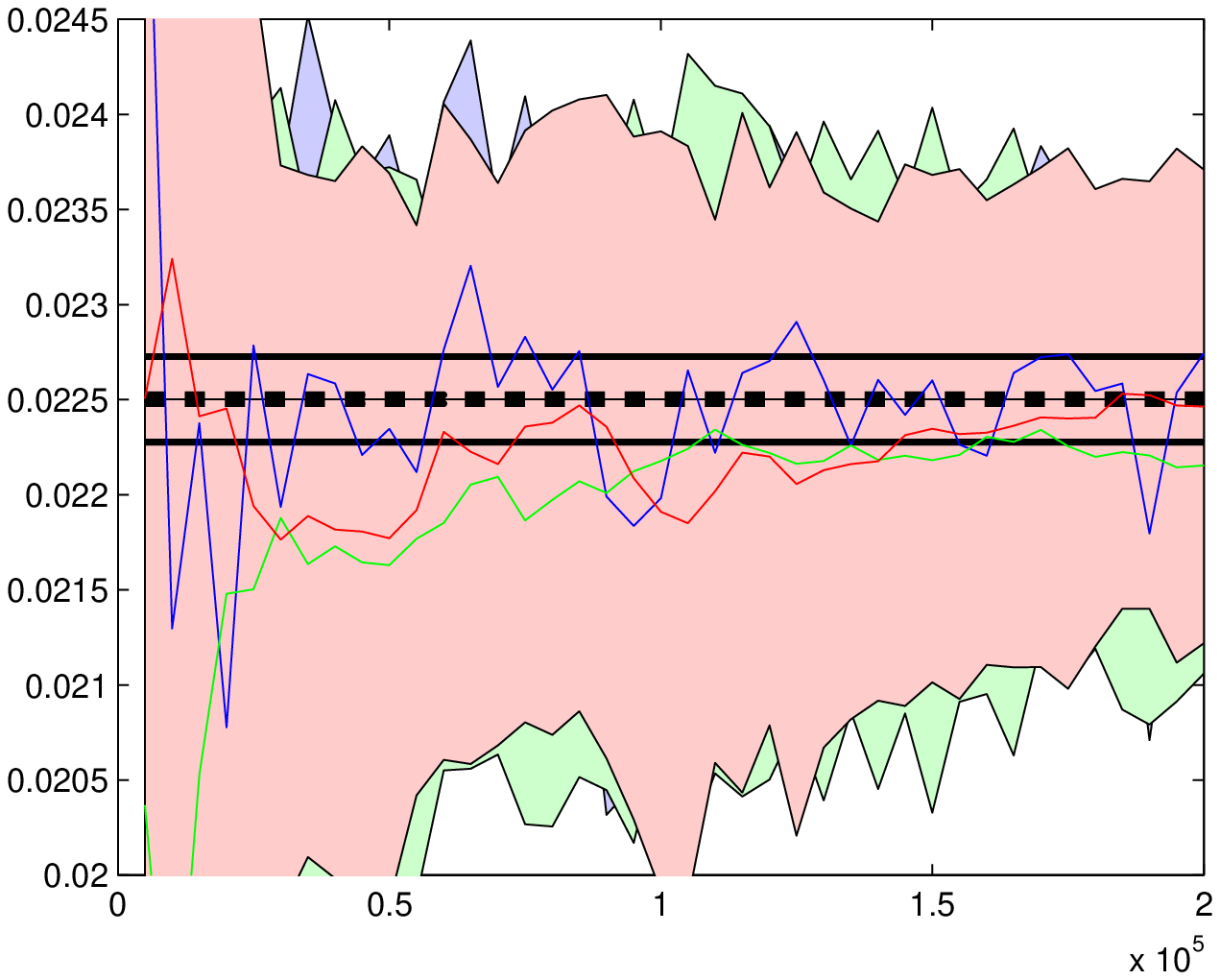}}
\subfigure[Vega]{\includegraphics[width=3.1in,height=2.4in,keepaspectratio=false]{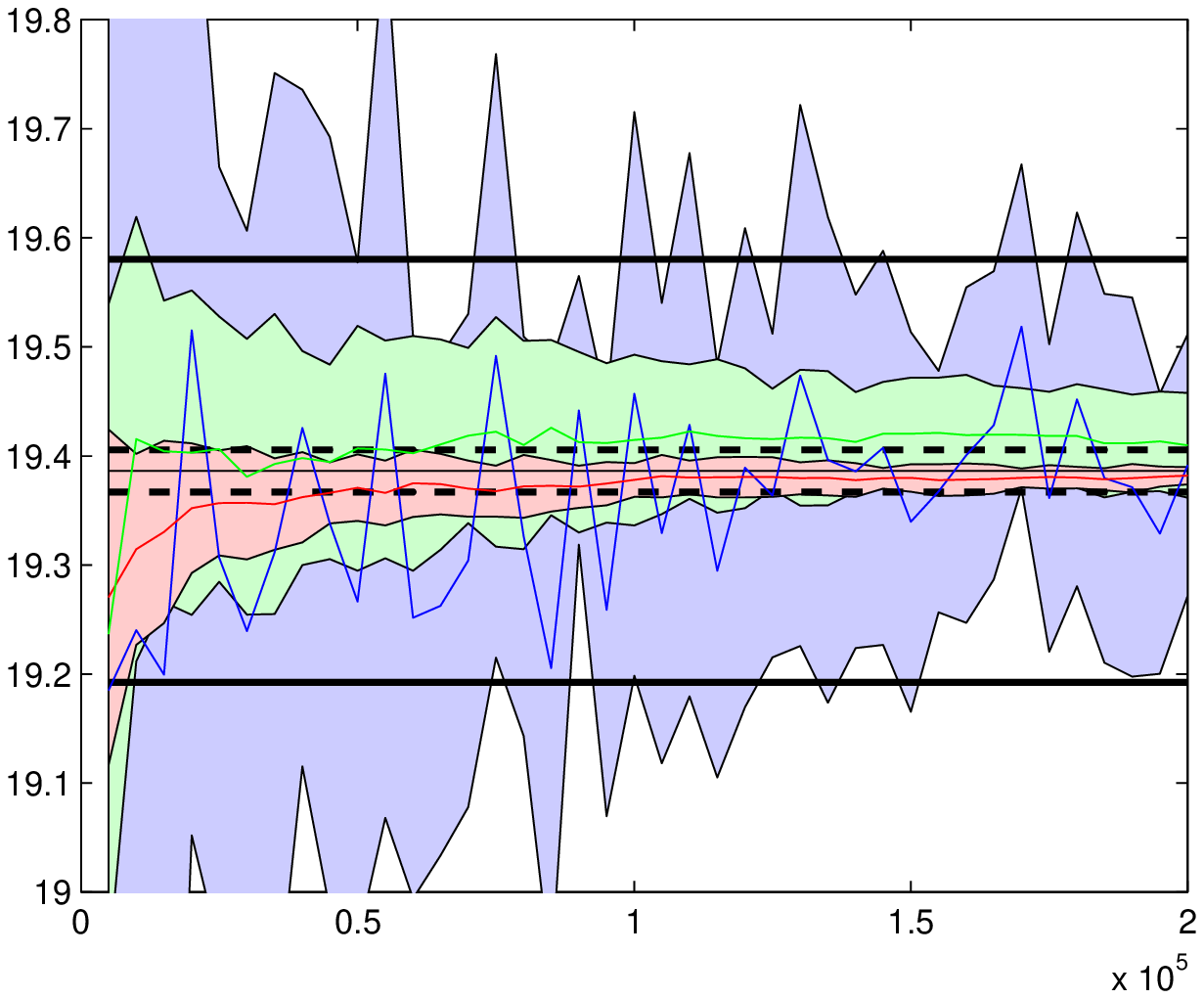}}
\caption{ Asian call option. $\epsilon = 5\times 10^{-3}$. Other details as in Figure \ref{fig:9}.}
\label{fig:10}
\end{figure}
\begin{figure}[ht]
\centering
\subfigure[Price]{\includegraphics[width=3.1in,height=2.4in,keepaspectratio=false]{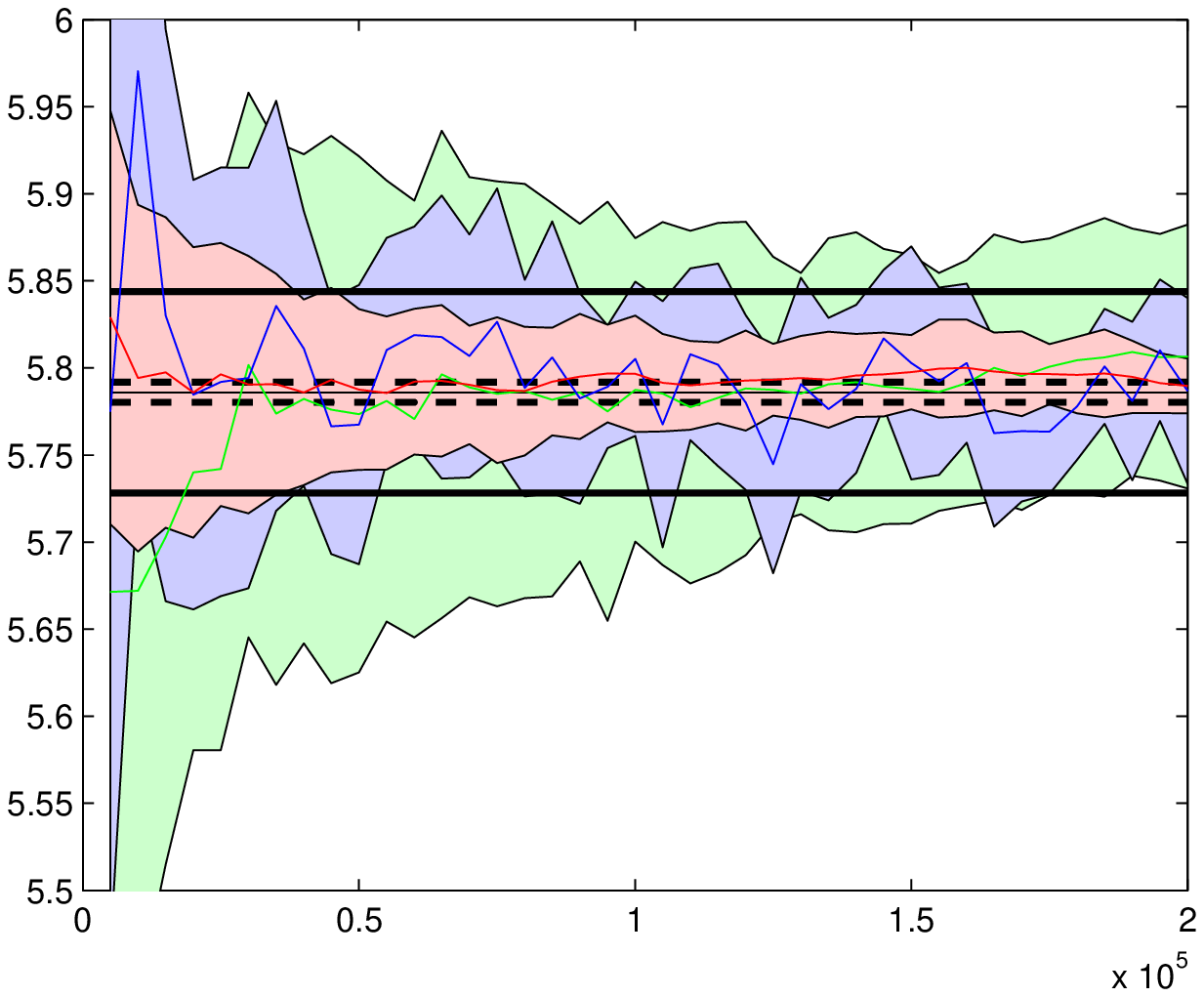}}
\subfigure[Delta]{\includegraphics[width=3.1in,height=2.4in,keepaspectratio=false]{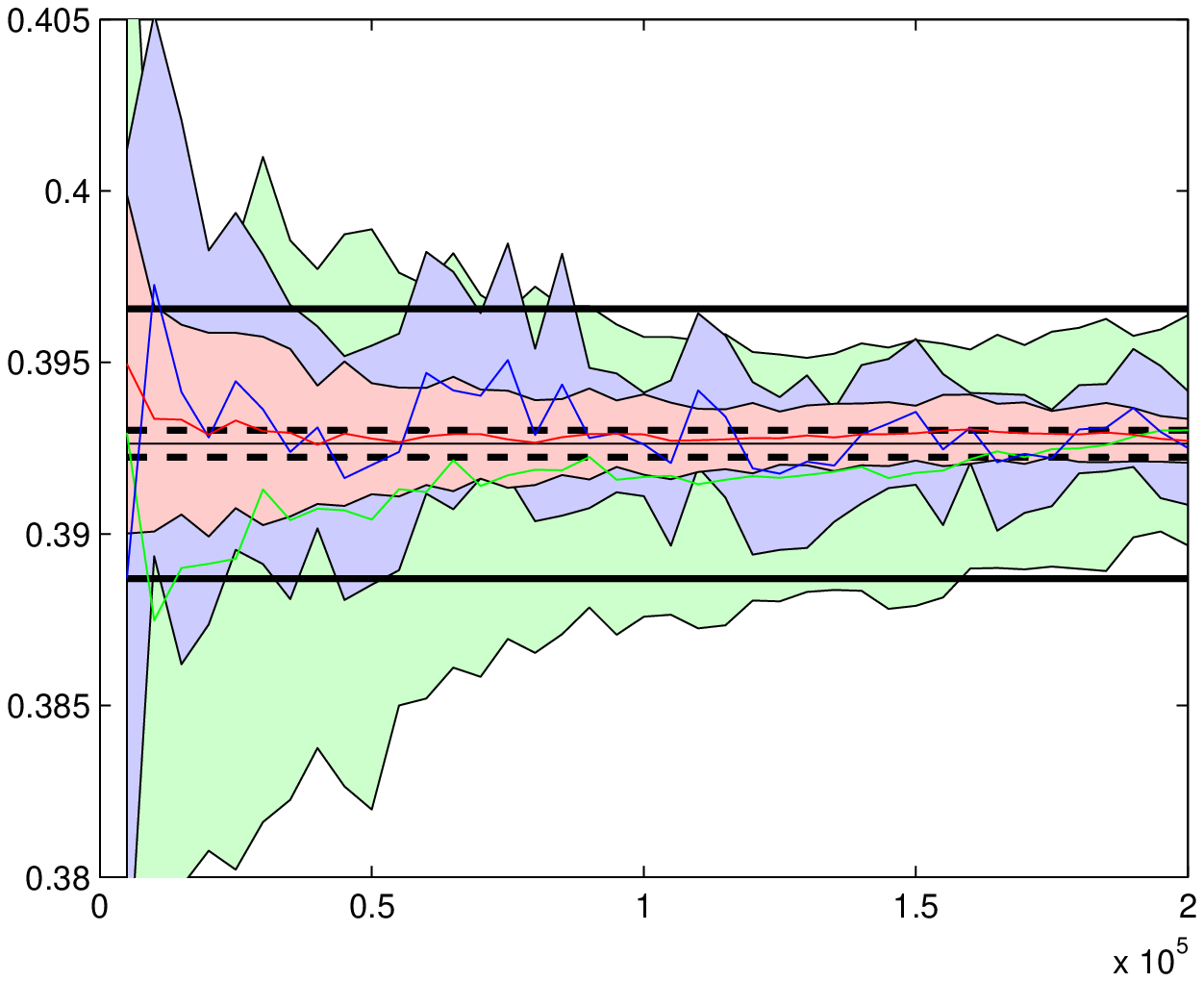}}
\subfigure[Gamma]{\includegraphics[width=3.1in,height=2.4in,keepaspectratio=false]{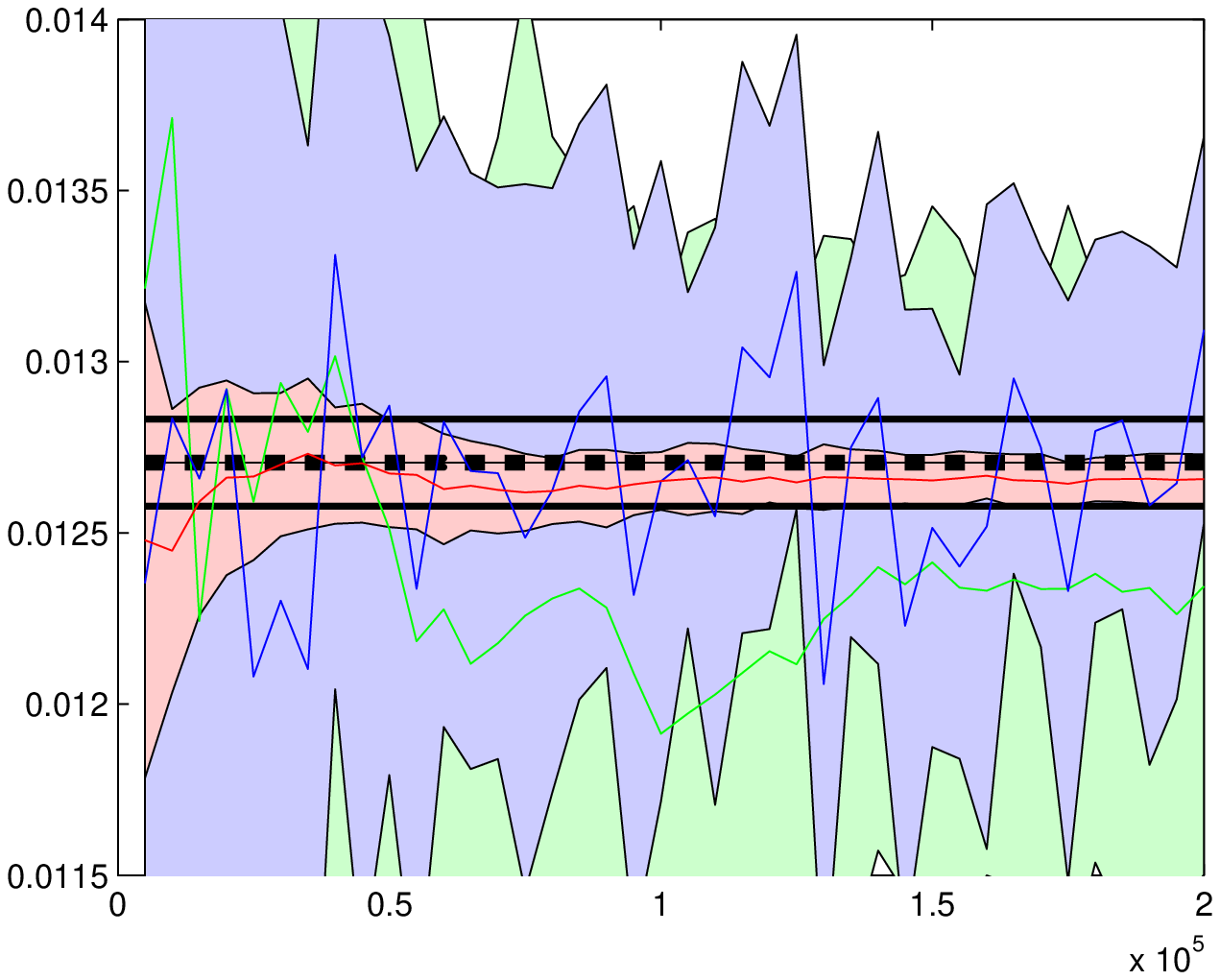}}
\subfigure[Vega]{\includegraphics[width=3.1in,height=2.4in,keepaspectratio=false]{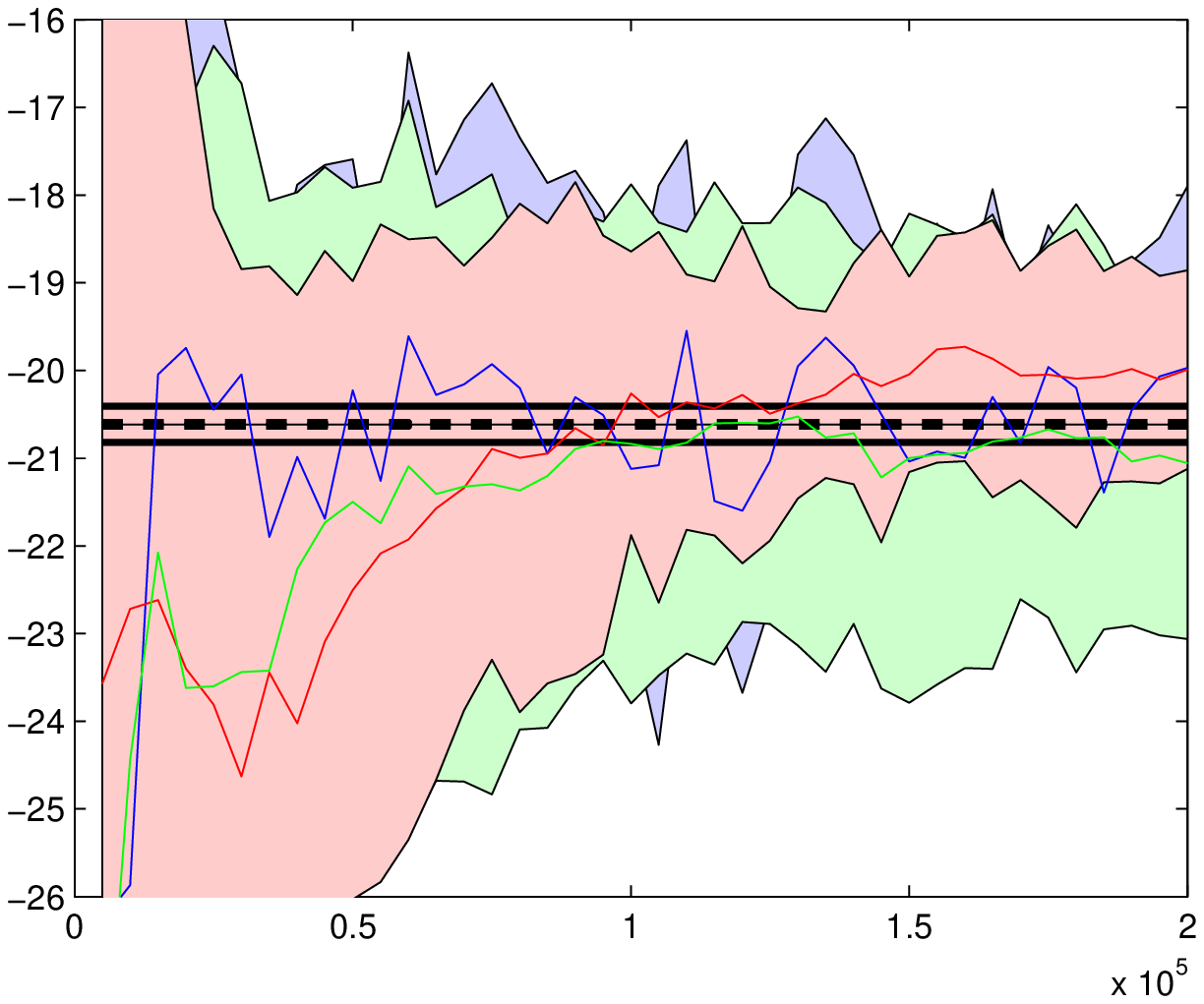}}
\caption{ Double Knock-out call option. Details as in Figure \ref{fig:10}.}
\label{fig:11}
\end{figure}
\begin{figure}[ht]
\centering
\subfigure[Price]{\includegraphics[width=3.1in,height=2.4in,keepaspectratio=false]{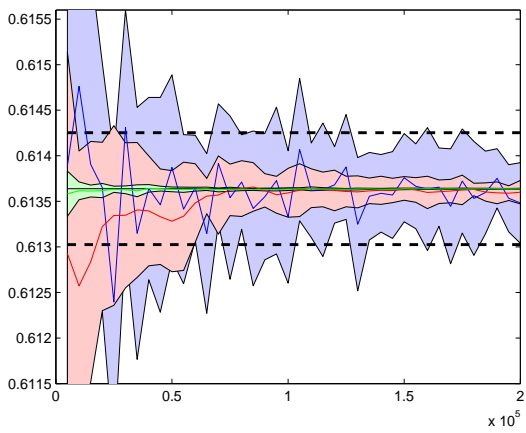}}
\subfigure[Vega]{\includegraphics[width=3.1in,height=2.4in,keepaspectratio=false]{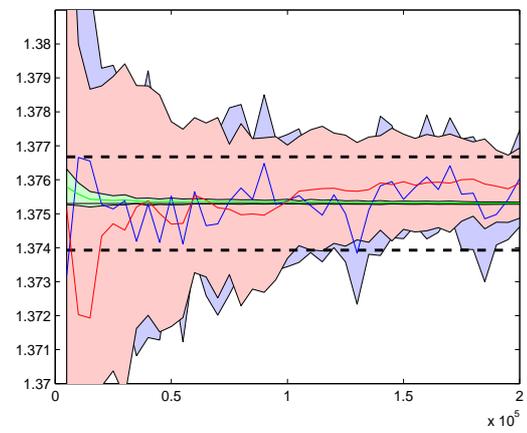}}
\caption{ Cliquet option. Details as in Figure \ref{fig:10}.}
\label{fig:12}
\end{figure}

Firstly, following the suggestion of \cite{Jac01}, Section 14.4,
we analyze convergence diagrams for prices and greeks, showing the
dependence of the MC simulation error upon the number of MC paths.
The results for the four payoffs are shown in Figures
\ref{fig:9}-\ref{fig:12}    .
\clearpage
We observe what follows.
\begin{enumerate}
\item European option (Figure \ref{fig:9}): QMC+BBD outperforms both QMC+SD and MC+SD in all cases (the 3-sigma regions for QMC+BBD are systematically smaller). We also note that for price, delta and vega, the QMC+BBD convergence is practically monotonic, which makes on-line error approximation possible. For gamma, the QMC+BBD convergence is much less oscillating than for MC+SD.
\item Asian option (Figure \ref{fig:10}): QMC+BBD outperforms both QMC+SD and MC+SD for price and vega. For delta, QMC with both SD and BBD is marginally better than MC+SD. For gamma, QMC with both SD and BBD has nearly the same efficiency as MC+SD. The QMC+BBD convergence is also smoother for price, delta and vega.
\item Double KO option (Figure \ref{fig:11}): QMC+BBD outperforms both QMC+SD and MC+SD in all cases.
\item Cliquet option (Figure \ref{fig:12}): QMC+SD outperforms QMC+BBD and MC+SD in all cases. QMC+BBD outperforms MC+SD only for price.
\end{enumerate}
\par
\begin{figure}[ht]
\centering
\subfigure[Price]{\includegraphics[width=2.5in,height=1.9in,keepaspectratio=false]{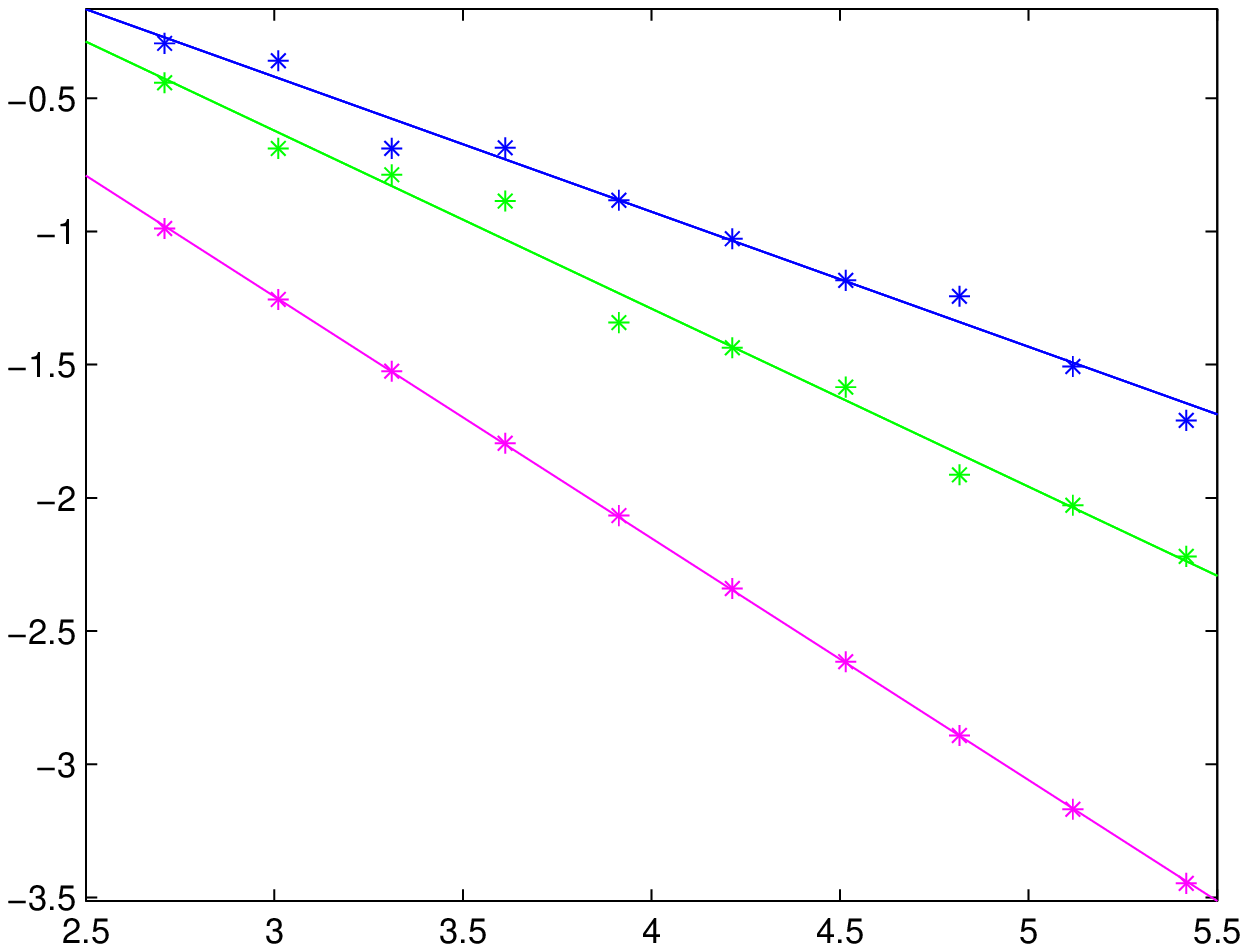}}
\subfigure[Delta]{\includegraphics[width=2.5in,height=1.9in,keepaspectratio=false]{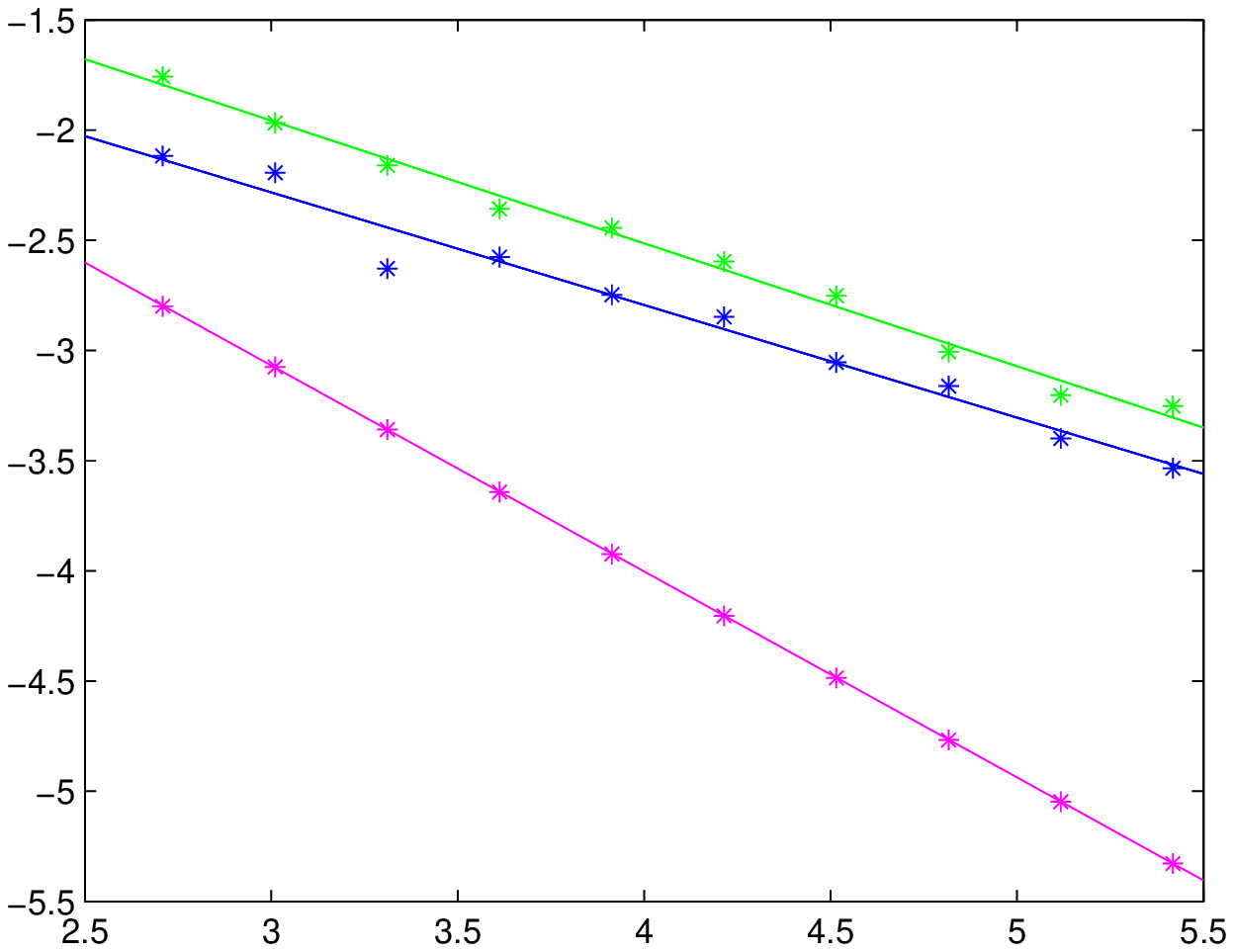}}
\subfigure[Gamma]{\includegraphics[width=2.5in,height=1.9in,keepaspectratio=false]{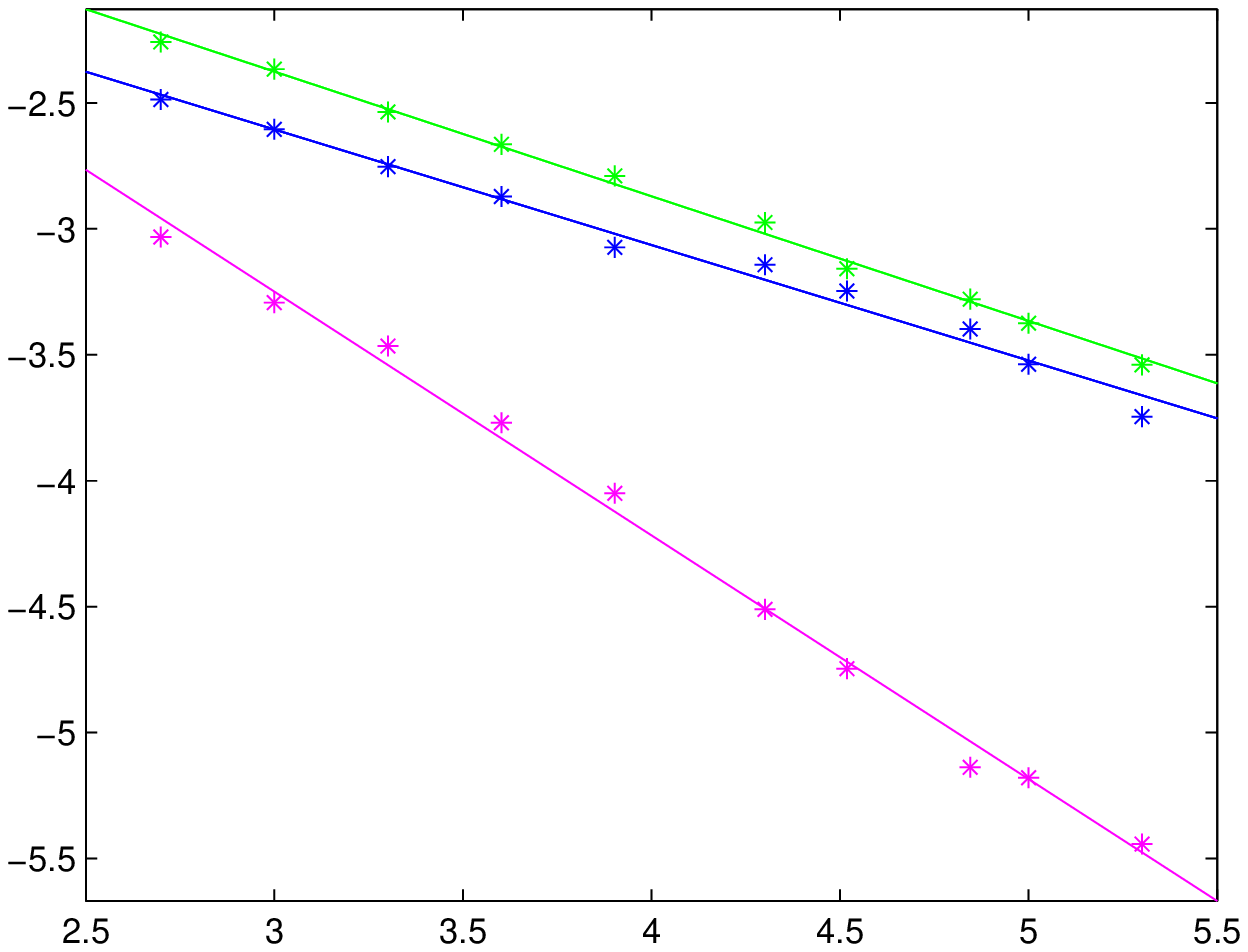}}
\subfigure[Vega]{\includegraphics[width=2.5in,height=1.9in,keepaspectratio=false]{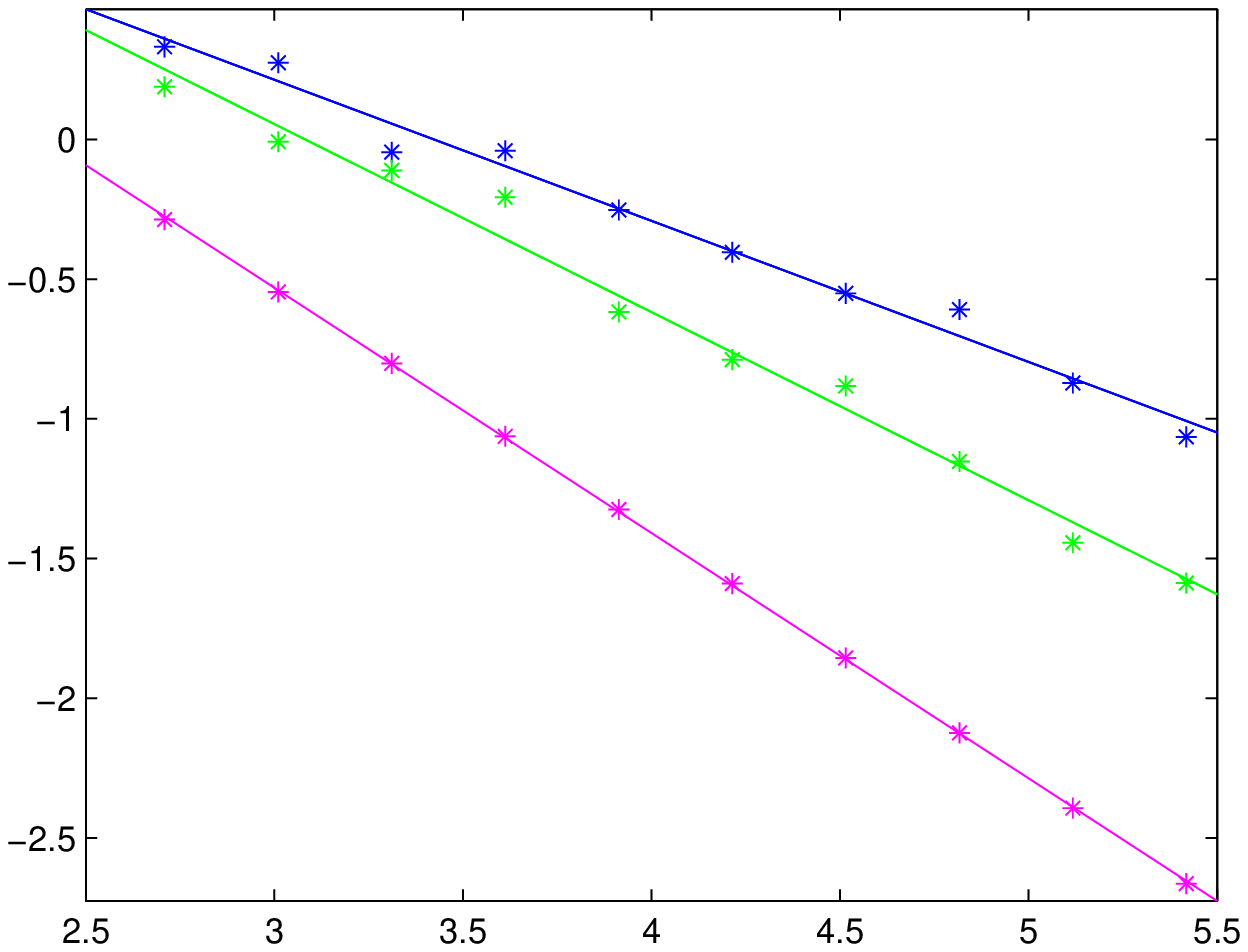}}
\caption{European call option price $(a)$ and greeks
$(b),(c),(d)$, Log-Log plots of error $\varepsilon_N$ versus
number of simulated paths $N=2^p,\;p=9,\ldots,18$, $D=32$,
$\epsilon=10^{-3}$, $L=30$ runs: MC+SD with antithetic variables
(blue), QMC+SD (green), QMC+BBD (magenta). Linear regression lines
are also shown.} \label{fig:13}
\end{figure}
\begin{figure}[ht]
\centering
\subfigure[Price]{\includegraphics[width=2.5in,height=1.9in,keepaspectratio=false]{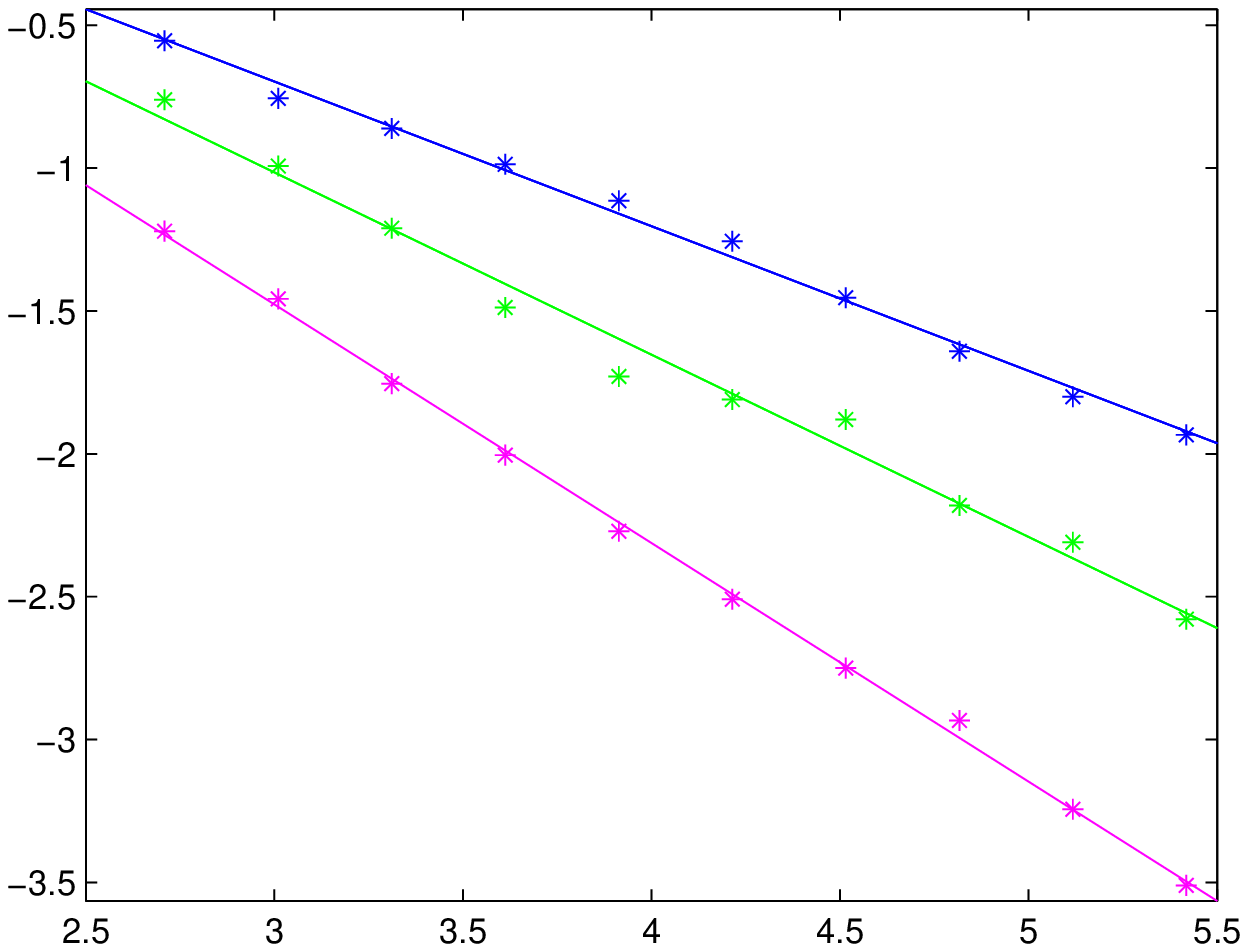}}
\subfigure[Delta]{\includegraphics[width=2.5in,height=1.9in,keepaspectratio=false]{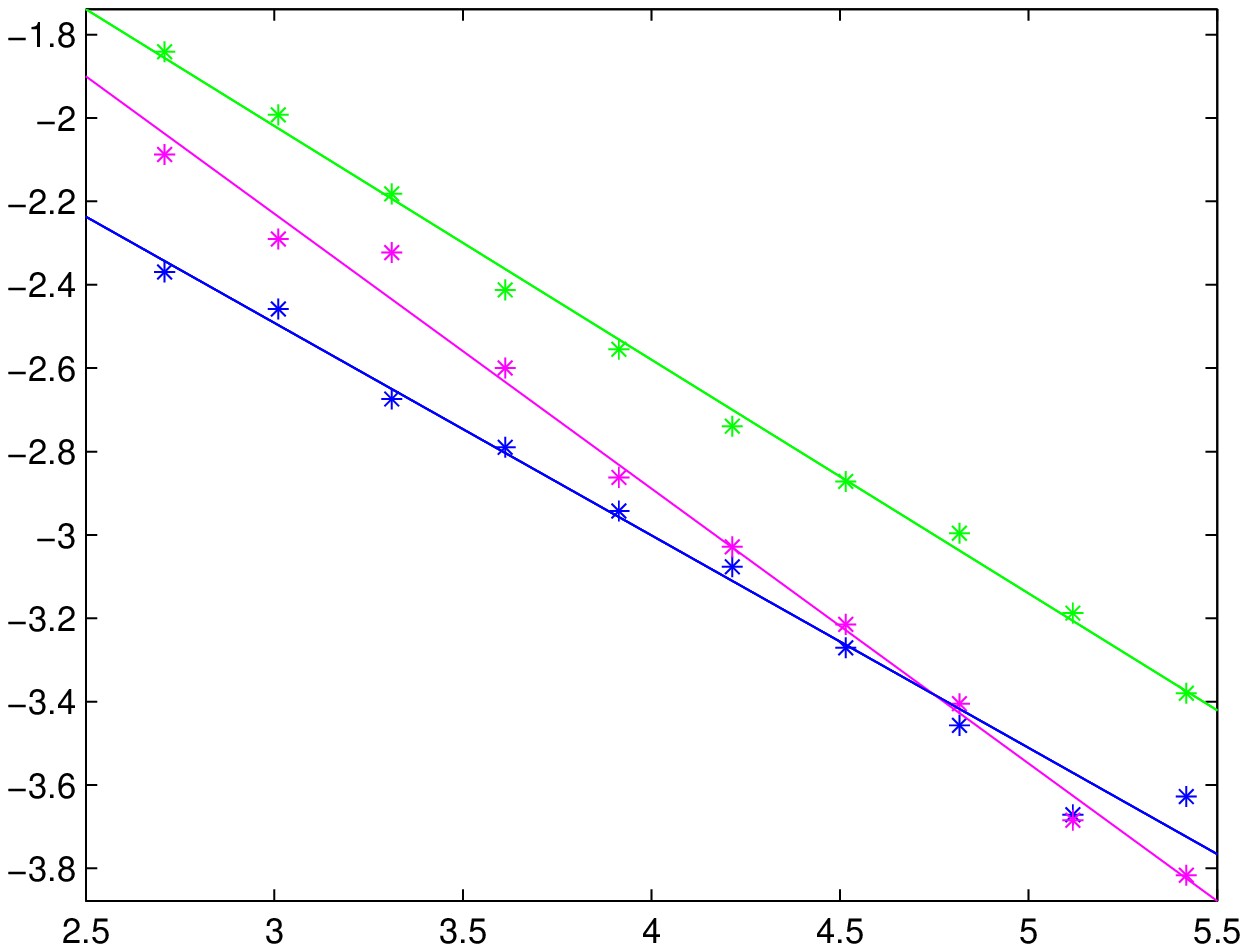}}
\subfigure[Gamma]{\includegraphics[width=2.5in,height=1.9in,keepaspectratio=false]{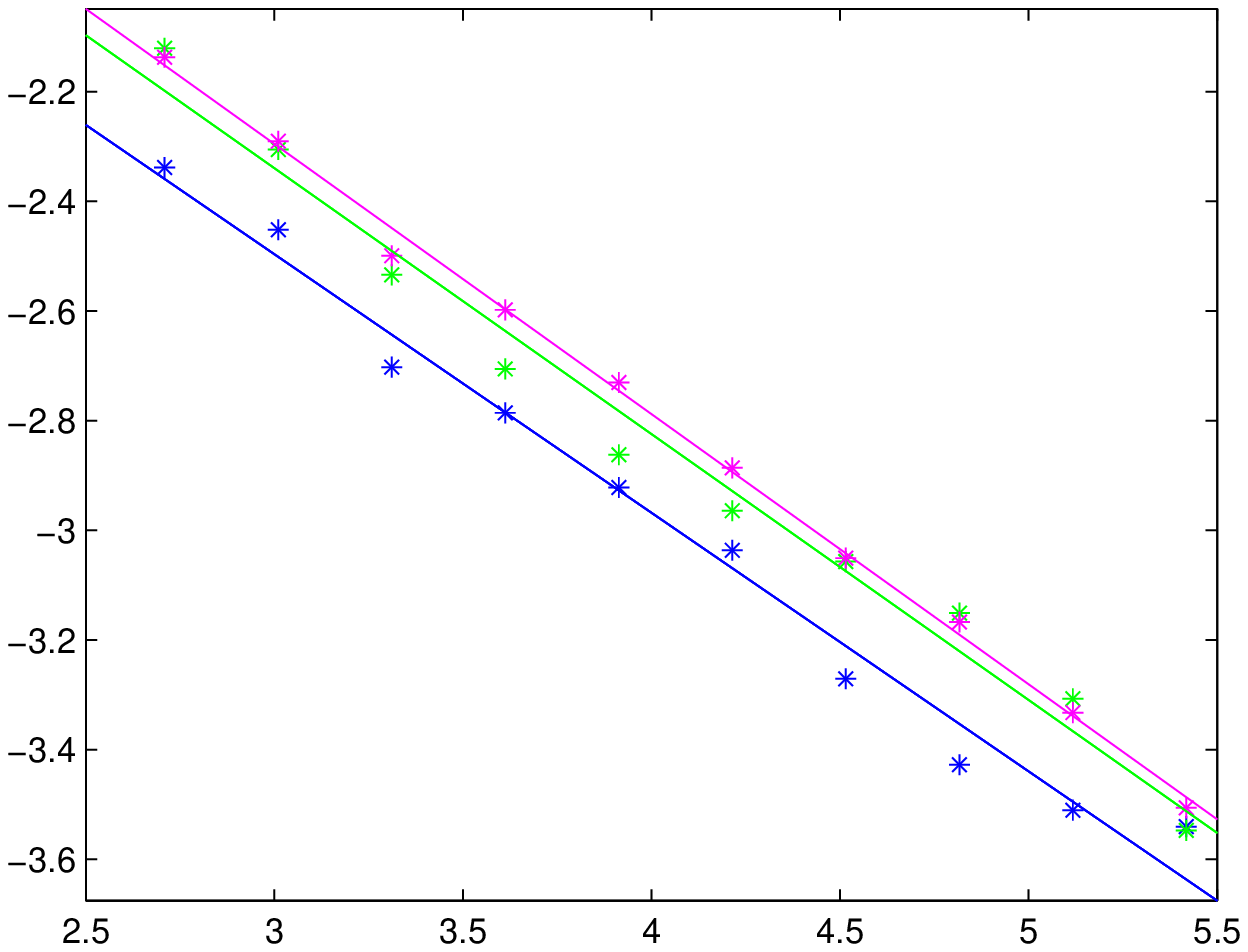}}
\subfigure[Vega]{\includegraphics[width=2.5in,height=1.9in,keepaspectratio=false]{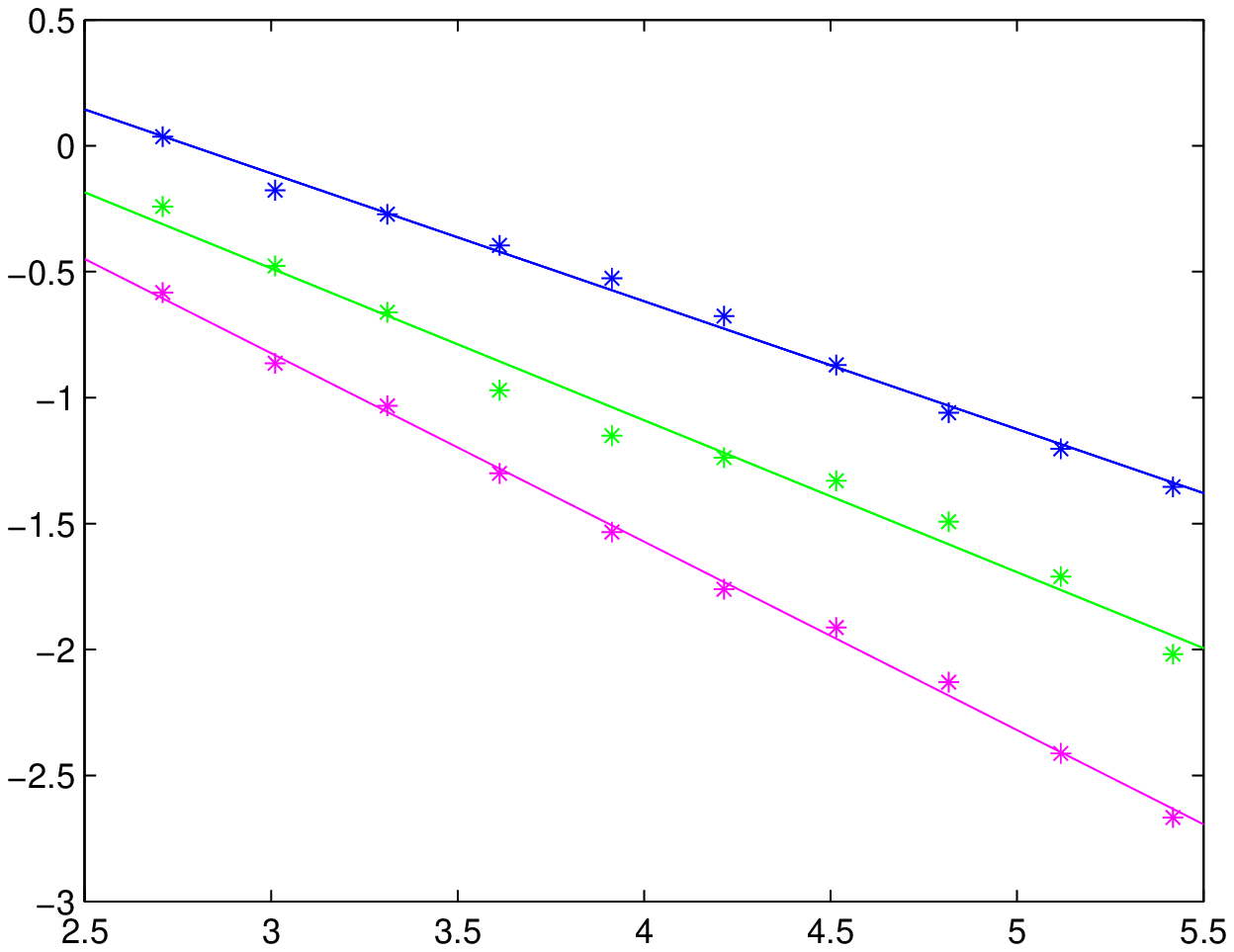}}
\caption{Asian call option. $\epsilon=5\times 10^{-3}$. Other details as in Figure \ref{fig:13}.}
\label{fig:14}
\end{figure}
\begin{figure}[ht]
\centering
\subfigure[Price]{\includegraphics[width=2.5in,height=1.9in,keepaspectratio=false]{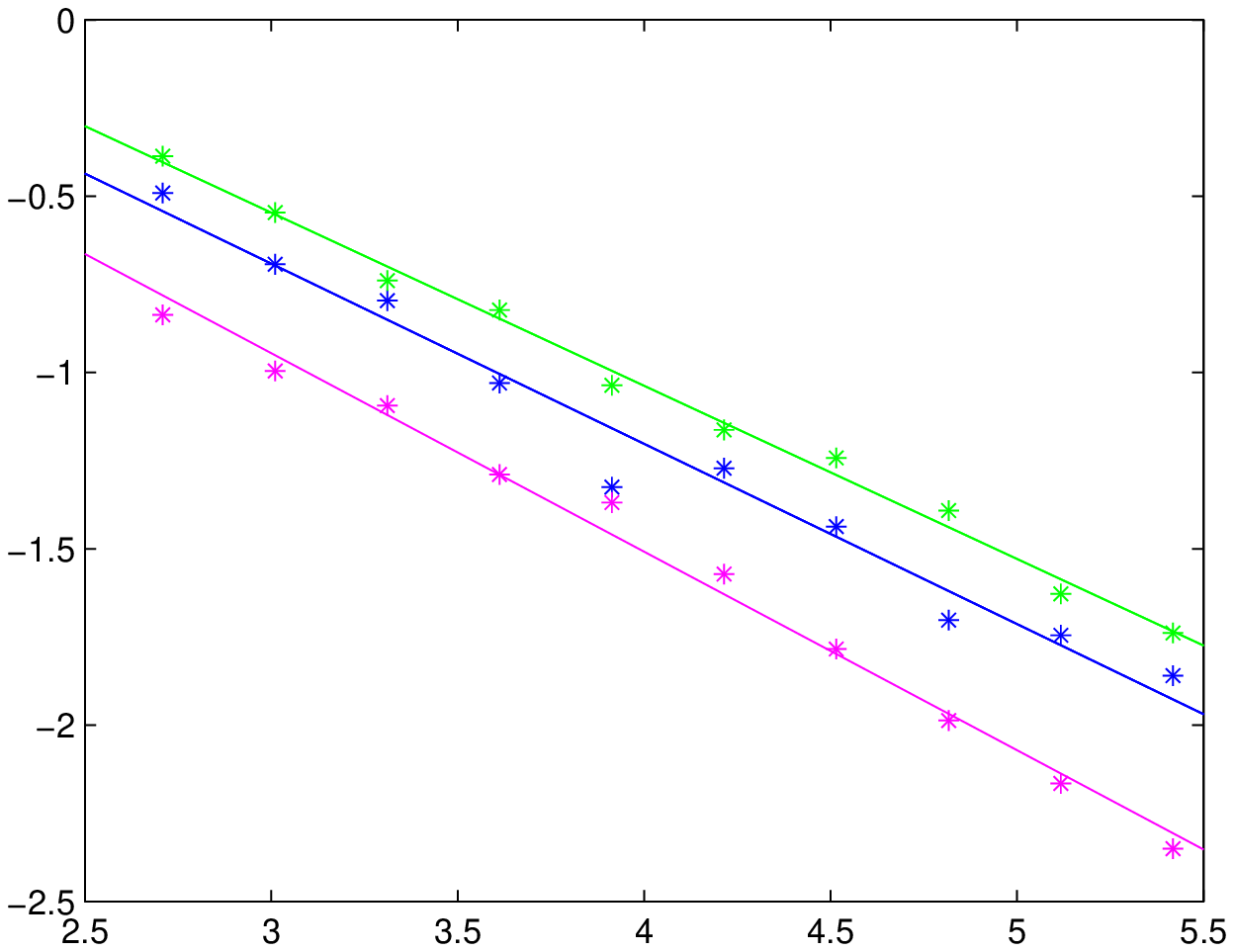}}
\subfigure[Delta]{\includegraphics[width=2.5in,height=1.9in,keepaspectratio=false]{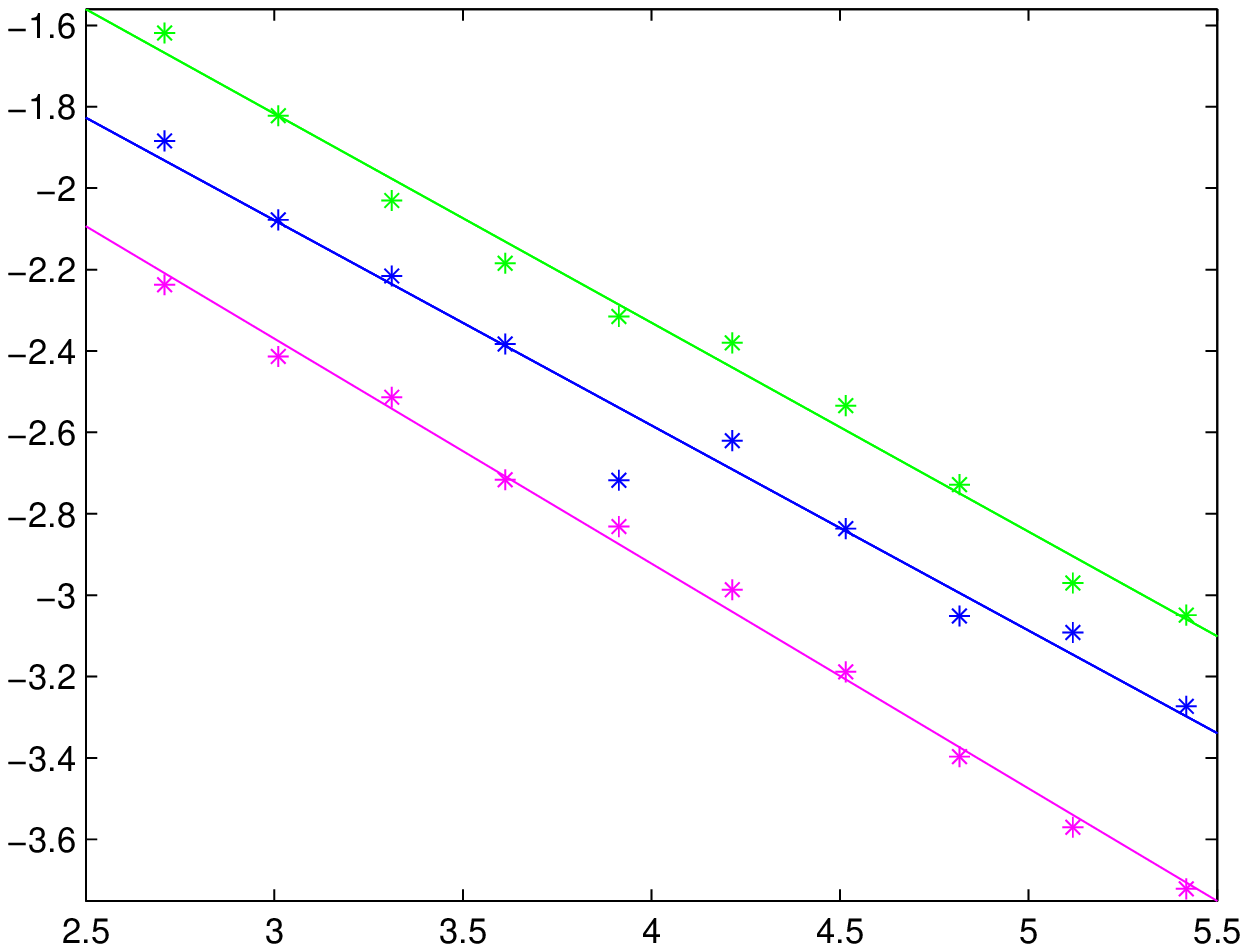}}
\subfigure[Gamma]{\includegraphics[width=2.5in,height=1.9in,keepaspectratio=false]{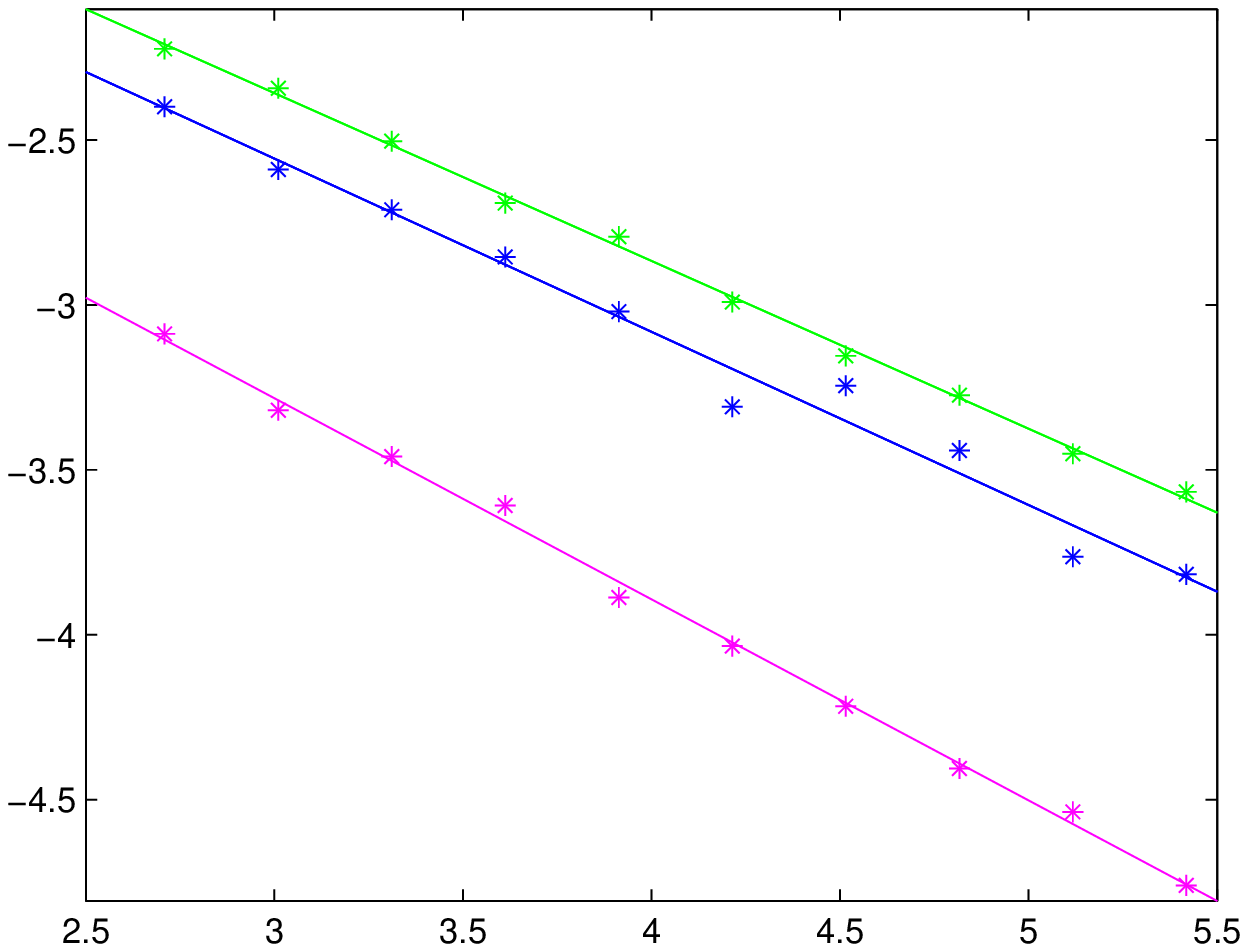}}
\subfigure[Vega]{\includegraphics[width=2.5in,height=1.9in,keepaspectratio=false]{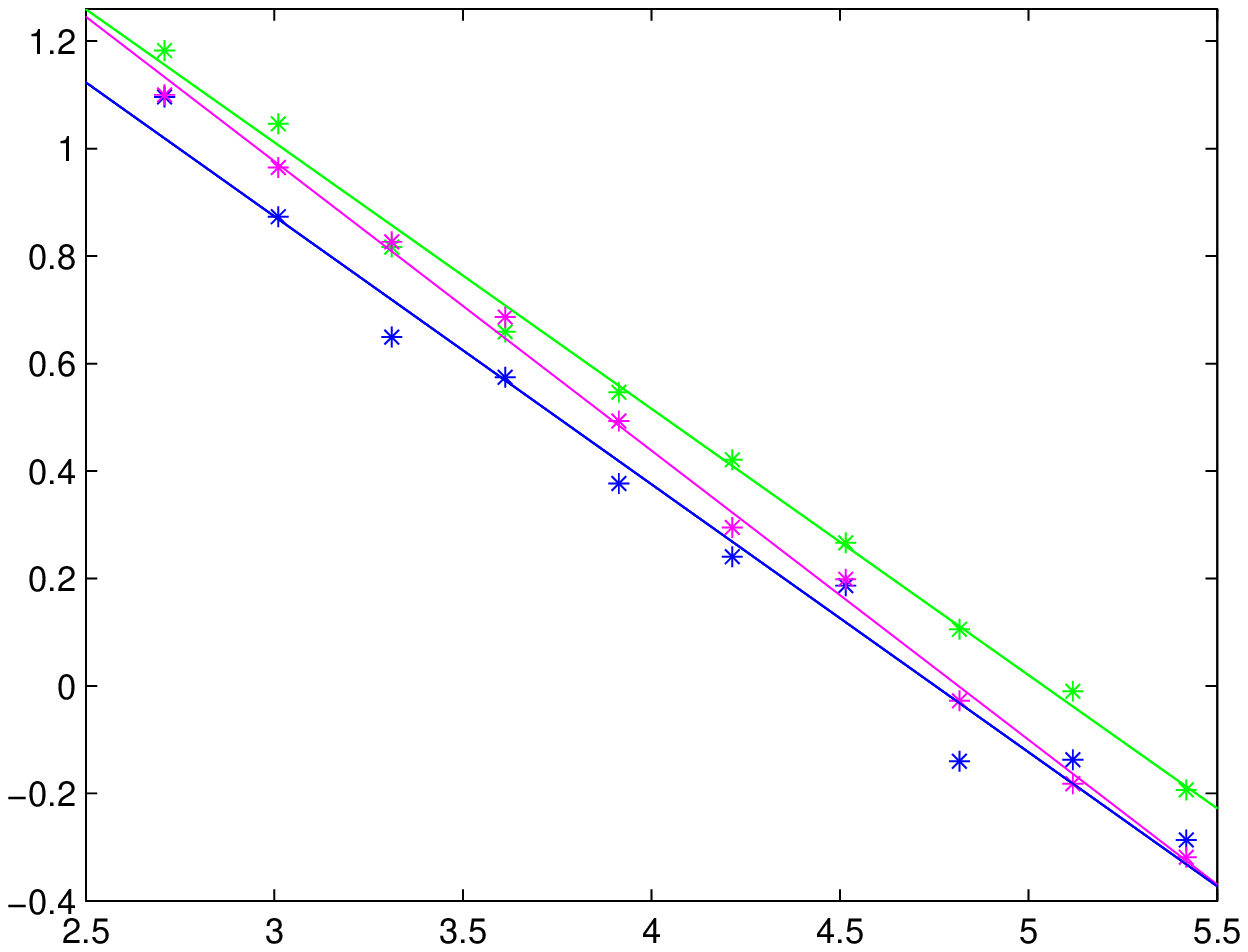}}
\caption{Double Knock-out call option. Details as in Figure \ref{fig:14}.}
\label{fig:15}
\end{figure}
\begin{figure}[ht]
\centering
\subfigure[Price]{\includegraphics[width=2.5in,height=1.9in,keepaspectratio=false]{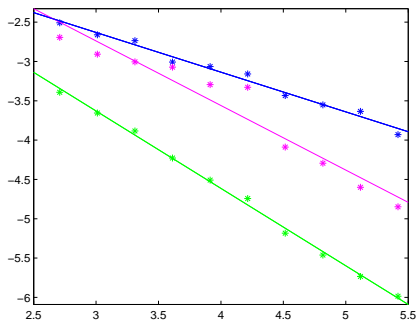}}
\subfigure[Vega]{\includegraphics[width=2.5in,height=1.9in,keepaspectratio=false]{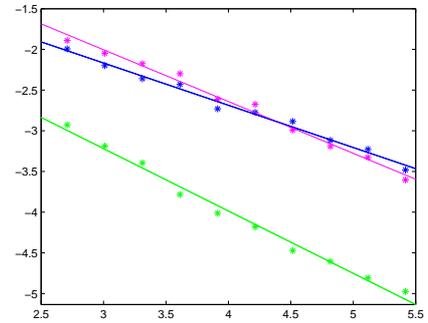}}
\caption{Cliquet option. Details as in Figure \ref{fig:14}.}
\label{fig:16}
\end{figure}

Next, we analyze the relative performance of QMC vs MC in terms of convergence rate. We plot in Figures \ref{fig:13}-\ref{fig:16} the root mean square error, eq. (\ref{error}), versus the number of MC scenarios $N$ in Log-Log scale. In all our tests we have chosen an appropriate range for $N$ such that, in the computation of greeks, the bias term is negligible with respect to the variance term (see Appendix \ref{App:GreekErr} for details). Hence, the observed relations are, with good accuracy, linear, therefore the power law (\ref{err:QMC}) is confirmed, and the convergence rates $\alpha$ can be extracted as the slopes of the regression lines.
Furthermore, also the intercepts of regression lines provide useful information about the efficiency of the QMC and MC methods: in fact, lower intercepts mean that the simulated value starts closer to the exact value. The resulting slopes and intercepts from linear regression are presented in tabs. \ref{tab:3} and \ref{tab:4} for all test cases.
\begin{table}[ht]
\small
\centering
\subtable{%
\begin{tabular}{c c c c c}
  \toprule
  \textbf{Payoff} & \textbf{Function} & \textbf{MC+SD} & \textbf{QMC+SD} & \textbf{QMC+BBD} \\
  \midrule
  European & Price & -0.3 $\pm$ 0.1 & -0.2 $\pm$ 0.1 & -0.84 $\pm$ 0.01 \\
           & Delta & -2.0 $\pm$ 0.1 & -1.6 $\pm$ 0.1 & -2.64 $\pm$ 0.01 \\
           & Gamma & -2.0 $\pm$ 0.1 & -2.1 $\pm$ 0.1 & -2.8 $\pm$ 0.2 \\
           & Vega  & 0.4 $\pm$ 0.1 & 0.4 $\pm$ 0.1 & -0.13 $\pm$ 0.01 \\
  \hline
  Asian    & Price & -0.5 $\pm$ 0.1 & -0.5 $\pm$ 0.1 & -1.0 $\pm$ 0.1 \\
           & Delta & -2.2 $\pm$ 0.1 & -1.7 $\pm$ 0.1 & -1.9 $\pm$ 0.1 \\
           & Gamma & -2.1 $\pm$ 0.2 & -2.1 $\pm$ 0.1 & -2.0 $\pm$ 0.1 \\
           & Vega  & 0.1 $\pm$ 0.1 & -0.1 $\pm$ 0.1 & -0.4 $\pm$ 0.1 \\
  \hline
  Double KO& Price & -0.4 $\pm$ 0.1 & -0.3 $\pm$ 0.1 & -0.7 $\pm$ 0.1 \\
           & Delta & -1.8 $\pm$ 0.1 & -1.6 $\pm$ 0.1 & -2.1 $\pm$ 0.1 \\
           & Gamma & -2.4 $\pm$ 0.1 & 2.1 $\pm$ 0.1 & -2.9 $\pm$ 0.1 \\
           & Vega  & 1.1 $\pm$ 0.1 & 1.3 $\pm$ 0.1 & 1.3 $\pm$ 0.2 \\
  \hline
  Cliquet  & Price & -2.4 $\pm$ 0.1 & -3.2 $\pm$ 0.1 & -2.5 $\pm$ 0.3 \\
           & Vega  & -2.0 $\pm$ 0.1 & -2.7 $\pm$ 0.1 & -1.7 $\pm$ 0.2 \\
  \bottomrule
\end{tabular}}
\caption{Intercepts from linear regression with their errors, for
MC+SD with antithetic variables, QMC+SD and QMC+BBD, $L=30$ runs.
Results are shown for $N=10^{2.5}$ paths.} \label{tab:3}
\end{table}
\begin{table}[ht]
\small
\centering
\subtable{%
\begin{tabular}{c c c c c}
  \toprule
  \textbf{Payoff} & \textbf{Function} & \textbf{MC+SD} & \textbf{QMC+SD} & \textbf{QMC+BBD} \\
  \midrule
  European & Price & -0.46 $\pm$ 0.03 & -0.71 $\pm$ 0.03 & -0.901 $\pm$ 0.003 \\
           & Delta & -0.49 $\pm$ 0.03 & -0.56 $\pm$ 0.02 & -0.926 $\pm$ 0.004 \\
           & Gamma & -0.51 $\pm$ 0.02 & -0.51 $\pm$ 0.01 & -0.98  $\pm$ 0.04 \\
           & Vega  & -0.45 $\pm$ 0.03 & -0.69 $\pm$ 0.03 & -0.869 $\pm$ 0.003 \\
  \hline
  Asian    & Price & -0.50 $\pm$ 0.02 & -0.70 $\pm$ 0.03 & -0.85 $\pm$ 0.01 \\
           & Delta & -0.49 $\pm$ 0.03 & -0.59 $\pm$ 0.02 & -0.61 $\pm$ 0.03 \\
           & Gamma & -0.53 $\pm$ 0.05 & -0.49 $\pm$ 0.03 & -0.50 $\pm$ 0.03 \\
           & Vega  & -0.51 $\pm$ 0.02 & -0.64 $\pm$ 0.04 & -0.75 $\pm$ 0.01 \\
  \hline
  Double KO& Price & -0.49 $\pm$ 0.03 & -0.49 $\pm$ 0.02 & -0.56 $\pm$ 0.03 \\
           & Delta & -0.49 $\pm$ 0.02 & -0.52 $\pm$ 0.03 & -0.55 $\pm$ 0.02 \\
           & Gamma & -0.45 $\pm$ 0.03 & -0.51 $\pm$ 0.02 & -0.61 $\pm$ 0.02 \\
           & Vega  & -0.50 $\pm$ 0.03 & -0.53 $\pm$ 0.02 & -0.57 $\pm$ 0.04 \\
  \hline
  Cliquet  & Price & -0.51 $\pm$ 0.02 & -1.00 $\pm$ 0.03 & -0.72 $\pm$ 0.09 \\
           & Vega  & -0.48 $\pm$ 0.02 & -0.86 $\pm$ 0.04 & -0.62 $\pm$ 0.04 \\
  \bottomrule
\end{tabular}
}%\qquad\qquad
\caption{Slopes from linear regression with their errors, as in Table \ref{tab:3}. In a few cases we show three decimals since the MC error is lower.}
\label{tab:4}
\end{table}
\clearpage
\par
We observe what follows.
\begin{enumerate}
\item European option (Figure \ref{fig:13}): QMC+BBD outperforms other methods, having the highest rate of convergence $\alpha$ and the smallest intercept. QMC+SD for price and vega has higher rate of convergence $\alpha$ but also somewhat higher intercepts than MC+SD. It performs marginally better for delta and as good as MC+SD for gamma in terms of $\alpha$ values.
\item Asian option (Figure \ref{fig:14}): for price and vega QMC+BBD and QMC+SD have higher $\alpha$ than MC+ SD, with QMC+BBD being the most efficient. They also have slightly higher $\alpha$ for delta, but both have lower intercepts than MC. For gamma all methods show similar convergence.
\item Double KO option (Figure \ref{fig:15}): QMC+BBD has the highest $\alpha$ although its highest value $\alpha=0.61$ (for gamma) is lower than $\alpha$ for European and Asian options (with an exception of gamma of Asian option). Its intercepts for price, delta and gamma also have the lowest values among all methods. The QMC+SD is as efficient as MC.
\item Cliquet option (Figure \ref{fig:16}): QMC+SD has the highest $\alpha$, close to 1.0. It also has the lowest intercepts among all methods. QMC+BBD has higher $\alpha$ but similar intercepts in comparison with MC+SD.
\end{enumerate}
We stress that slopes and intercepts shown in the previous figures \ref{fig:13}-\ref{fig:16} do not depend on the details of the simulations, in particular the MC seed or the LDS starting point, since we are averaging over $L=30$ runs.
\par
In conclusion, QMC+BBD generally outperforms the other methods, except for asian gamma where all methods show similar convergence properties and Cliquet option for which QMC+SD is the most efficient method.

\subsection{Speed-Up Analysis}
\label{SecSpeedUp} A typical question with Monte Carlo simulation
is \QuoteDouble{how many scenarios are necessary to achieve a
given precision?}. When comparing two numerical simulation
methods, the typical question becomes \QuoteDouble{how many
scenarios may I save using method B instead of method A,
preserving the same precision?}.
\par
A useful measure of the relative computational performance of two
numerical methods is the so called speed-up $S_*(a)$
\cite{KreMer1998a,PapTrau1996}. It is defined as
\begin{equation}
\label{EqSpeedUp}
S_*^{(i,j)}(a) = \frac{N_*^{(j)}(a)}{N_*^{(i)}(a)}\, ,
\end{equation}
where, in our context, $N_*^{(i)}(a)$ is the number of scenarios using the \textit{i-th} computational method (MC+SD, QMC+SD, or QMC+BBD) needed to reach and maintain a given accuracy $a$ \wrt exact or almost exact results.
Thus, the speed-up $S_*(a)$ quantifies the computational gain of method $i$ \wrt method $j$.
\par
The speed-up $N_*$ could be evaluated through direct simulation, but this would be extremely computationally expensive. Thus we resort to the much simpler algorithm described in Appendix \ref{App:SU}.
\begin{table}[ht]
\small
\centering
\subtable{%
\begin{tabular}{c c c c c c c c}
\toprule
  \textbf{Payoff} & \textbf{Function} & \multicolumn{2}{c}{\textbf{QMC+SD}} & \multicolumn{2}{c}{\textbf{QMC+BBD}} & \multicolumn{2}{c}{\textbf{QMC+BBD}} \\
  & & \multicolumn{2}{c}{\textbf{vs MC+SD}} & \multicolumn{2}{c}{\textbf{vs MC+SD}} & \multicolumn{2}{c}{\textbf{vs QMC+SD}} \\
        & & $a=1\%$ & $a=0.1\%$ & $a=1\%$ & $a=0.1\%$ & $a=1\%$ & $a=0.1\%$\\
  \midrule
  European  & Price & 3 & 6 & 30 & 140 & 10 & 20\\
            & Delta & 0.3 & 0.5 & 20 & 100 & 60 & 200\\
            & Gamma & 0.5 & 0.5 & 200 & 1000 & 600 & 5000\\
            & Vega  & 5 & 6 & 50 & 140 & 10 & 20\\
  \hline
  Asian     & Price & 5 & 10 & 30 & 100 & 5 & 10\\
            & Delta & 0.2 & 0.5 & 0.4 & 2 & 2 & 5\\
            & Gamma & 0.5 & - & 0.5 & - & 1 & -\\
            & Vega  & 6 & 10 & 30 & 100 & 5 & 10\\
  \hline
  Double KO & Price & 0.5 & 0.8 & 5 & 10 & 10 & 15\\
            & Delta & 0.5 & 1 & 5 & 20 & 10 & 20\\
            & Gamma & 0.7 & 1.3 & 110 & 650 & 150 & 500\\
            & Vega  & 0.5 & 0.5 & 1.5 & 1.5 & 3 & 3\\
  \hline
  Cliquet   & Price & 10 & 100 & 1 & 10 & 0.1 & 0.1\\
            & Vega  & 20 & 100 & 0.5 & 1 & 0.02 & 0.01\\
  \bottomrule
\end{tabular}
}%\qquad\qquad
\caption{Speed-up $S_*(a)$ of the various numerical methods \wrt each other (see columns), for different option types. The shift $\epsilon$ for finite differences is the same as used in the previous sections. Missing values of $S_*$ mean that the required accuracy cannot be reached since it is smaller than the bias.}
\label{tab:5}
\end{table}
\par
We show in Table \ref{tab:5} the results of the speed-up analysis obtained for all methods and for all option types described in the previous sections.
The speed-up measure clearly shows the relative efficiencies of the methods considered for each case. In general, QMC+BBD largely outperforms the other methods, with a speed-up factor up to $10^3$ (European and Barrier gamma) and a few exceptions (Asian delta and gamma, Cliquet). QMC+SD is the best method for Cliquet. We notice in particular that, in most cases, a ten-fold increase of the accuracy $a$ results in a two-fold increase of speed-up $S_*(a)$. However, in a few cases (gamma for European and Cliquet options), such an increase can result in up to ten-folds increase of $S_*(a)$.
\par
The difficulty with speed-up is the possible non-monotonicity of
the convergence plot for a given numerical methods. Unfortunately,
our algorithm to estimate speed-up in Appendix \ref{App:SU} cannot
capture unexpected fluctuations of the convergence plot, which
could lead to underestimate $N_*(a)$. However, we believe that the
choice of the 3-sigma confidence interval in eq. (\ref{Nstardef})
makes our speed-up analysis reliable, at least when coupled with
the stability analysis described in the next Section
\ref{SecStability}.

\subsection{Stability Analysis}
\label{SecStability} 
\begin{figure}[t!]
\centering
\subfigure[Price]{\includegraphics[width=3.1in,height=2.5in,keepaspectratio=false]{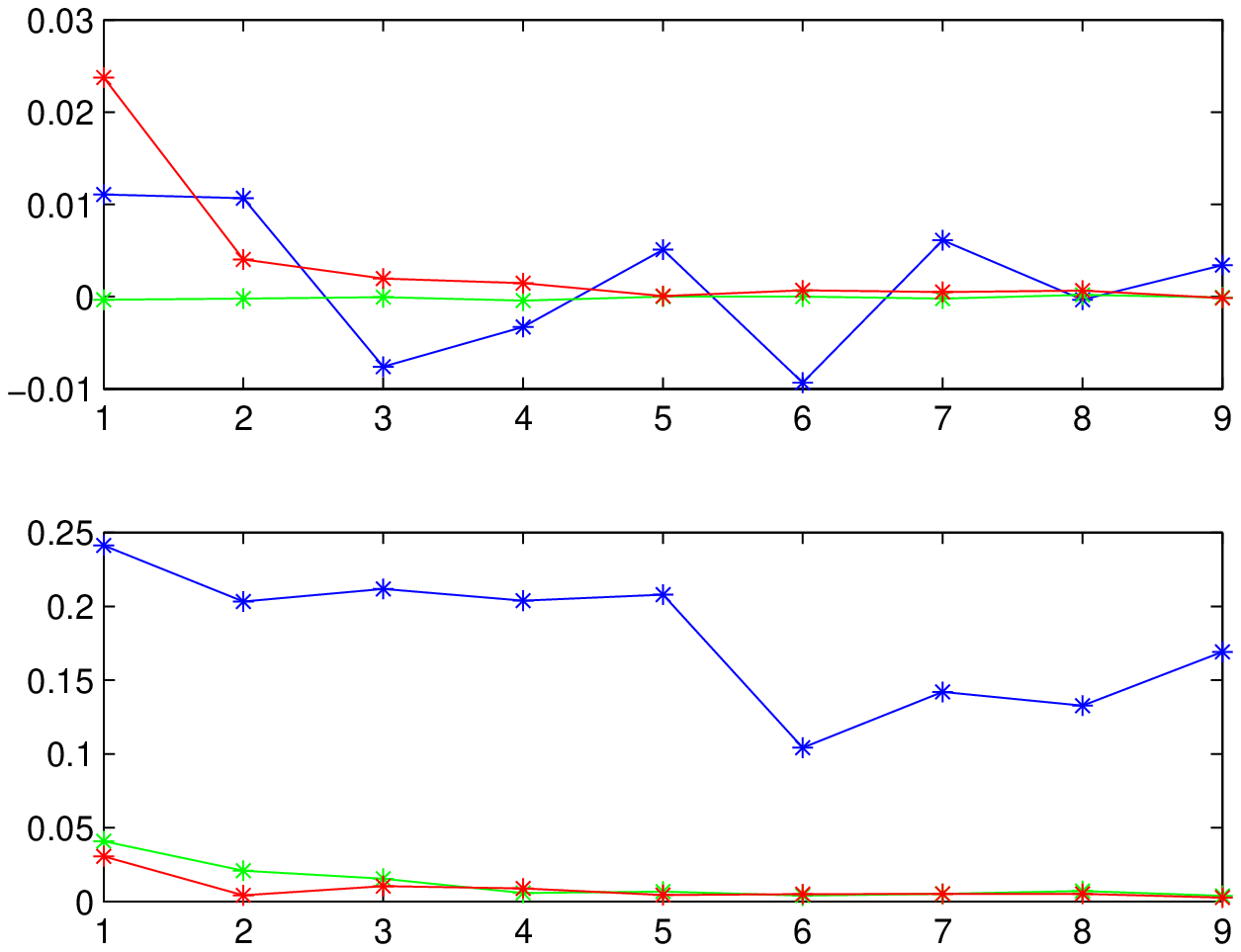}}
\subfigure[Delta]{\includegraphics[width=3.1in,height=2.5in,keepaspectratio=false]{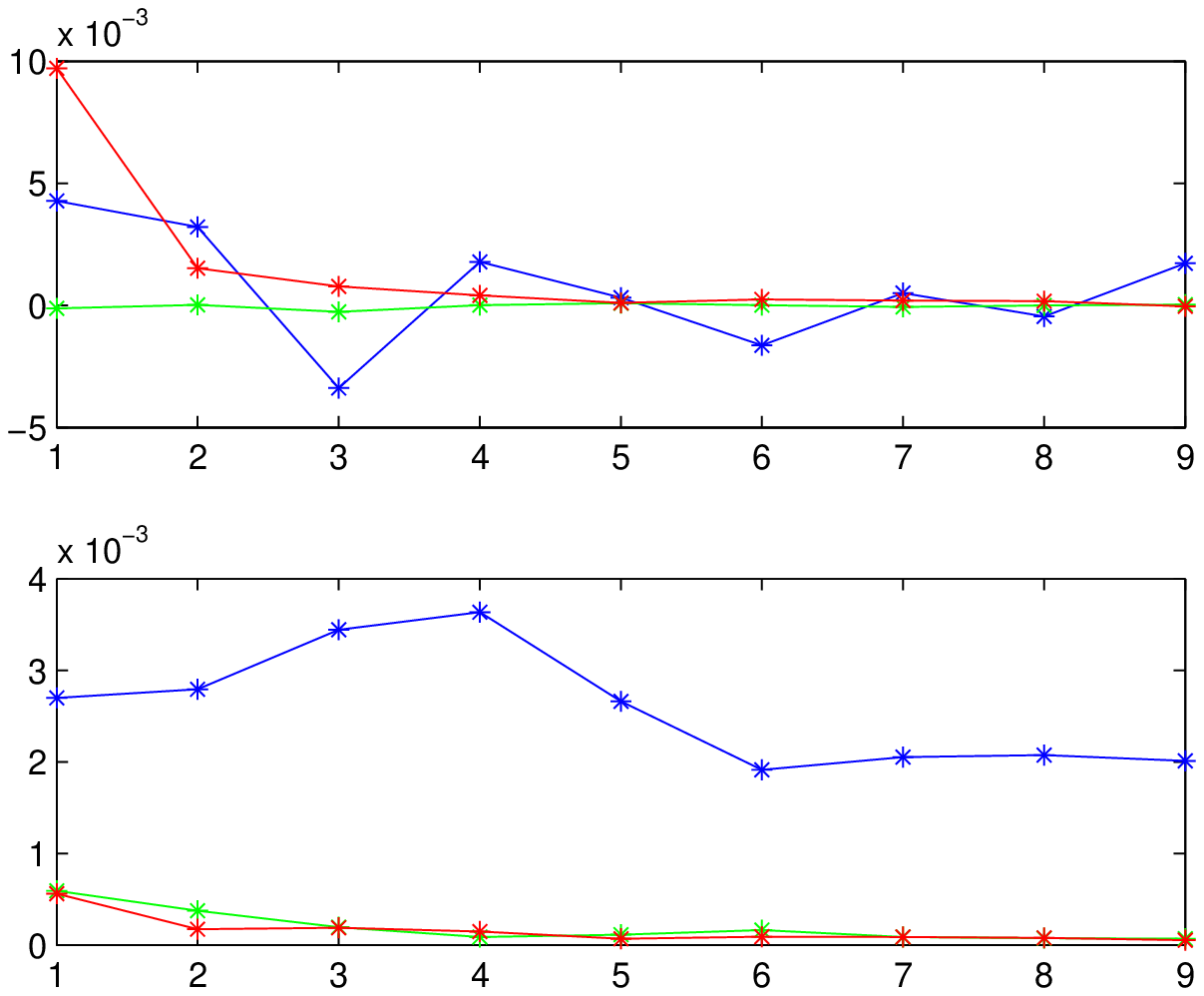}}
\subfigure[Gamma]{\includegraphics[width=3.1in,height=2.5in,keepaspectratio=false]{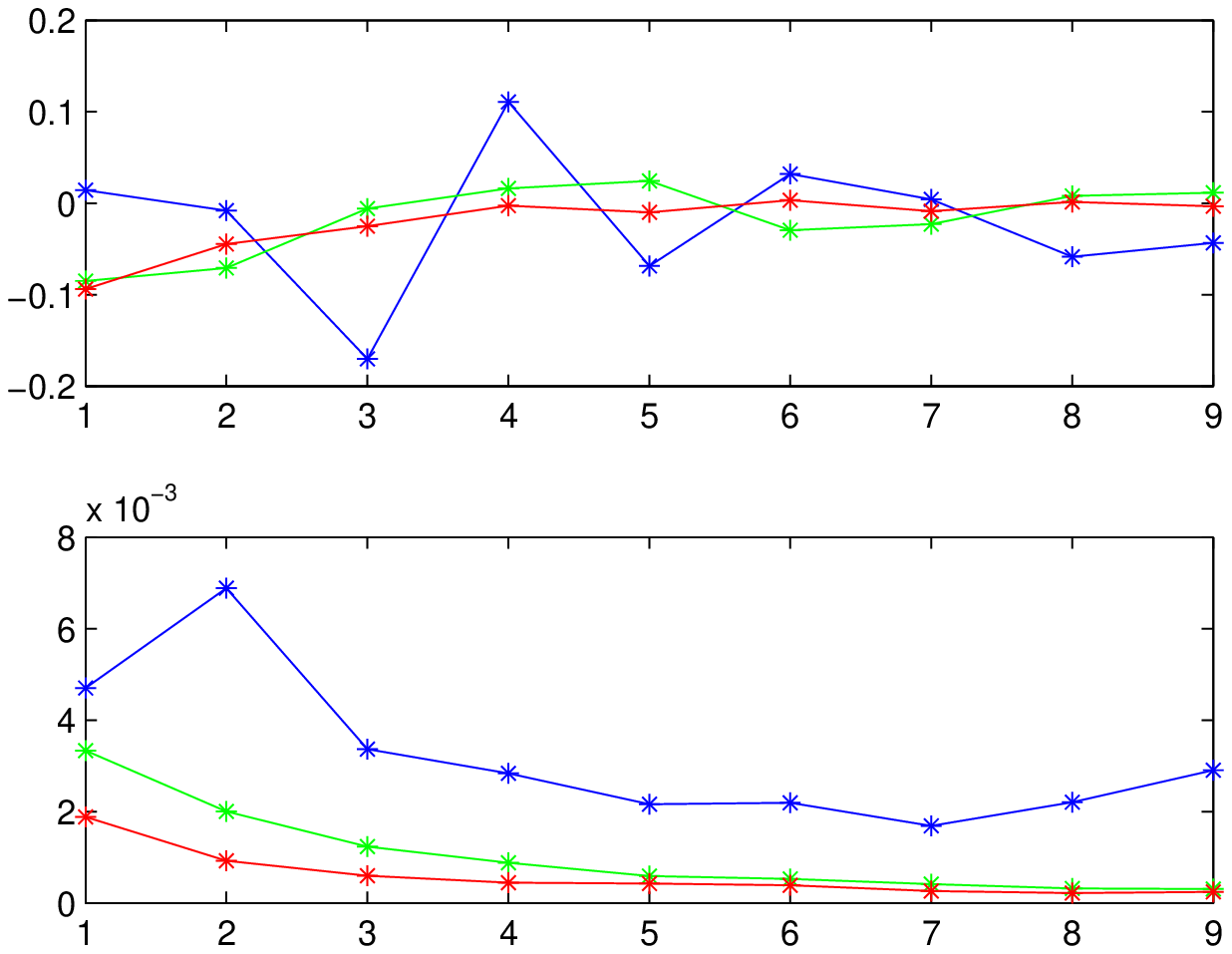}}
\subfigure[Vega]{\includegraphics[width=3.1in,height=2.5in,keepaspectratio=false]{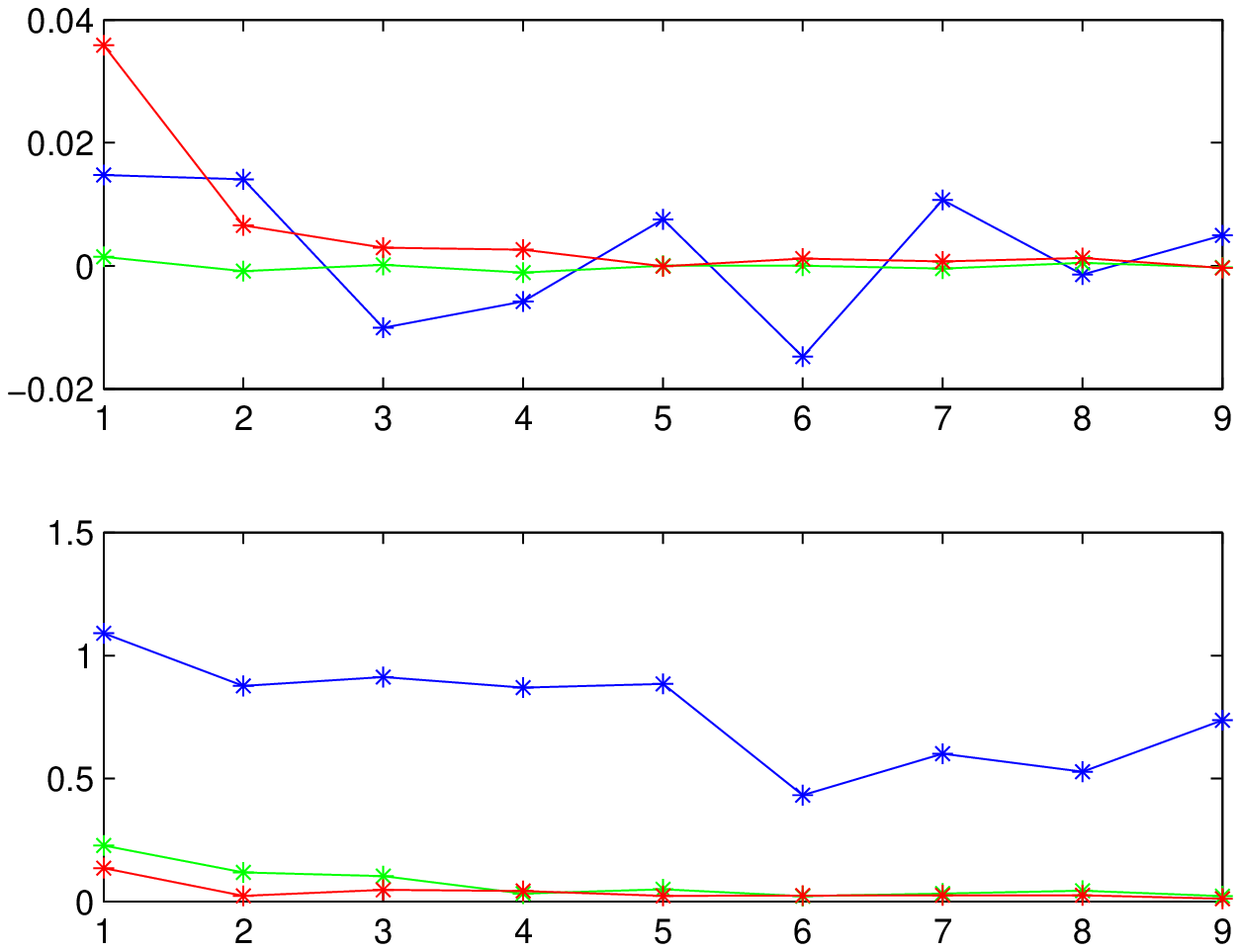}}
\caption{Log-returns (upper plots) and volatilities (lower plots)
of European call option price $(a)$ and greeks $(b),(c),(d)$, for
$D=32$, $\epsilon=10^{-3}$, MC+SD (blue), rQMC+BBD (green) and
pure QMC+BBD (red). The number of simulation paths ranges from 100
to 10,000 grouped in 10 windows each containing 10 samples
(x-axis).} \label{fig:17}
\end{figure}

\begin{figure}[t!]
\centering
\subfigure[Price]{\includegraphics[width=3.1in,height=2.5in,keepaspectratio=false]{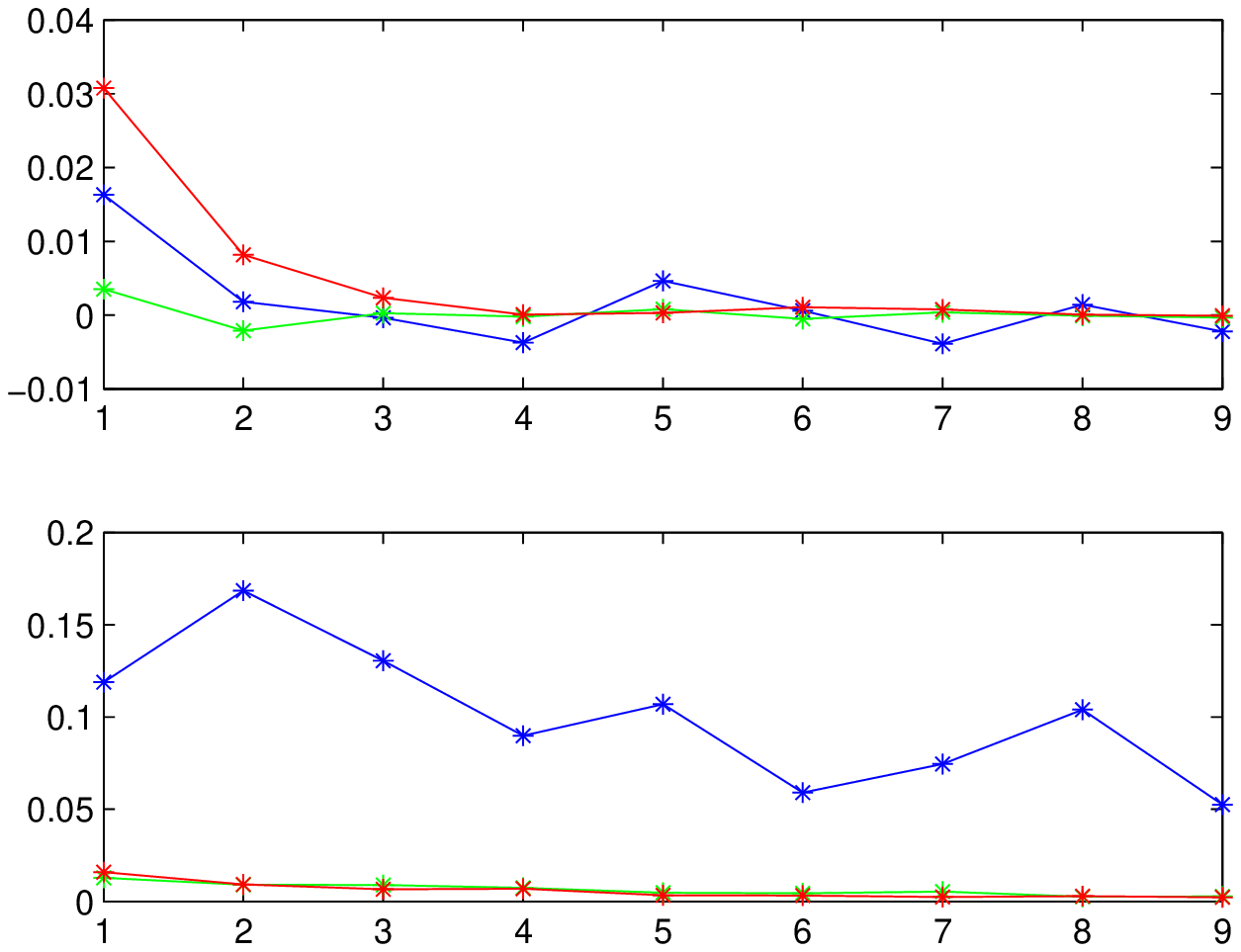}}
\subfigure[Delta]{\includegraphics[width=3.1in,height=2.5in,keepaspectratio=false]{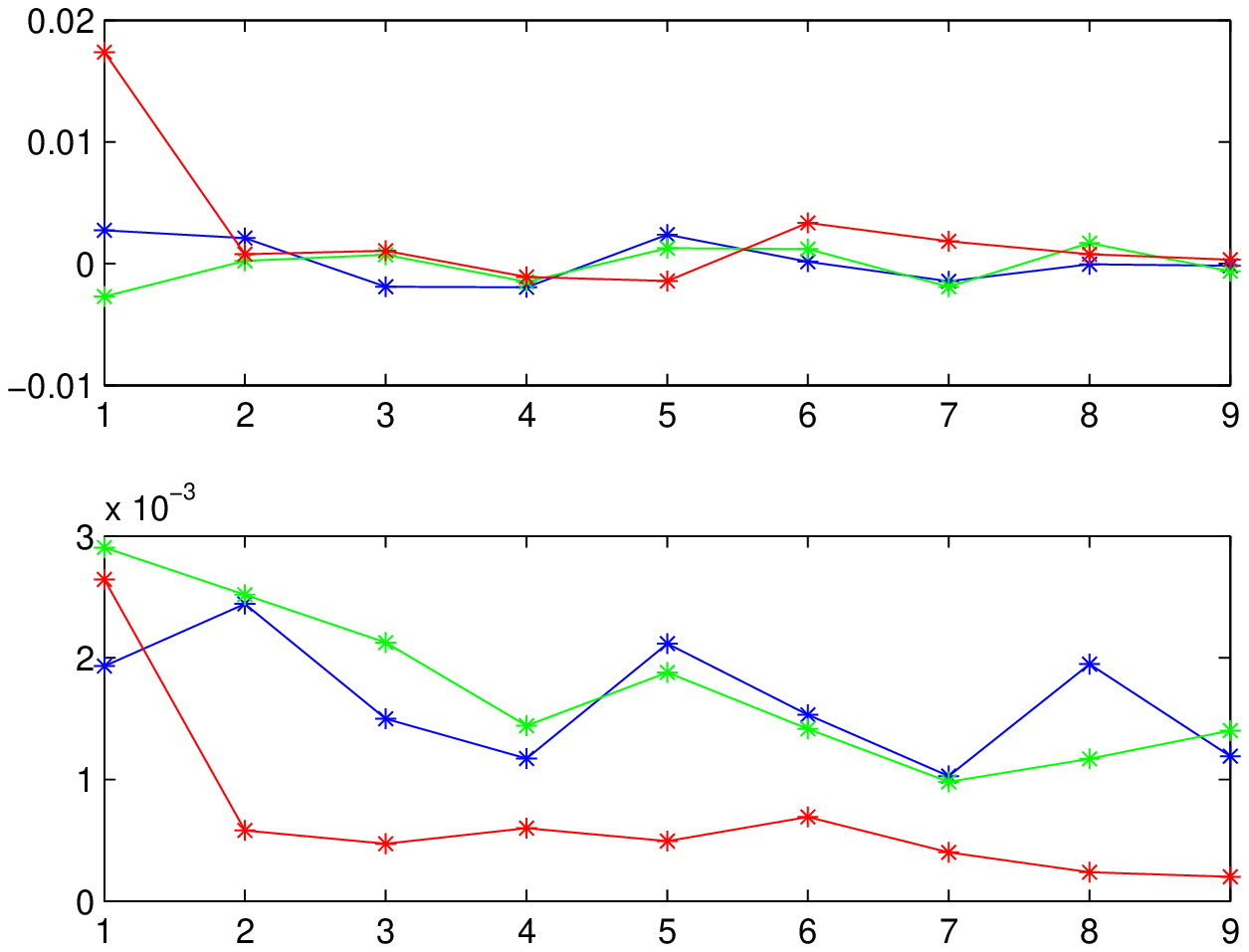}}
\subfigure[Gamma]{\includegraphics[width=3.1in,height=2.5in,keepaspectratio=false]{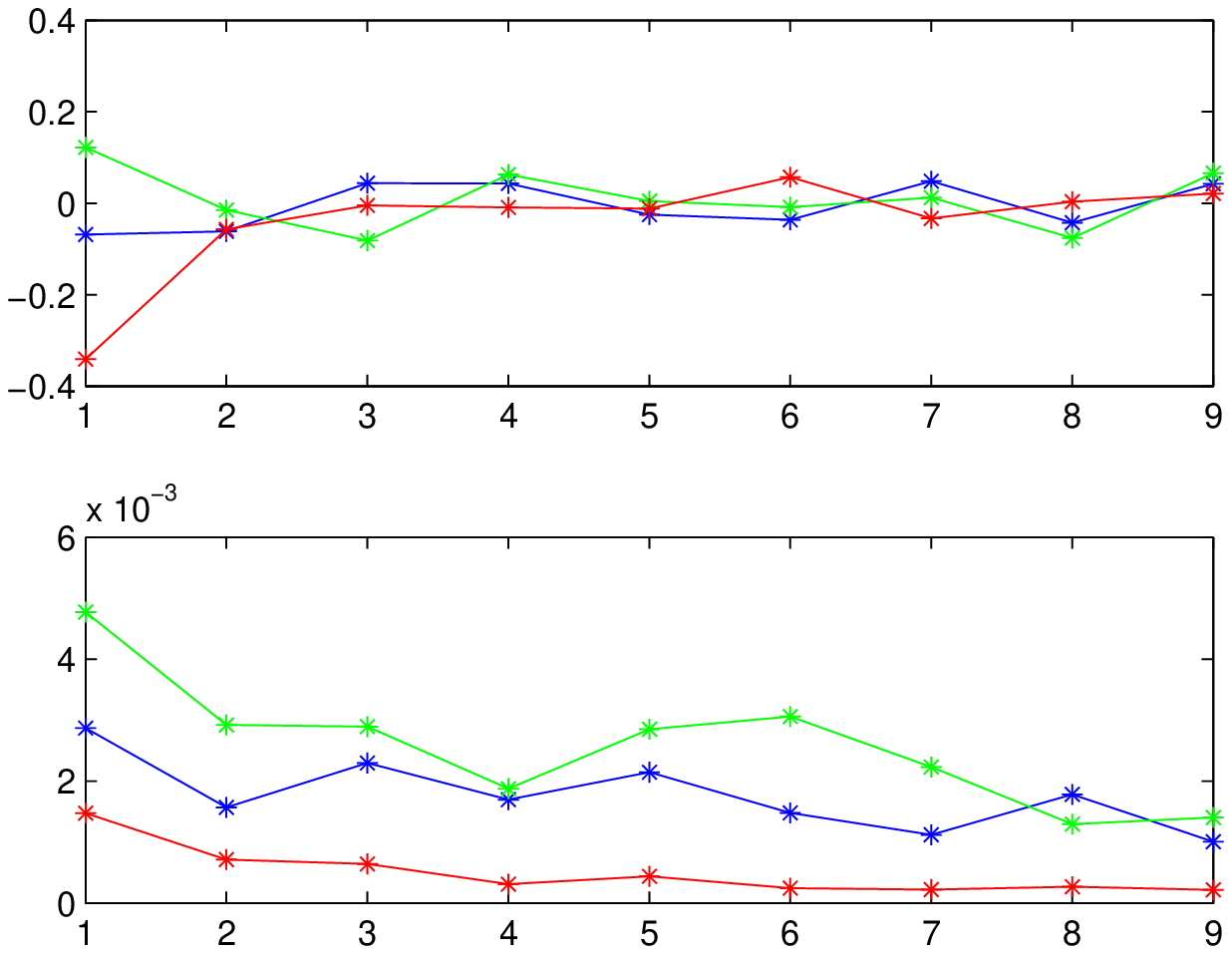}}
\subfigure[Vega]{\includegraphics[width=3.1in,height=2.5in,keepaspectratio=false]{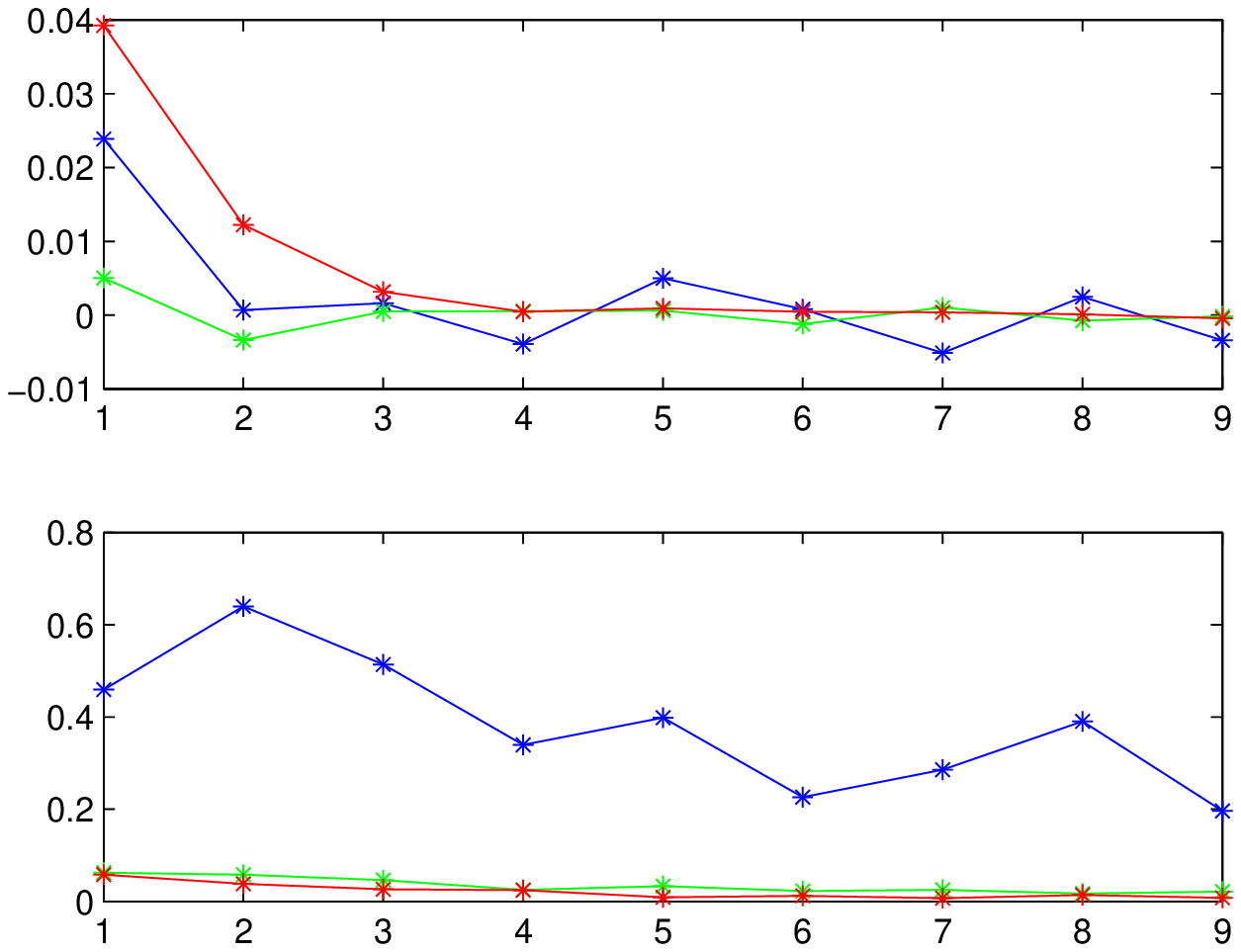}}
\caption{Asian call option. $\epsilon=5\times 10^{-3}$. Other
details as in Figure \ref{fig:17}.} \label{fig:18}
\end{figure}

\begin{figure}[t!]
\centering
\subfigure[Price]{\includegraphics[width=3.1in,height=2.5in,keepaspectratio=false]{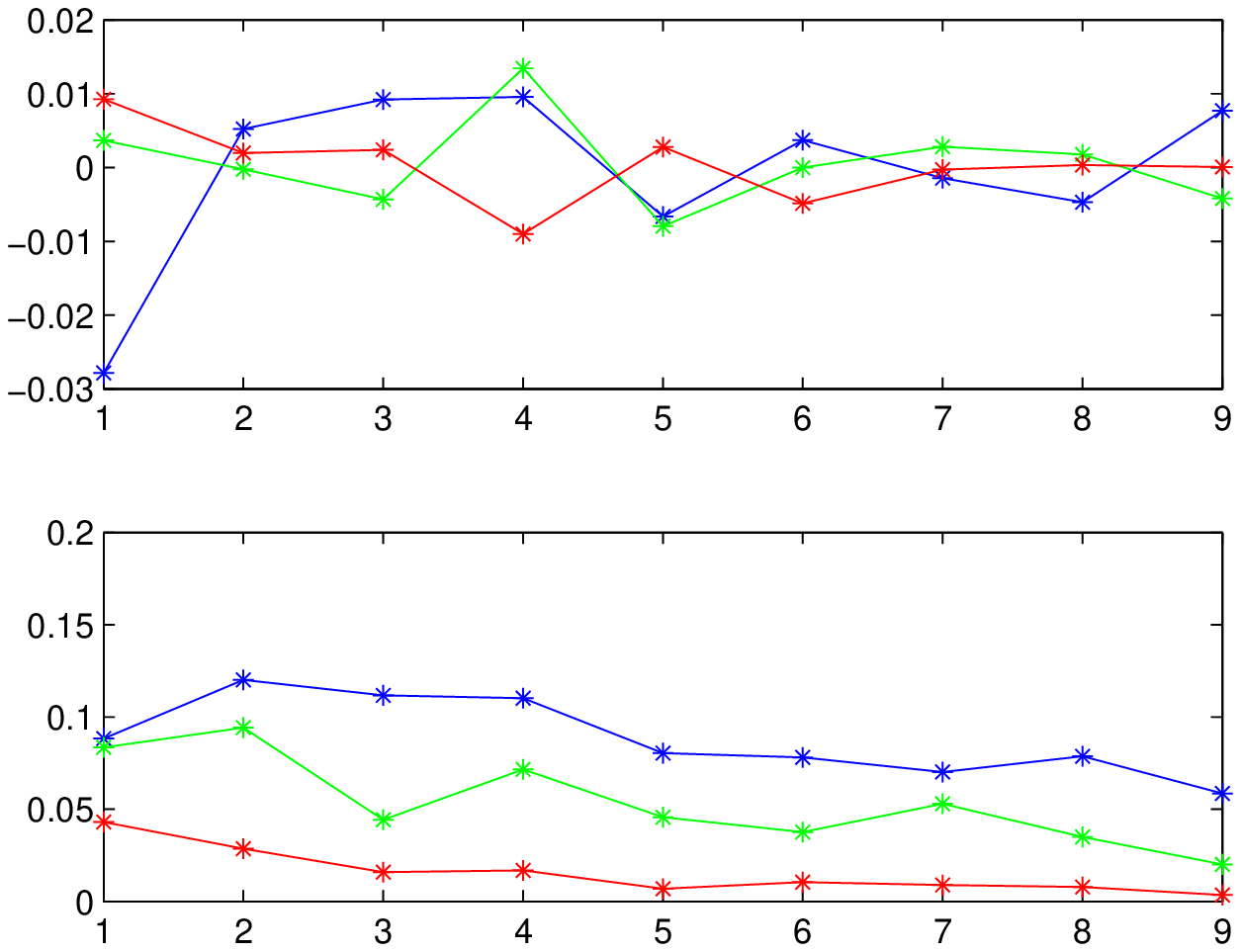}}
\subfigure[Delta]{\includegraphics[width=3.1in,height=2.5in,keepaspectratio=false]{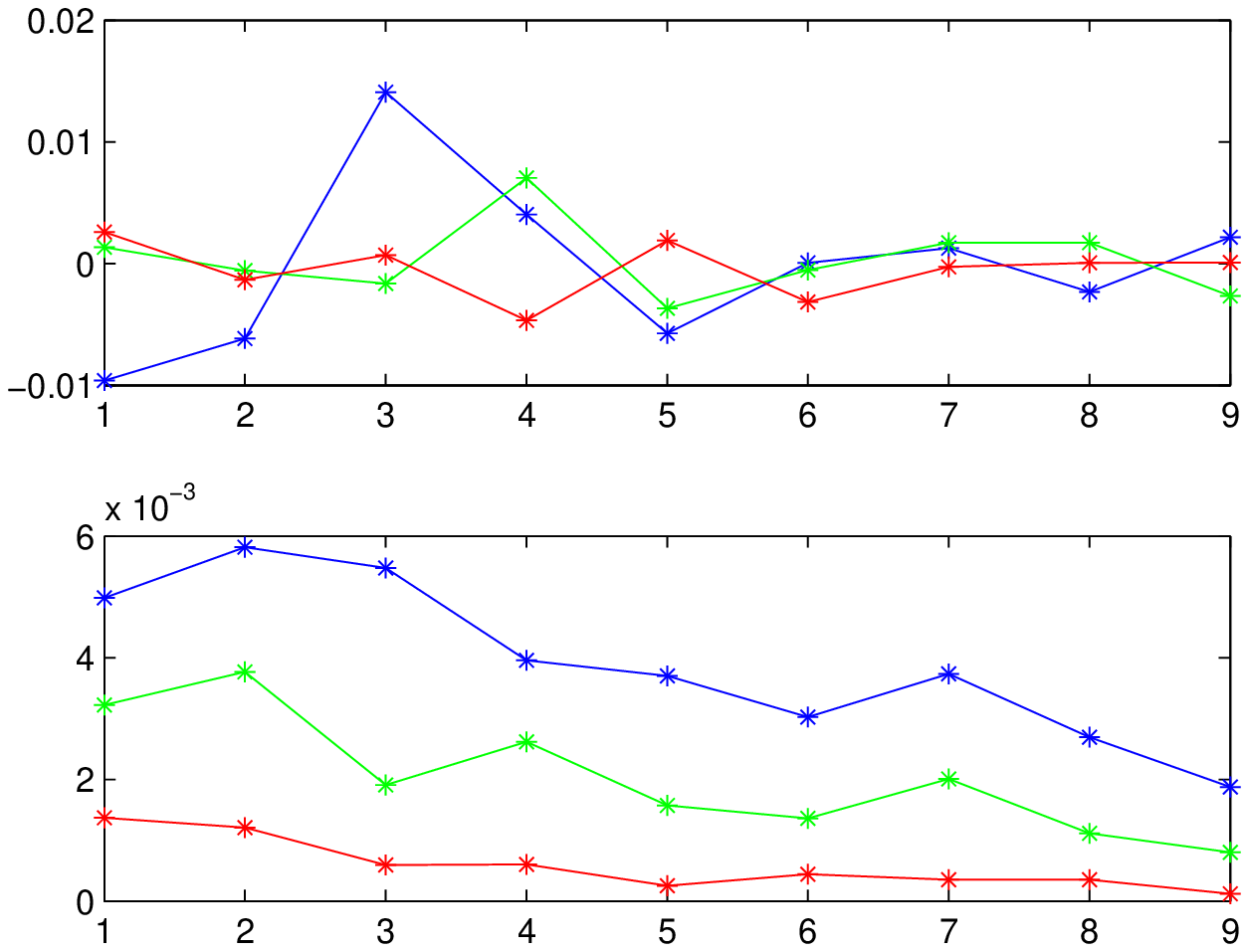}}
\subfigure[Gamma]{\includegraphics[width=3.1in,height=2.5in,keepaspectratio=false]{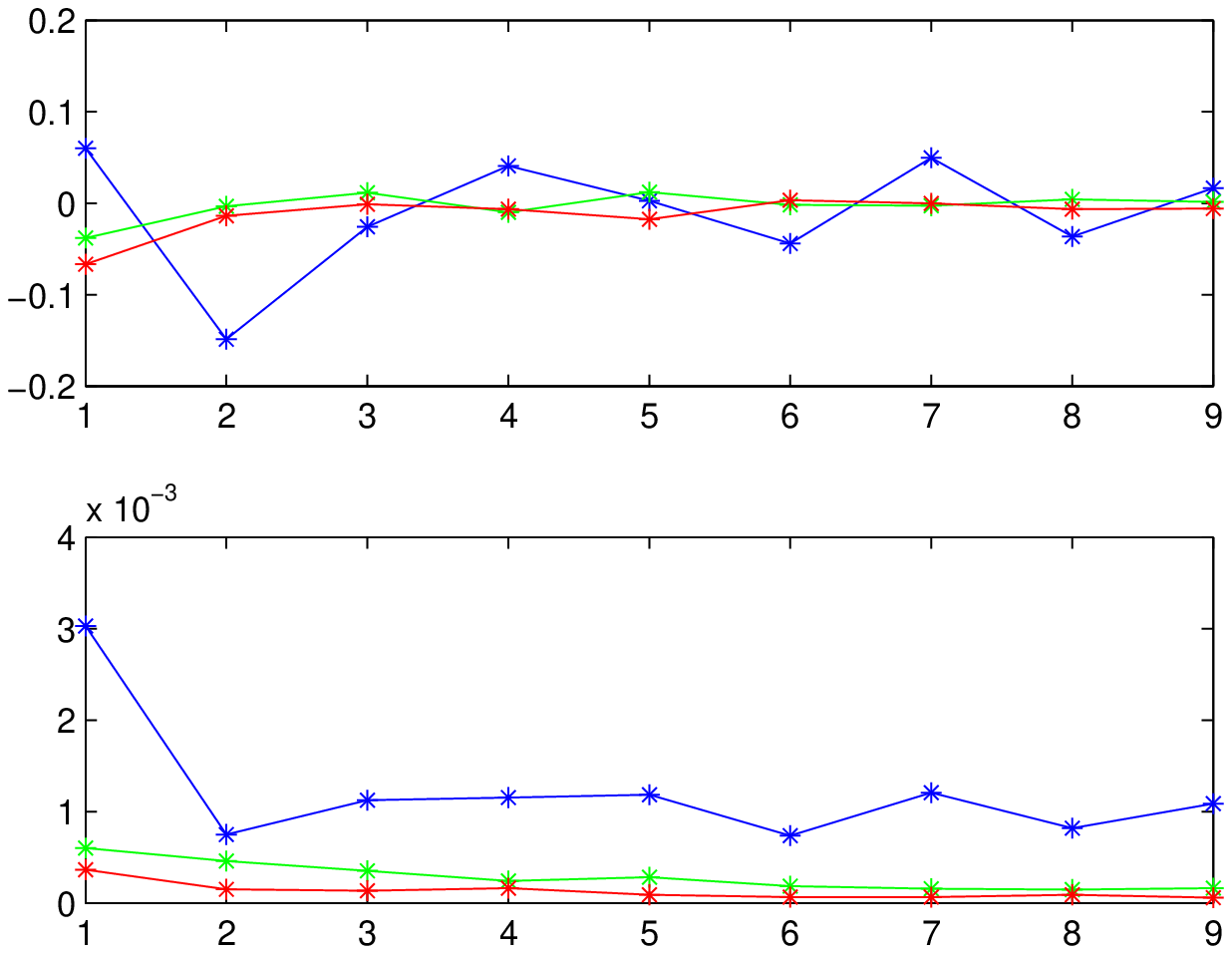}}
\subfigure[Vega]{\includegraphics[width=3.1in,height=2.5in,keepaspectratio=false]{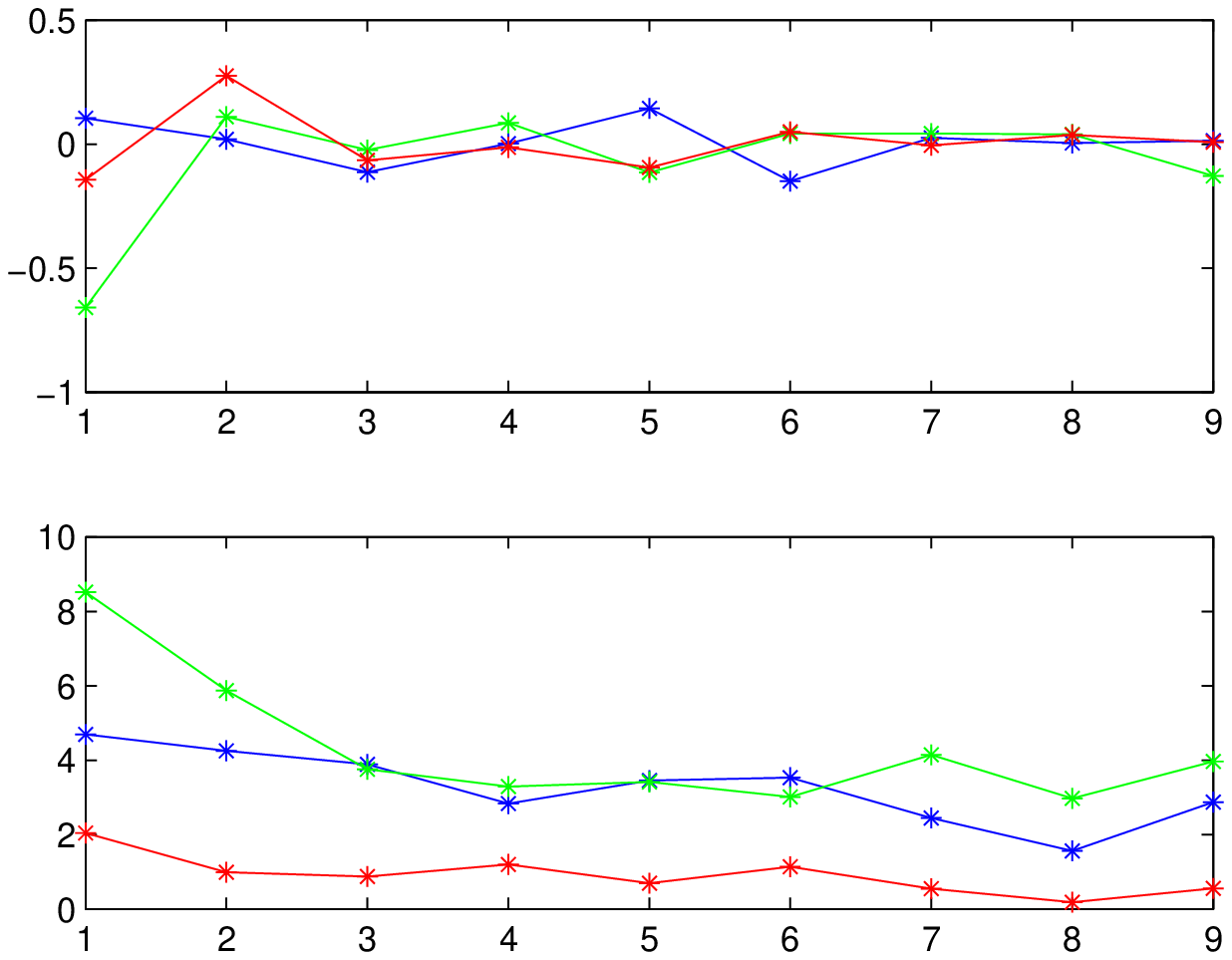}}
\caption{Double Knock-out call option. Details as in Figure
\ref{fig:18}.} \label{fig:19}
\end{figure}

\begin{figure}[t!]
\centering
\subfigure[Price]{\includegraphics[width=3.1in,height=2.5in,keepaspectratio=false]{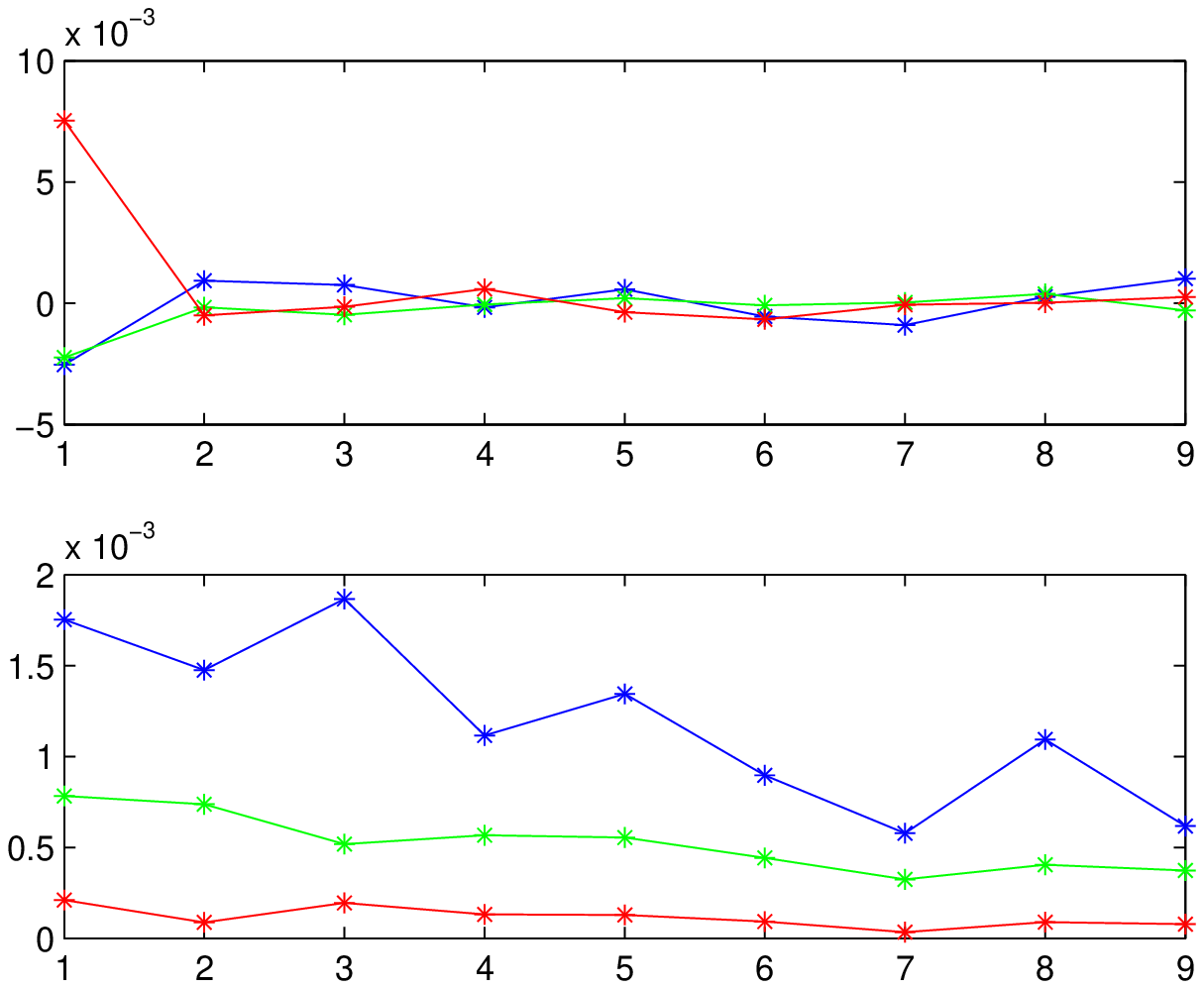}}
\subfigure[Vega]{\includegraphics[width=3.1in,height=2.5in,keepaspectratio=false]{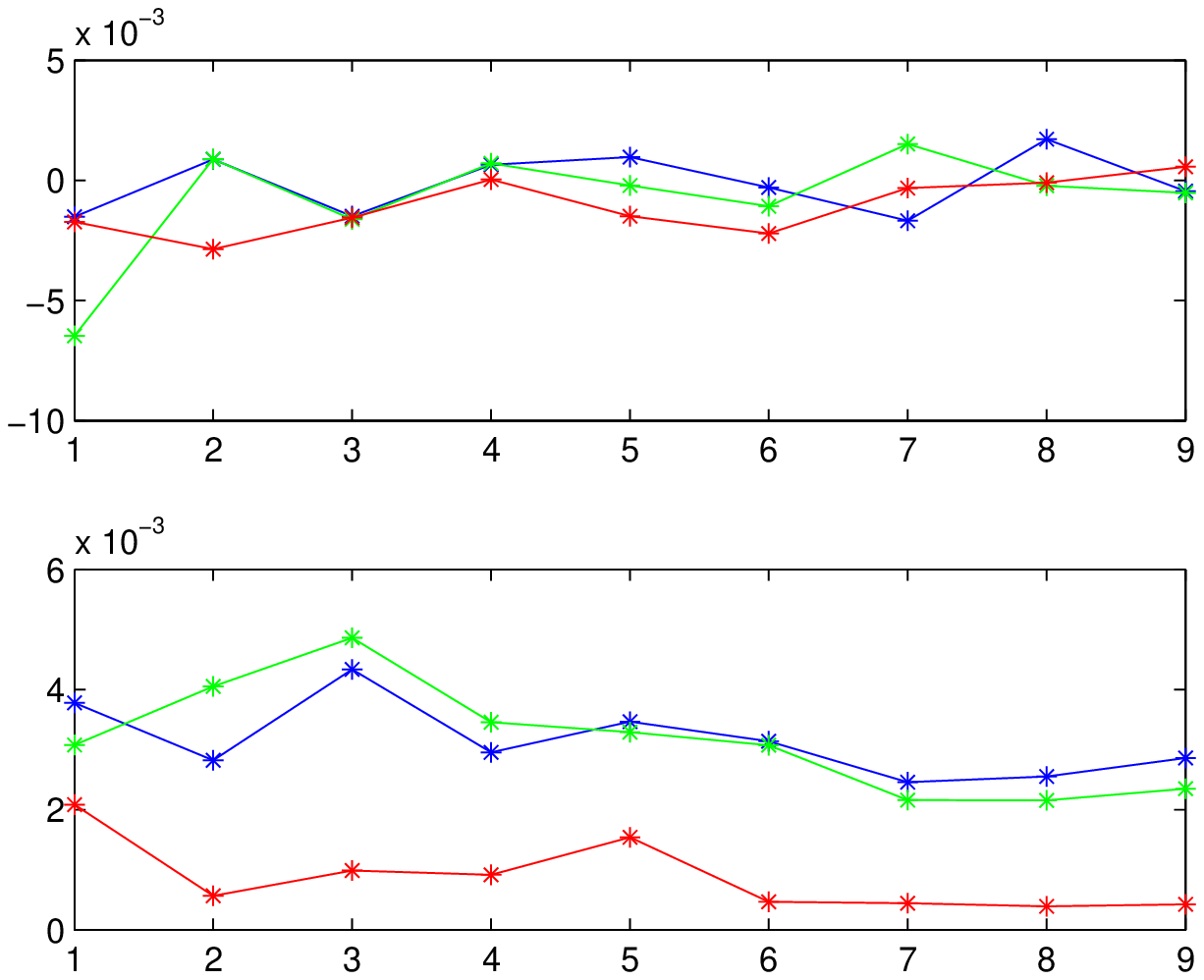}}
\caption{Cliquet option. QMC and rQMC with SD were used here.
Other details as in Figure \ref{fig:18}.} \label{fig:20}
\end{figure}

We have already observed that QMC convergence
is often smoother than MC (see Figures \ref{fig:9}-\ref{fig:12}):
such monotonicity and stability guarantee better convergence for a
given number of paths $N$. In order to quantify monotonicity and
stability of the various numerical techniques, the following
strategy is used: we divide the range of path simulations $N$ in
10 windows of equal length, and we compute the sample mean $m_i$
and the sample standard deviation (``volatility'') $s_i$ for each
window $i$. Then, log-returns $\log(m_i/m_{i-1})$ and volatilities
$s_i$, for $i=2,\ldots,10$, are used as measures of, respectively,
monotonicity and stability: ``monotonic'' convergence will show
non oscillating log-returns converging to zero, ``stable''
convergence will show low and almost flat volatility. We performed
stability analysis for MC and QMC methods. For QMC we used two
different generators: pure QMC with \texttt{Broda} generator and
randomized Quasi Monte Carlo (rQMC) with \texttt{Matlab}
generator\footnote{Matlab Function \texttt{sobolset} with the
\texttt{MatousekAffineOwen} scrambling method was used.}. The
results are shown in Figures \ref{fig:17}-\ref{fig:20}.
\begin{figure}[t!]
\centering
\subfigure[Price]{\includegraphics[width=3.1in,height=2.5in,keepaspectratio=false]{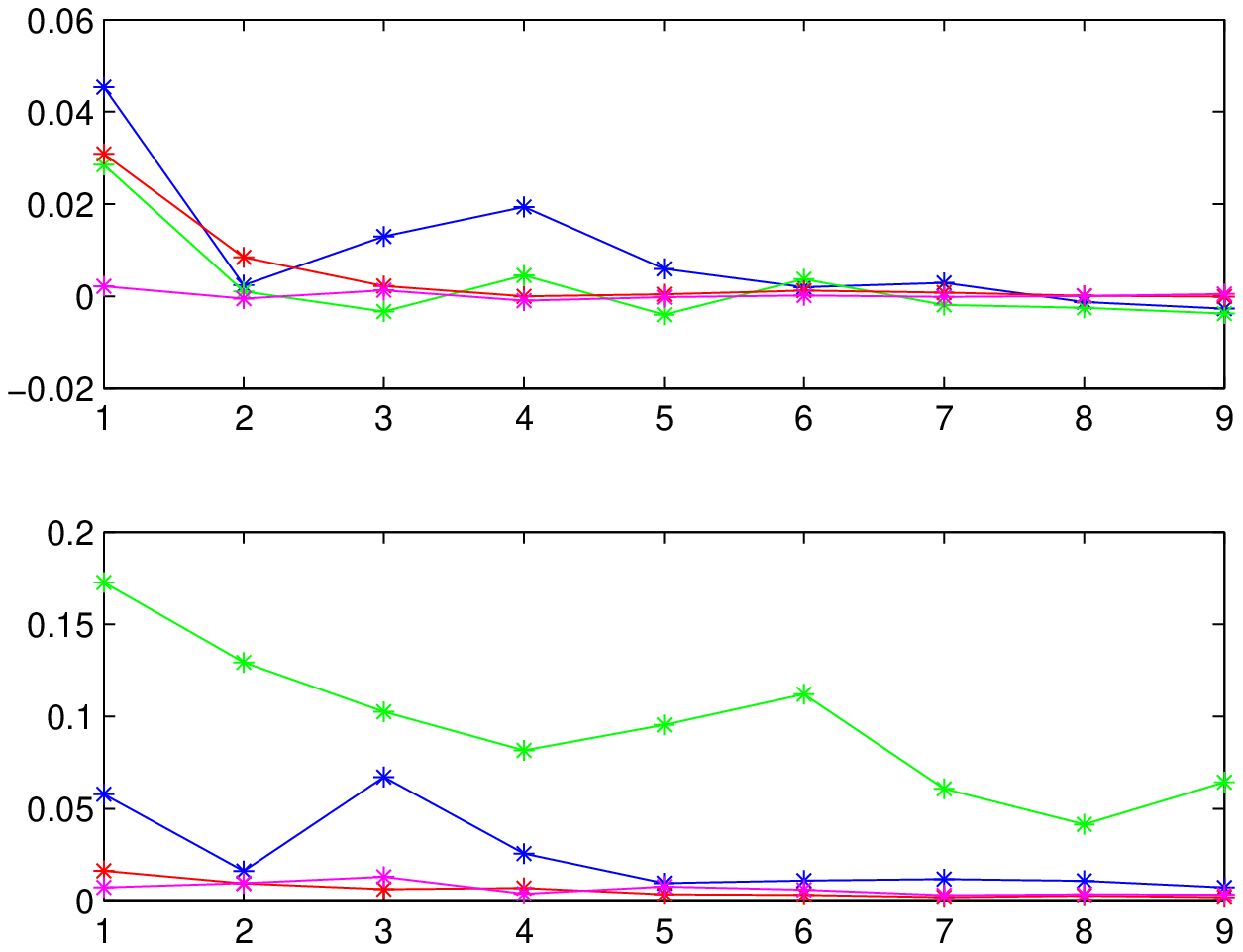}}
\subfigure[Delta]{\includegraphics[width=3.1in,height=2.5in,keepaspectratio=false]{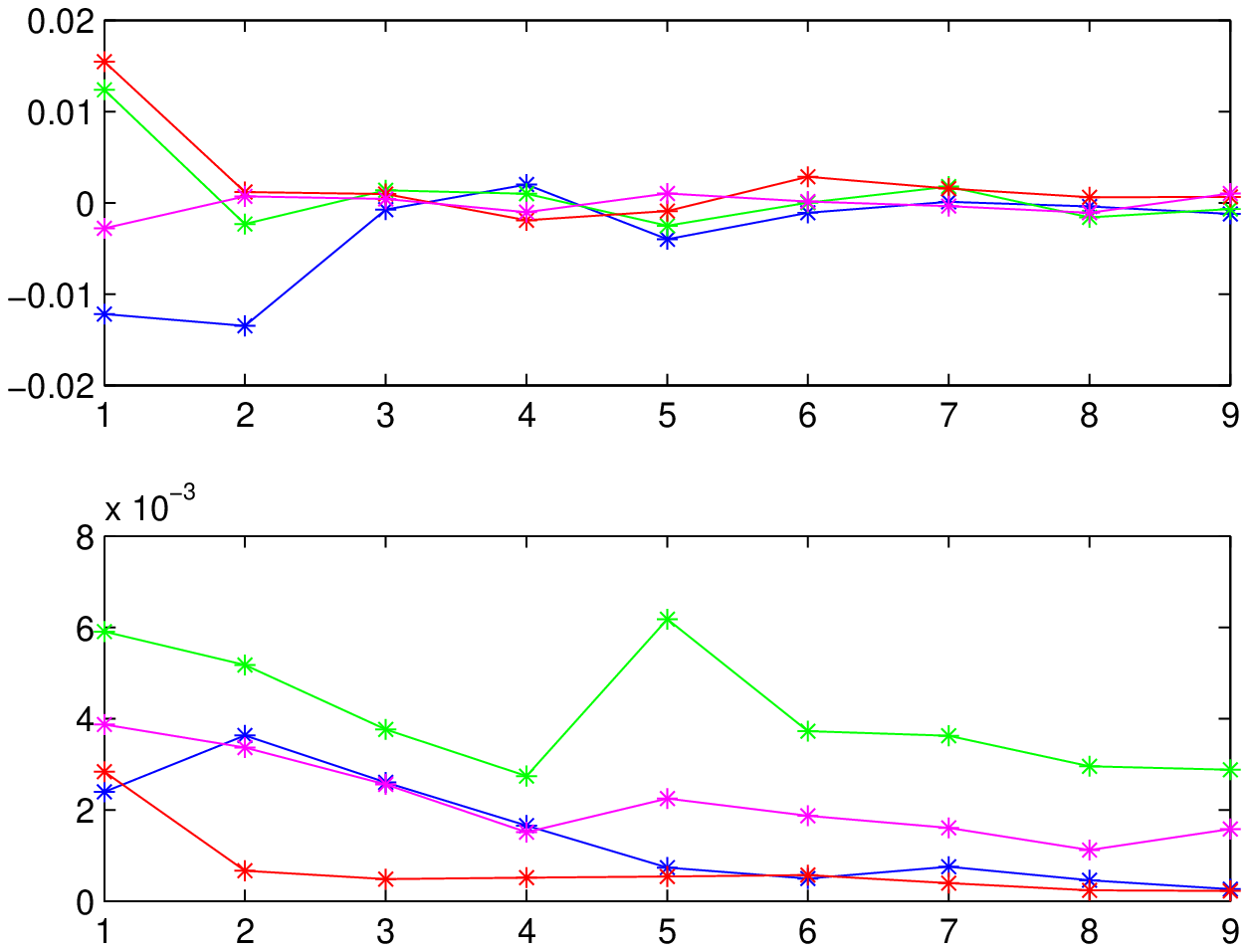}}
\subfigure[Gamma]{\includegraphics[width=3.1in,height=2.5in,keepaspectratio=false]{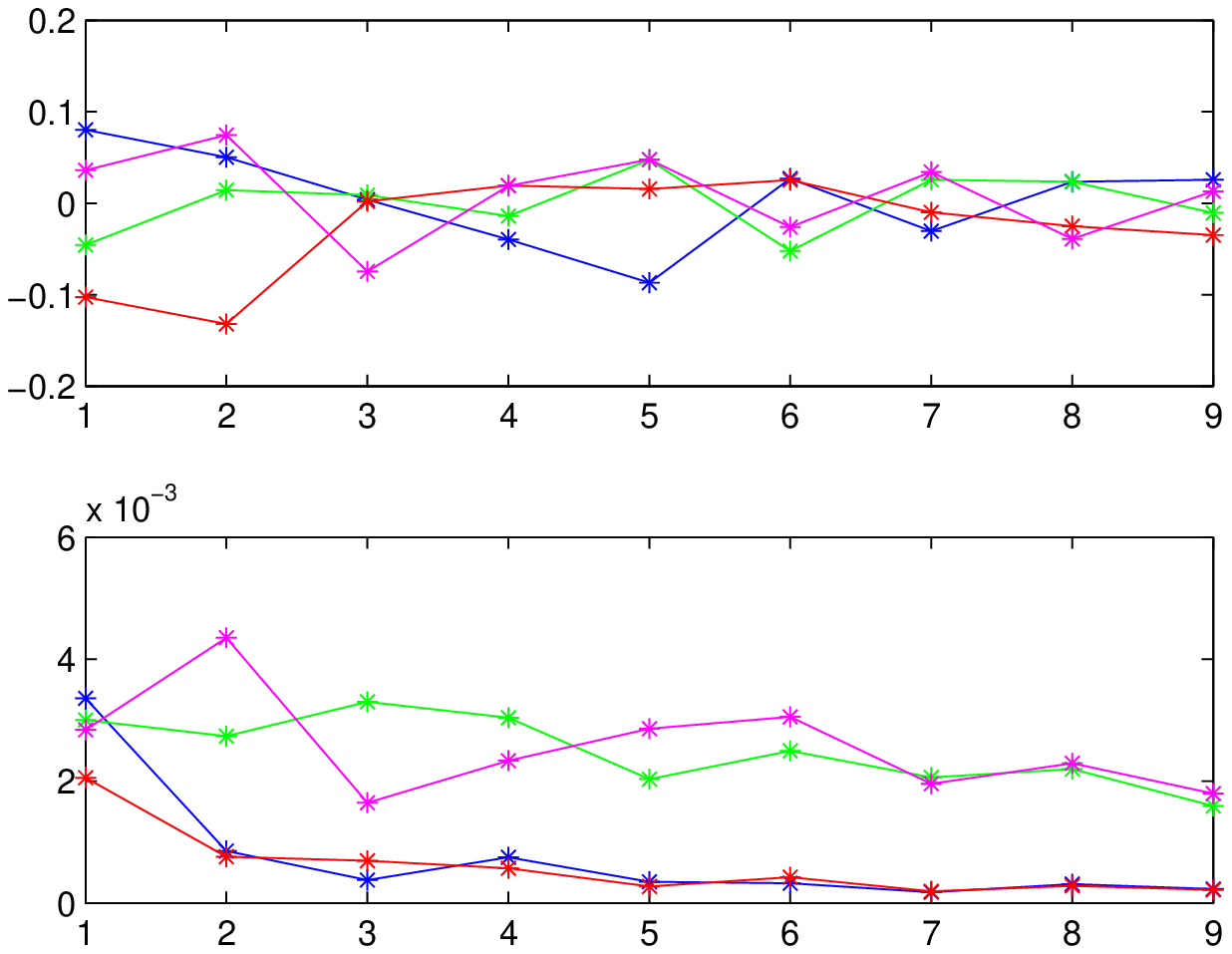}}
\subfigure[Vega]{\includegraphics[width=3.1in,height=2.5in,keepaspectratio=false]{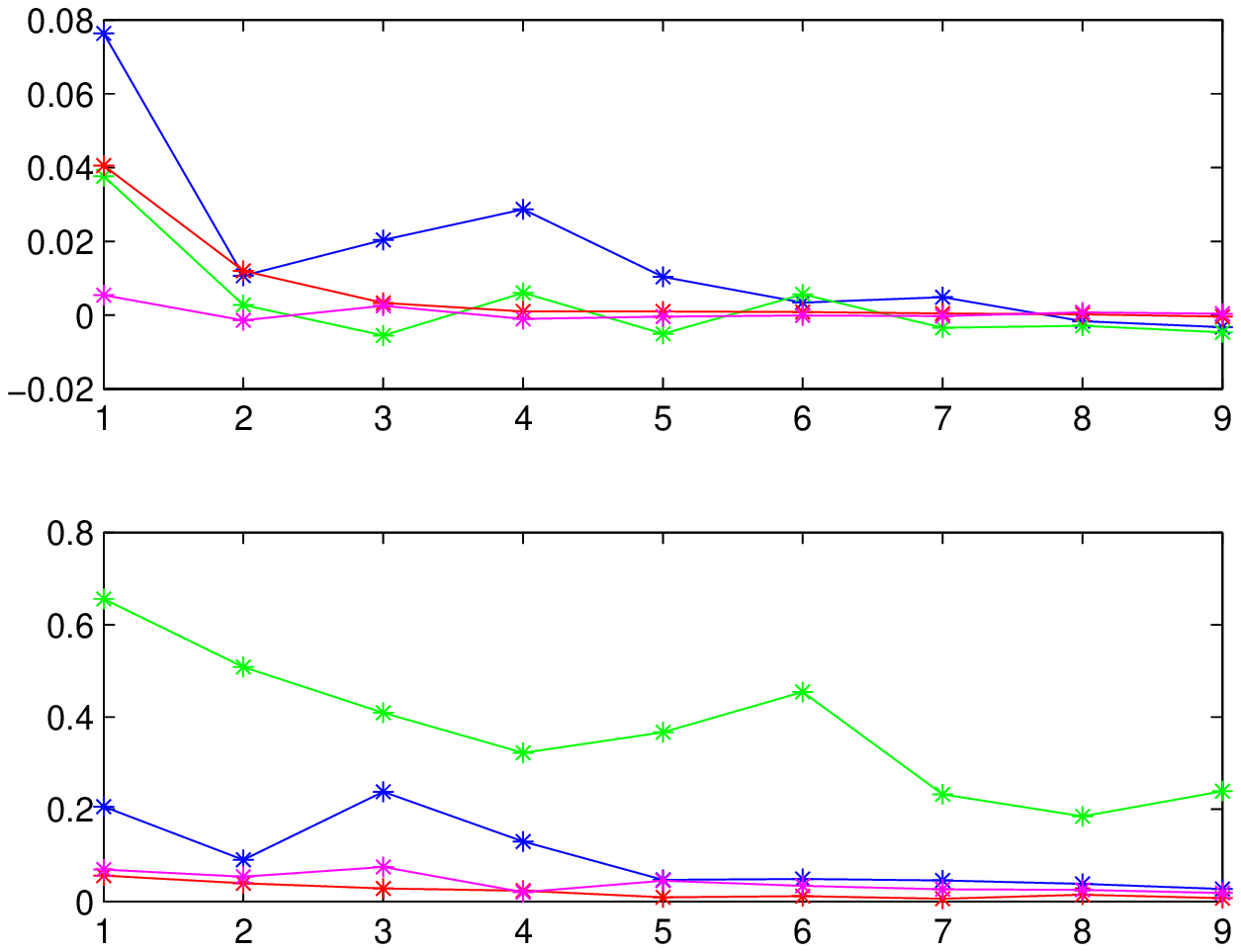}}
\caption{Asian call option with $D=252$, $\epsilon=5\times 10^{-3}$. Results are shown for rQMC+SD+\texttt{Matlab} (green) and rQMC+BBD+\texttt{Matlab} (magenta), and QMC+SD+\texttt{Broda} (blue) and QMC+BBD+\texttt{Broda} (red).}
\label{fig:21}
\end{figure}
\clearpage
\par
We observe that, in general, QMC+\texttt{Broda} and rQMC+\texttt{Matlab} are more monotonic and stable than MC+SD. However, this fact is less evident for Asian delta and gamma, where QMC lacks monotonicity and stability \wrt MC, with QMC+BRODA being slightly more stable than rQMC+\texttt{Matlab}. As we know from the results of GSA for this case, higher order interactions are present and the effective dimensions are large (see Table \ref{tab:2}).
\par
In order to understand also the effect of dimension $D$ on monotonicity and stability, we run a similar experiment for an Asian option with $D=252$ fixing dates using both QMC and rQMC with SD and BBD. The results are shown in Figure \ref{fig:21}. We observe that pure QMC with \texttt{Broda} generator preserves monotonicity and stability much more than randomized QMC based on \texttt{Matlab} generator for all cases including delta and gamma, with QMC+BBD+\texttt{Broda} showing the best stability.
It is also interesting to note that the increase in dimension resulted in the decrease in the effective dimensions for the case of the BBD (but not for the SD).
\par
We conclude that good high-dimensional LDS generators are crucial to obtain a smooth monotonic and stable convergence of the Monte Carlo Simulation in high effective dimensional problems.

\section{Conclusions}
\label{SecConclusions}
In this work we presented an updated overview of the application of Quasi Monte Carlo (QMC) and Global Sensitivity Analysis (GSA) methods in Finance, \wrt standard Monte Carlo (MC) methods.
In particular, we considered prices and greeks (delta, gamma, vega) for selected payoffs with increasing degree of complexity and path-dependency (European Call, Geometric Asian Call, Double Barrier Knock-Out, Cliquet options).
We compared standard discretization (SD) vs Brownian bridge discretization (BBD) schemes of the underlying stochastic diffusion process, and different sampling of the underlying distribution using pseudo random vs high dimensional Sobol' low discrepancy sequences.
We applied GSA and we performed detailed and systematic analysis of convergence diagrams, error estimation, performance, speed-up and stability of the different MC and QMC simulations.
\par
The GSA results in Section \ref{SecGSAresults} revealed that effective dimensions associated to QMC+BBD simulations are generally lower than those associated to MC+SD simulations, and how much such dimension reduction acts for different payoffs and greeks (Figures \ref{fig:1}-\ref{fig:8} and Tables \ref{tab:1}-\ref{tab:2}).
Effective dimensions, being linked with the structure of ANOVA decompositions (the number of important inputs, importance of high order interactions) fully explained the superior efficiency of QMC+BBD due to the specifics of Sobol' sequences and BBD.
The BBD is generally more efficient than SD, but with some exceptions, Cliquet options in particular.
\par
The performance analysis results in Section \ref{SecPerformance} showed that QMC+BBD outperforms MC+SD in most cases, showing faster and more stable convergence to exact or almost exact results (Figures \ref{fig:9}-\ref{fig:12}, \ref{fig:13}-\ref{fig:16}, and Tables \ref{tab:3}-\ref{tab:4}), with some exceptions such as Asian option gamma where all methods showed similar convergence properties.
\par
The speed-up analysis results in Section \ref{SecSpeedUp} confirmed that the superior performance of QMC+BBD allows significative reduction of the number of scenarios to achieve a given accuracy, leading to significative reduction of computational effort (Table \ref{tab:5}). The size of the reduction scales up to $10^3$ (European and Double KO gamma), with a few exceptions (Asian delta and gamma, Cliquet).
\par
Finally, the stability analysis results in \ref{SecStability} confirmed that QMC+BBD simulations are generally more stable and monotonic than MC+SD, with the exception of Asian delta and gamma (Figures \ref{fig:17}-\ref{fig:21}).
\par
We conclude that the methodology presented in this paper, based on
Quasi Monte Carlo, high dimensional Sobol' low discrepancy
generators, efficient discretization schemes, global sensitivity
analysis, detailed convergence diagrams, error estimation,
performance, speed-up and stability analysis, is a very promising
technique for more complex problems in finance, in particular,
credit/debt/funding/capital valuation adjustments
(CVA/DVA/FVA/KVA) and market and counterparty risk
measures\footnote{Some of these metrics, such as EPE/ENE or
expected shortfall, are defined as means or conditional means,
while some other metrics, such as VaR or PFE, are defined as
quantiles of appropriate distributions.}, based on
multi-dimensional, multi-step Monte Carlo simulations of large
portfolios of trades. Such simulations can run, in typical real
cases, $\sim 10^2$ time simulation steps, $\sim 10^3$ (possibly
correlated) risk factors, $\sim 10^3-10^4$ MC scenarios, $\sim
10^4-10^5$ trades, $60$ years maturity, leading to a nominal
dimensionality of the order $D \sim 10^5$, and to a total of
$10^9-10^{11}$ evaluations. Unfortunately, a fraction $\sim 1\%$
of exotic trades may require distinct MC simulations for their
evaluation, nesting another set of $\sim 10^3-10^5$ MC scenarios,
thus leading up to $10^{14}$ evaluations. Finally, hedging
CVA/DVA/FVA/KVA valuation adjustments \wrt to their underlying
risk factors (typically credit/funding curves) also requires the
computation of their corresponding greeks \wrt each term structure
node, adding another $\sim 10^2$ simulations. This is the reason
why the industry is continuously looking for advanced techniques
to reduce computational times: grid computing, GPU computing,
adjoint algorithmic differentiation (AAD), etc. (see \eg
\cite{She15a}).
\par
We argue that, using QMC sampling (instead of MC) to generate the
scenarios of the underlying risk factors and to price exotic
trades may significantly improve the accuracy, the performance and
the stability of such monster-simulations, as shown by preliminary
results on real portfolios \cite{BiaKuc14}. Furthermore, GSA
should suggest how to order the risk factors according to their
relative importance, thus reducing the effective dimensionality.
Such applications will need further research.

\begin{appendices}
\section{Error Optimization in Finite Difference Approximation}
\label{App:GreekErr} There are two contributions to the root mean
square error when greeks are computed by MC/QMC simulation via
finite differences: variance and bias \cite{Gla03}. The first
source of uncertainty comes from the fact that we are computing
prices through simulation over a finite number of scenarios, while
the latter is due to the approximation of a derivative with a
finite difference. In order to minimize the variance, we use the
same set of (quasi)random numbers for the computation of
$V(\theta)$, $V(\theta+h)$ and $V(\theta-h)$, where $V$ is the
option price, the parameter $\theta$ is the spot for delta and
gamma or the volatility for vega and $h$ is the increment on
$\theta$. In order to minimize the bias of the finite differences
we use central differences, so that it is of the order $h^2$. The
increments $h$ are chosen to be $h=\epsilon S_0$, for $\Delta$ and
$\Gamma$, and $h=\epsilon$, for $\mathcal{V}$, for a given ``shift
parameter'' $\epsilon$. The choice of the appropriate $\epsilon$
is guided by the following considerations. The MC/QMC root mean
square error estimate of finite differences is given by
\cite{Gla03}:
\begin{equation}\label{RMSEGreeks}
\varepsilon=\sqrt{\frac{c}{N^{2\alpha}h^{\beta}}+b^2h^4}\, .
\end{equation}
The first term in the square root is a ``statistical'' error
related to the variance $c$. It depends on $N$ as well as on
$\epsilon$. $\alpha=0.5$ for MC and, usually, $0.5<\alpha<1$ for
QMC, while $\beta=1$ for first derivatives and $\beta=3$ for
second derivatives. The second term is the systematic error due to
the bias of finite differences: it is independent of $N$ but it
depends on $\epsilon$. The constant $b$ is given by
$b=\frac{1}{6}\frac{\partial^3 V}{\partial \theta^3}(\theta)$ for
central differences of the first order (delta and vega) and
$b=\frac{1}{12}\frac{\partial^4 V}{\partial \theta^4}(\theta)$ for
central differences of the second order (gamma). One can see that,
when $h$ is decreasing, the bias term decreases as well while the variance term
increases, therefore we fine tune $h$ in such a way that the
variance term is not too high in the relevant range for $N$, while
the bias term remains negligible so that (\ref{RMSEGreeks})
follows approximately a power law. We note that the optimal value of $h$
is not observed to vary too much with $N$ in the range used for
our tests. Indeed, it can be computed analytically from
(\ref{RMSEGreeks}) as the minimum of $\varepsilon$:
\begin{equation}
h_N = \left(\frac{\beta
c}{4b^2N^{2\alpha}}\right)^{\frac{1}{\beta+4}}\, .
\end{equation}
We see that the powers $\frac{1}{5}$ and $\frac{1}{7}$
(corresponding to $\beta=1$ and $\beta=3$ respectively) largely
flatten $h_N$ as a function of $N$.

\section{Speed-Up Computation}
\label{App:SU}
We identify the number of scenarios $N_*^{(i)}(a)$ in eq. (\ref{EqSpeedUp}) using the \textit{i-th} computational method needed to reach and maintain a given accuracy $a$ as the first number of simulated paths such that, for any $N>N_*$
\begin{equation}
\label{Nstardef}
V-a\leq V_N\pm 3\, \varepsilon\leq V+a\, ,
\end{equation}
where $V$ and $V_N$ are respectively the exact and simulated
values of prices or greeks and $\varepsilon$ is the standard
error. The threshold $N_*$ could be evaluated through direct
simulation, but this would be extremely computationally expensive.
Extracting $N_*$ from plots defined by (\ref{err:QMC}) can't be
applied directly because, in the case of greeks, such plots are
correct only for a limited range of values of $N$,  \ie as long as
the bias term in (\ref{RMSEGreeks}) does not become dominant.
Extrapolating $N_*$ from plots to high values of $N$ is necessary
to compute speed-up, but the relation between RMSE and $N$ would
not be linear anymore. We therefore follow a different procedure
to determine $N_*$. Equation (\ref{RMSEGreeks}) can be rewritten
as
\begin{equation}
\log\varepsilon=k-\alpha\log N,
\end{equation}
where $k=\frac{1}{2}\log\frac{c}{h^\beta}$ and $\alpha$ are, respectively, the intercept and the slope computed from linear regressions on $\varepsilon_N$ given by (\ref{error}). Therefore, $N_*$ is found by imposing
\begin{equation}
a=3\,\sqrt{\frac{e^{2k}}{N_*^{2\alpha}}+b^2h^4},
\end{equation}
and is given by
\begin{equation}
\label{Nstar}
N_*(h,a) = \left(\frac{9\,e^{2k_h}}{a^2-9b^2h^4}\right)^{\frac{1}{2\alpha_h}}.
\end{equation}
We have written $k_h$ and $\alpha_h$ in order to stress that they
also depend on the choice of $h$ made while carrying out the tests
in Section \ref{SecRes}: this dependence on $h$ can be stronger
than the explicit dependence in (\ref{Nstar}). Constant $b$ is
computed from derivatives of $V$ (see discussion after presenting
equation (\ref{RMSEGreeks}) ); $k$ and $\alpha$ are the intercepts
and slopes correspondingly taken from plots (Figures
\ref{fig:13}-\ref{fig:16}): this is possible, since these plots
are obtained for a range of $N$ such that the second term in
(\ref{RMSEGreeks}) is negligible. It is clear that the domain of
$N_*$ is limited to $a>3bh^2$. In the case of prices, equation
(\ref{Nstar}) simplifies to
\begin{equation}
N_*(a) = \left(\frac{3e^{k_h}}{a}\right)^{\frac{1}{\alpha_h}}\, .
\end{equation}

\end{appendices}

\bibliographystyle{alpha}
\bibliography{FinanceBibliography}

\end{document}